%% file: draft_WR_v4.tex
\documentclass[longauth,traditabstract]{aa}
\usepackage{graphicx}
\usepackage{natbib}
\usepackage{enumerate}  %enumerate with capital roman numbers
\usepackage{amssymb}
\usepackage{longtable,lscape}
\usepackage{rotating}
\usepackage{textcomp}
\usepackage{balance}
\usepackage{placeins}
\usepackage{amsmath}
\usepackage{txfonts}
\usepackage{color}
\usepackage[bf]{caption}
\usepackage{afterpage}
\usepackage{systeme}

\input{journals}

\newcommand{\ariv}{[Ar{\footnotesize{IV}}]~}

\newcommand{\hiiexplorer}{HII{\sc{explorer}}~}
\newcommand{\bpass}{{\small{BPASS}}~}
\newcommand{\cloudy}{{\small{CLOUDY}}~}
\newcommand{\hb}{H$\beta$~}
\newcommand{\ha}{H$\alpha$~}

\newcommand{\msun}{M$_{\odot}$~}
\newcommand{\onespace}{\hspace{-3pt}}
\newcommand{\starlight}{{\small{STARLIGHT}}~}

\newcommand{\popstar}{{\small{POPSTAR}}~}

\newcommand{\zsun}{$Z_{\odot}$~}
\newcommand{\reff}{$r_{\mathrm{eff}}$~}

\DeclareRobustCommand{\ion}[2]{%
\relax\ifmmode
\ifx\testbx\f@series
{\mathbf{#1\,\mathsc{#2}}}\else
{\mathrm{#1\,\mathsc{#2}}}\fi
\else\textup{#1\,{\mdseries\textsc{#2}}}%
\fi}

\newcommand{\ciii}{\ion{C}{iii}~}
\newcommand{\civ}{\ion{C}{iv}~}
\newcommand{\feiii}{[\ion{Fe}{iii}]~}

\newcommand{\hii}{\ion{H}{ii}~}
\newcommand{\hh}{\ion{H}{ii}~}
\newcommand{\hei}{\ion{He}{i}~}
\newcommand{\heii}{\ion{He}{ii}~}
\newcommand{\nii}{[\ion{N}{ii}]}
\newcommand{\neiv}{[\ion{Ne}{iv}]~}
\newcommand{\niii}{\ion{N}{iii}~}
\newcommand{\nv}{\ion{N}{v}~}

\newcommand{\oii}{[\ion{O}{ii}]}
\newcommand{\oiii}{[\ion{O}{iii}]}
\newcommand{\ovi}{\ion{O}{vi}~}

\makeatletter
\newcounter{subsubsubsection}[subsubsection]
\renewcommand\thesubsubsubsection{\thesubsubsection .\@arabic\c@subsubsubsection}
\newcommand\subsubsubsection{\@startsection{subsubsubsection}{4}{\z@}%
                                     {-3.25ex\@plus -1ex \@minus -.2ex}%
                                     {1.5ex \@plus .2ex}%
                                     {\normalfont\normalsize}}
\newcommand*\l@subsubsubsection{\@dottedtocline{3}{10.0em}{4.1em}}
\newcommand*{\subsubsubsectionmark}[1]{}
\makeatother

\begin{document}
 
\title{
%% First survey of Wolf-Rayet stars in the full extension of galaxies using CALIFA
%% populations in galaxies observed with integral field spectroscopy}
% 1) Exploring the spatial distribution of Wolf-Rayet stars in external galaxies from the CALIFA survey
First survey of Wolf-Rayet star populations over the full %\\
 extension of nearby galaxies observed with CALIFA\thanks{Based on
  observations collected at the Centro Astron\'omico Hispano-Alem\'an (CAHA) at
  Calar Alto, operated jointly by the Max-Planck Institut f\"ur Astronomie and
  the Instituto de Astrof\'isica de Andaluc\'ia (CSIC).}}

\authorrunning{Miralles-Caballero et al.}

\titlerunning{First survey of WR-rich regions in nearby galaxies using CALIFA}

\author{D. Miralles-Caballero \inst{1}\thanks{\email{thorius@gmail.com}}
  \and  
  A. I. D\'iaz \inst{\ref{uam}}
  \and 
  \'A. R. L\'opez-S\'anchez \inst{\ref{aao},\ref{macq}}
  \and
  F.\,F. Rosales-Ortega \inst{\ref{inaoe}}
  \and
  A. Monreal-Ibero \inst{\ref{oparis}}
  \and
  E. P\'erez-Montero \inst{\ref{iaa}}
  \and
  C. Kehrig \inst{\ref{iaa}}
  \and
  R. Garc\'ia-Benito\inst{\ref{iaa}}
  \and
  S.\,F. S\'anchez \inst{\ref{unam}, \ref{iaa}}
  \and
  C.\,J. Walcher \inst{\ref{aip}}
  \and 
  L. Galbany \inst{\ref{mill}, \ref{chile}}
  \and
  J. Iglesias-P\'aramo \inst{\ref{iaa}}
  \and
  J. M. V\'ilchez \inst{\ref{iaa}}
  \and
  R. M. Gonz\'alez Delgado \inst{\ref{iaa}}
  %
  % CALIFA-Extension and technical team
  %
  \and
  G. van de Ven \inst{\ref{mpia}}
  \and
  J. Barrera-Ballesteros \inst{\ref{iac}}
  \and
  M. Lyubenova \inst{\ref{mpia}}
  \and
  S. Meidt \inst{\ref{porto}}
  \and
  J. Falcon-Barroso \inst{\ref{iac}}
  \and
  D. Mast\inst{\ref{caha},\ref{iaa}}
  \and
  M. A. Mendoza\inst{\ref{iaa}}
  \and
  the CALIFA collaboration.}
\institute{
  \label{uam}Departamento de F\'isica Te\'orica, Universidad Aut\'onoma de Madrid, 28049 Madrid, Spain.
  \and
  \label{aao}Australian Astronomical Observatory, PO Box 915, North Ryde, NSW 1670, Australia.
  \and
  \label{macq}Department of Physics and Astronomy, Macquarie University, NSW 2109, Australia.
  \and
  \label{inaoe}Instituto Nacional de Astrof{\'i}sica, {\'O}ptica y Electr{\'o}nica, Luis E. Erro 1, 72840 Tonantzintla, Puebla, Mexico.
  \and
  \label{oparis}GEPI, Observatoire de Paris, CNRS, Universit\'e Paris-Diderot, Place Jules Janssen, 92190 Meudon, France.
  \and
  \label{iaa}Instituto de Astrof\'isica de Andaluc\'ia (CSIC), Glorieta de la Astronom\'ia s/n, Aptdo. 3004, E18080-Granada, Spain.
  \and
  \label{unam}Instituto de Astronom\'i a,Universidad Nacional Auton\'oma de Mexico, A.P. 70-264, 04510, M\'exico, D.F.
  \and
  \label{aip}Leibniz-Institut f\"ur Astrophysik Potsdam (AIP), An der Sternwarte 16, D-14482 Potsdam, Germany.
  \and
%  \label{centra}CENTRA - Instituto Superior Tecnico, Av. Rovisco Pais, 1, 1049-001 Lisbon, Portugal. 
  \label{mill}Millennium Institute of Astrophysics MAS, Nuncio Monse\~nor S\'otero Sanz 100, Providencia, 7500011 Santiago, Chile.
  \and
  \label{chile}Departamento de Astronom\'ia, Universidad de Chile, Camino El Observatorio 1515, Las Condes, Santiago, Chile.
  \and
  \label{mpia}Max Planck Institute for Astronomy, K\"onigstuhl 17, 69117 Heidelberg, Germany.
  \and
  \label{iac}Instituto de Astrof\'isica de Canarias (IAC), E-38205 La Laguna, Tenerife, Spain 
  \and
  \label{porto}Centro de Astrof\'isica and Faculdade de Ciencias, Universidade do Porto, Rua das Estrelas, 4150-762 Porto, Portugal.
  \and 
  \label{caha}Centro Astron\'omico Hispano Alem\'an, Calar Alto, (CSIC-MPG), C/Jes\'{u}s Durb\'{a}n Rem\'{o}n 2-2, E-04004 Almer\'{\i}a, Spain.
}

%\date{Received \ldots ; accepted \ldots}

%\abstract{}{}{}{}{} 
\abstract{
%Building large catalogues of 
The search of extragalactic regions with conspicuous presence of
Wolf-Rayet (WR) stars outside the Local Group is challenging task due to the
difficulties in detecting their faint spectral features. 
In this exploratory work, we develop a methodology to perform an automated
search of WR signatures through a pixel-by-pixel analysis of integral field
spectroscopy (IFS) data belonging to the \emph{Calar Alto Legacy Integral Field Area} survey, CALIFA.
This procedure has been applied to a sample of nearby galaxies spanning a
wide range of physical, morphological and environmental properties. 
This technique allowed us to build the first catalogue of Wolf-Rayet rich
regions with spatially-resolved information, allowing to study the properties of
these complexes in a 2D context. 
%% Here we present the resulting catalogue of a systematic search for WR
%% features in galaxies , observed with integral field spectroscopy (IFS). These
%% data allow us to search spatially the location of  WR stars within galaxies
%% and therefore to study the properties of these WR-rich regions
The detection technique is based on the identification of
the blue WR bump (around \heii~$\lambda$4686 \AA, mainly associated to nitrogen-rich
WR stars, WN) and the red WR bump (around \ion{C}{iv}~$\lambda$5808 \AA\ and associated to
carbon-rich WR stars, WC) using a pixel-by-pixel analysis, which
maximizes the number of independent regions within a given galaxy.
We identified 44 WR-rich regions with blue bumps distributed in 25 galaxies of a
total of 558. The red WR bump was identified only in 5 of those regions. 
%% The oxygen abundance of the ionised gas of the WR regions ranges between 0.25\,\zsun and
%% \zsun. 
Most of the WR regions are located within one effective radius from the
galaxy centre, and around 1/3 are located within  \mbox{$\sim$1~kpc} or less
from the centre. 
We found that the majority of the galaxies hosting WR populations in our sample are
involved in some kind of interaction process. Half of the host galaxies share
some properties with gamma-ray burst (GRB) hosts where WR stars, as potential
candidates to being the progenitors of GRBs, are found.
%% Special attention has been given to dilution effects on the WR features in
%% spatially non-resolved studies or when a large aperture is extracted. We have
%% found that this can dramatically affect the actual fluxes within a factor of two
%% or even larger. Therefore, caution must be taken when subtracting the underlying
%% stellar continuum if the signal-to-noise of the WR features is low.
We also compared the WR properties derived from the CALIFA data with
stellar population synthesis models, and confirm that simple star models are
generally not able to reproduce the observations. We conclude that other
effects, such as the binary star channel (which could extend the WR phase up to
10~Myr), fast rotation or other physical processes that causes the loss of
observed Lyman continuum photons, are very likely affecting the derived WR
properties, and hence should be considered when modelling the evolution of
massive stars.
}


\maketitle

\section{Introduction}

Despite their relatively low number, massive stars dominate the stellar feedback
to the local interstellar medium (ISM) through their stellar winds and
subsequent death as supernovae (SNe). The most massive stars (\mbox{$M \geq 25$
  \msun}for $Z_\odot$) will undergo the Wolf-Rayet (WR) phase, starting 2--3 Myr
after their birth~\citep{Meynet95}. These stars have typical wind densities
which are an order of magnitude higher than massive O stars, and hence play
a key role to the chemical enrichment of galaxies. 
Although evading direct detection, WR stars are likely to be the
progenitors of Type Ib and Type Ic core-collapse SNe, likely linked
to the on-going star-formation \citep{Galbany14}.
These SN types are characterised by neither showing Hydrogen (SNIb) nor
Helium (SNIc) in their spectra, suggesting that the external
layers of the progenitor star have been removed prior to explosion.
Moreover, a small fraction of SNe Ic showing broad-line features in
the spectra (SNIc-bl, $\sim$30,000~km s$^{-1}$) are associated to
long-duration gamma-ray bursts (GRBs,
\citealt{Galama98,Stanek03,Hjorth03,Modjaz06}).
Indeed, WR stars have been suggested to be candidates to being progenitors of long, soft
GRBs in regions of low metallicity~\citep{Woosley06}.

The emission features that characterize the spectra of WR stars are often
observed in extragalactic \hii regions. WR winds are sufficiently dense that an
optical depth of unity in the continuum arises in the outflowing material. The
spectral features are formed far out in the wind and are seen primarily in
emission~\citep{Crowther07}. Generally WR stars are identified
in galaxies  \citep[first renamed as ``WR galaxies'' by][]{Osterbrock82}
whenever their integrated spectra show a broad \heii emission feature centred at
4686\,\AA, the so-called blue WR bump. Some few other broad features, as the red
WR bump around  \ion{C}{iv}~$\lambda$5808, are also used to identify WR stars in
galaxies, but these features are in most cases much fainter than the blue WR
bump. In any case, it is important to note that the term ``WR galaxy'' may be
confusing. Depending on the distance to the observed galaxy and the spatial
resolution and extension of the extracted spectrum, it may well refer to 
extragalactic \hii regions and quite frequently to the nucleus of a powerful
starburst. Galaxies showing a significant population of WR stars have been known
for several decades, beginning with the first  WR detection in the blue compact
dwarf galaxy He 2-10~\citep{Allen76}. Since then a number of the WR galaxies
have been reported, though generally through a serendipitous detection
\citep[e.g.][]{Kunth81,Kunth83,Ho95,Heckman97}.

The investigation of the WR content in galaxies provides important constrains to
stellar evolutionary models. This is particularly important at low metallicities
because only few data of sub-solar WR stars are available. Several studies have
attempted to reproduce the number of WR stars needed to account for the observed
stellar emission features. The disagreement between observations and models on
simple calculations, such as the flux ratio between the blue WR bump and
H$\beta$ or the WR/O number ratio at sub-solar metallicities (which is a
sensitive test of evolutionary models), has lead to the development of
sophisticated models which include rotation \citep{Meynet05} or binary evolution
of massive stars \citep{vanBever03,vanBever07,Eldridge08}. 
Therefore, studies of large samples of galaxies showing WR features, especially
in the intermediate- and low-metallicity regime, are needed to better constrain
such models.

Ideally, detailed, spatially-resolved studies of WR populations would be needed
to fill the gap between early detailed works and large surveys of integrated
properties. Several studies in nearby starbursts
\citep[e.g.][]{Gonzalez-Delgado94,Perez-Montero07b,Perez-Montero10,Lopez-Sanchez10a,Karthick14}
and resolved knots of star formation \citep[e.g.][]{Gonzalez-Delgado95,Castellanos02,Hadfield06} have
been published during the last few years, in conjunction with studies of
individual WR stars in galaxies of the Local Group
\citep[e.g.][]{Massey98,Massey03,Crowther06a,Crowther06b,Neugent12,Hainich14,Sander14}.
However, spatially-resolved and detailed studies of extragalactic regions
showing a significant WR population in large galaxy samples have not been
accomplished yet.

The advent of the integral field spectroscopy (IFS) allows to obtain
simultaneously both spectral and spatial information of galaxies. Therefore, IFS
techniques provide a more efficient way to study the spatial distribution of the
WR population in nearby galaxies. Indeed, such studies has been conducted in
some individual galaxies
\citep[e.g.][]{James09,Lopez-Sanchez11,Monreal13,Kehrig13}. In particular,
\cite{Miralles-Caballero14b} (henceforth, MC14b) used data from the PPAK
Integral Field Spectroscopy (IFS) Nearby Galaxies Survey
\citep[PINGS][]{Rosales-Ortega10} to locate the WR-rich regions in the nearby
spiral galaxy NGC~3310. PINGS was specifically designed to study the
spatially-resolved properties of a sample of 17 nearby spiral galaxies. Among
the almost 100 \hii regions identified throughout the disc of NGC~3310 by
\cite{Miralles-Caballero14}, MC14b reported the detection of WR features in 18
of them. The methodology developed by MC14b using IFS data mitigates aperture
effects and allows to spatially resolve the emission of the WR population in
local galaxies~\citep{Kehrig13}. Indeed, their technique  can be applied to
large galaxy samples. IFS observations also provide powerful tools to minimize
the WR bump dilution and find WR stars in extragalactic \hii regions where they
were not detected before
\citep{Kehrig08,Cairos10,Garcia-Benito10,Monreal10,Monreal12,Perez-Montero13}.
In addition, spatially resolved studies can also be used to explore the
connection between the environment and the WR emission. 
The study of the environment is of particular interest to get clues
about the connection between WR stars and SNIc-bl/GRBs, since these
massive stars may actually be progenitors of these explosions.
Although a direct link between GRBs to WR stars cannot be made, it is possible
to compare the properties of GRB host galaxies with those of WR stars and
indirectly obtain information on the GRB progenitors in this way.
\citet{Kelly08}, expanding the work by \citet{Fruchter06},
showed that both SNIc-bl and GRB tend to occur in the brightest regions of their
host galaxies. \citet{Leloudas10} correlated the locations of WR stars
with GRBs and different SN types finding that SNe Ibc and WR showed a high
degree of association, and that the connection between WR and GRBs could not be
excluded. Some other studies have reported that the host galaxies of nearby GRB
often host WR stars too, although the WR-rich areas do not necessarily
coincide with the location of the GRBs \citep{Cervinyo98,Hammer06,Christensen08,Han10,Levesque11,Thone14}.
Therefore, studies of WR galaxies and their relation with their environment may also
provide insight into the properties of GRB progenitors.

Previous searches for WR regions across individual galaxies outside the
Local Group have been performed, for example in NGC 300 \citep{Schild03}, M83
\citep{Hadfield05}, NGC 1313 \citep{Hadfield07}, NGC 7793
\citep{Bibby10}, NGC 5068 \citep{Bibby12}, and M101
\citep{Shara13}.
Nowadays a systematic search for the spatially resolved WR populations in sample
of nearby galaxies can be performed  with the aid of IFS. 
Examples include ~\cite{Brinchmann08a} (henceforth B08a), but also \cite{Shirazi12} who accomplished the task by
analysing a few hundred thousand spectra obtained with fibres. 
%% To adapt the searching and analysis tools to IFS data, we have to take
%% into account that in the nearby future huge data sets of tens of millions of
%% spectra will be produced with IFS instruments. As an example, 90000 spectra are
%% collected with
%% \textit{MUSE}\footnote{http://www.eso.org/sci/facilities/develop/instruments/muse.html}
%% in one simple shot, i.e., a sample of a few hundred galaxies will provide tens
%% of million spectra that will have to be analysed. Therefore, conducting this
%% kind of search is challenging but necessary. 
In this work we analyse a sample of 558 galaxies observed with IFS to compile
for the first time a homogeneous catalogue of galaxies showing regions with WR populations.
Galaxies of all different environments were considered, i.e. isolated galaxies,
compact dwarf galaxies, as well as targets showing past signs of recent and
ongoing interactions. Thanks to the 2D coverage of IFS
data we were are able to identify resolved \ha clumps in the galaxies, thus
avoiding restricting the search of WR-rich areas to only the brightest \hii
complexes, as it is usually imposed in other studies. 
The main goal of this article is to relate for the first time the WR populations
in this homogeneous galaxies sample with their spatially resolved properties, to
explore the influence of morphology in the recurrence of these features, and to
study the effect of binarity, photon leakage and metallicity in the observed
properties of WR emission lines in the Local Universe.

The paper is organized as follows: Sec.~\ref{sec:sample} presents our IFS
galaxy sample and gives some details about the data reduction process. We then
describe the methodology used to pinpoint regions showing WR emission using data
from IFS surveys in Sec.~\ref{sec:identification}. Sec.~\ref{sec:properties}
presents our catalogue of regions showing WR emission features. Here we
outline the overall properties of the host galaxies and describe the
procedure to measure the different components of the WR features. These
measurements are later used to estimate the number of WR stars in each
region. In Sec.~\ref{sec:discussion} we discuss the
influence of the environment and the GRB-WR connection, quantifying the effect of
the dilution of the WR features when integrating over larger apertures than the
observed spatial extent of the WR population, and comparing our observations with
the predictions provided by synthesis stellar models. Finally, we present our
conclusions in Sec.~\ref{sec:conclusions}.

\section{Initial sample and data reduction}
\label{sec:sample}

The galaxies of this study were mainly selected from the Calar Alto Legacy
Integral Field Area: CALIFA survey \citep{Sanchez12a}, an ongoing exploration of
the spatially resolved spectroscopic properties of galaxies in the Local
Universe (z $<$ 0.03) using wide-field IFS. CALIFA observations  cover the full
optical extent (up to $\sim 3-4$ effective radii) of around 600 galaxies of any
morphological type, distributed across the entire colour-magnitude
diagram~\cite{Walcher14}. Observations were carried out using the Potsdam
MultiAperture Spectrophotometer~\citep{Roth05}, mounted at the Calar Alto 3.5m
telescope, in the PPAK configuration~\citep{Kelz06}. The observations cover a
hexagonal field-of-view (FoV) of \mbox{74\arcsec\,$\times$\,64\arcsec,} which
allows to map the full optical extent of the galaxies up to two to three
effective radii. A dithering scheme of three pointings was adopted in order to
cover the complete FoV and properly sample the PSF. The observing details,
selection of the galaxy sample, observational strategy, and reduction processes
are thoroughly explained in \cite{Sanchez12a} and \cite{Walcher14}.
CALIFA has recently launched to the astronomical community its second data
release DR2\footnote{http://www.caha.es/CALIFA/public\_html/?q=content/califa-2nd-data-release},
a set of fully reduced, quality tested and scientifically useful data cubes for
200 galaxies \citep{Garcia-Benito14}.

%% The CALIFA sample was built
%% from the SDSS DR7~\citep{Abazajian09}, with such criteria that the selected
%% galaxies encompass a wide range of morphologies, luminosities, stellar masses
%% and colours~\citep{Walcher14}. 

%\begin{figure}
%\centering
%\includegraphics[angle=90,trim = 0cm 0cm 0cm 2cm,clip=true,width=1.0\columnwidth]{figs/sample_ini.eps}
%  \caption{Distribution of the currently observed CALIFA + extension-projects
%  galaxies in the colour-magnitude diagram.}
%  \label{fig:ini_sample}
%\end{figure}

%% CALIFA is an ongoing survey and observations of the galaxies belonging to the
%% extension projects are scheduled within the same frame, hence in all cases the
%% list of objects increases.

%Fig.~\ref{fig:ini_sample} shows the distribution of the colour-magnitude
%diagram of this sample, covering at least 3 galaxies per magnitude bin up to
%magnitude $M_{\mathrm{z}} < -16$ mag. With a redshift range of \mbox{0.005 $< z
%<$ 0.03}, above this luminosity, $M_{\mathrm{z}} =-16$,

For the purpose of our study, we used 448 galaxies from the CALIFA survey
observed by April 2014 in the V500 spectral set-up, that has a nominal
resolution of $\lambda/\Delta \lambda \sim 850$ at 5000 \AA\ and a nominal
wavelength range of 3749-7300 \AA. The dataset was reduced using the version
1.5 of the CALIFA pipeline \citep{Garcia-Benito14}. The usual reduction tasks
per pointing include cosmic rays rejection, optimal extraction, flexure
correction, wavelength and flux calibration, and sky subtraction. Finally, all
three pointings are combined to reconstruct a spatially $1\arcsec\times1\arcsec$
re-sampled data cube, that includes science data, propagated error vectors,
masks and weighting factors (see~\citealt{Sanchez12a} and~\citealt{Husemann13}
for more details). The average value of the PSF of the datacubes is around
2.5$\arcsec$~\citep{Garcia-Benito14}.
In addition, 110 galaxies from the ``CALIFA-Extension'' programs were also
included as part of our data. The {\it CALIFA-Extension} sample consist of 
galaxies not included in the main CALIFA sample that are part of specific
science programs using the same observing set-up as CALIFA. This sample includes
interactive, merging and low-mass galaxies (classes S0, Sb and dwarfs),
providing a complementary to main sample since CALIFA  becomes incomplete below
$M_\mathrm{r} > -19$ mag (which corresponds to a stellar mass of \mbox{$\sim10^{9.5}$\,\msun)}.

With a redshift range of \mbox{0.005 $< z <$ 0.03} and absolute magnitudes above
$M_{\mathrm{z}} =-16$, the sample is representative of the galaxy population in
the Local Universe \citep{Walcher14}. 
Note that this sample is not purposely selected to detect as
many WR features as possible, given that these populations can only be found
when young ionising stellar populations exist (i.e., typically within the first
million years after they are born). Yet, this is up to now the largest sample of 
galaxies observed using a wide field-of-view (\mbox{$>$ 1 arcmin$^2$}) IFS. It
contains hundreds of star-forming objects (including spiral, barred, irregular
and merging galaxies as well as blue compact dwarfs), and it is expected to host
a significant number of regions where populations of WR stars can be
detected. With these data we can resolve regions larger than about 100 pc in
radius in the closest galaxies and larger than about 600 pc in the furthest
systems. These sizes are typical for giant extragalactic \hii regions and \hii
complexes \citep{Kennicutt84,Gonzalez-Delgado97,Youngblood99,Hunt09,Lopez11,Miralles-Caballero11}.

\begin{figure*}
\centering
\includegraphics[angle=90,trim=-0.5cm 0cm 0cm -0.5cm,clip=true,width=0.9\textwidth]{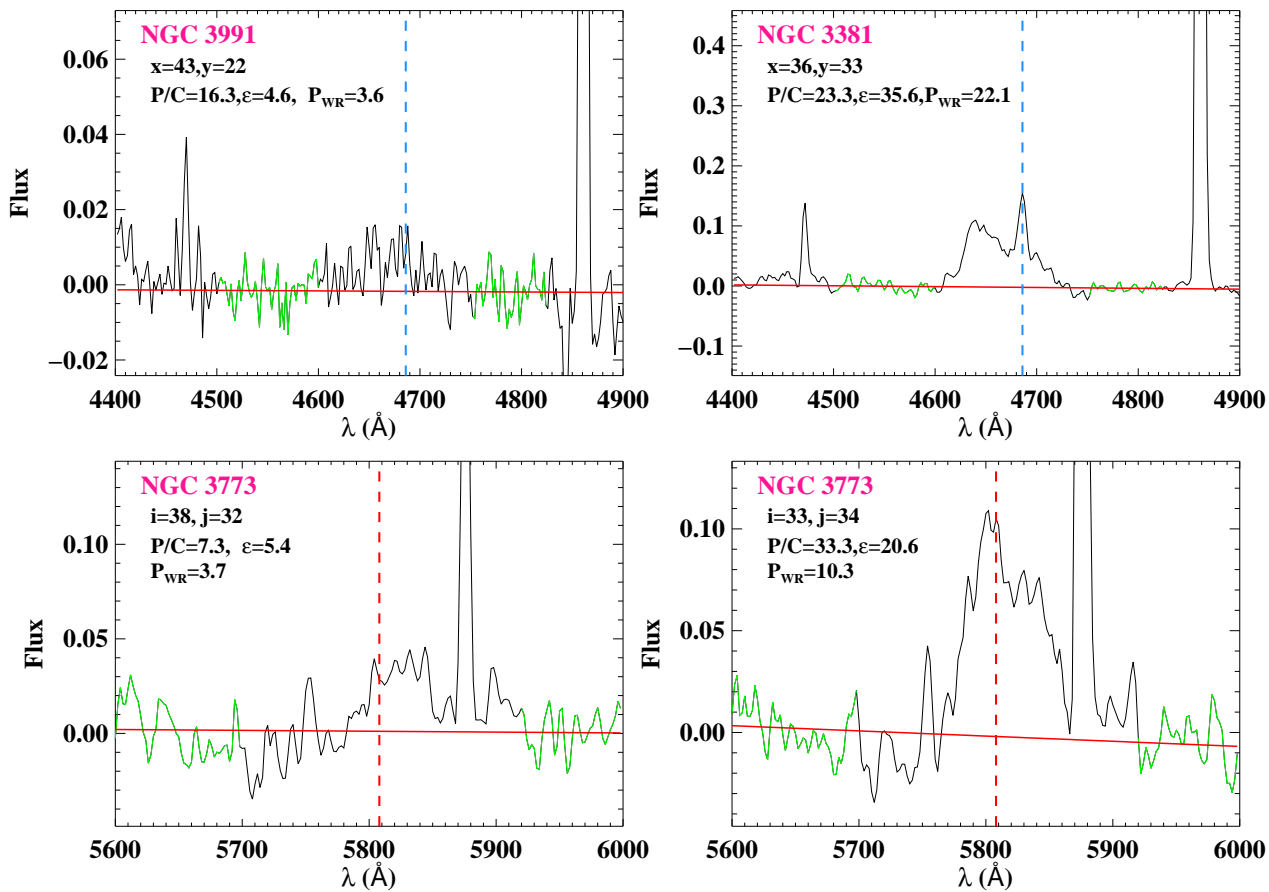}
  \caption{
    {\bf Top:} Examples of continuum-subtracted spectra for a given pixel in
    \mbox{NGC 3991} and \mbox{NGC 3381} around the blue WR bump spectral range. The
    blue vertical line shows the location of the \heii~$\lambda$4686 emission
    line. The spatial coordinates of the pixels (x and y) and the P/C, $\epsilon$,
    and P$_{\rm WR}$   parameters defined in the text are given on the left-top
    corner of each panel. A red continuous line indicates a linear fit to the
    continuum  (which ideally should be an horizontal line at $y$=0). The green
    sections of the spectrum correspond to those spectral ranges used to compute the
    $rms$. {\bf Bottom:} Examples of continuum-subtracted spectra for a given
    pixel in \mbox{NGC 3773} around the red WR bump spectral range. The red vertical
    line  indicates the location of the  \civ~$\lambda$5808 emission line. The
    coordinates of the pixels and their associated parameters are also given on the
    left-top corner. In both the  blue and the red WR bump cases one spectrum at the
    limit of the detection level (left) and another with a very good detection level
    (right) are shown. The flux density is given in units of \mbox{$10^{-16}$ erg
      s$^{-1}$ cm$^{-2}$ \AA{}$^{-1}$}.
    \label{fig:detection_plots}
  }
\end{figure*}

\section{Identification of WR features in IFS surveys}
\label{sec:identification}

WR stars have been typically discovered via techniques sensitive to their
unusually broad emission-line spectra. As mentioned before, the most prominent
WR emission feature observed in optical spectra corresponds to the so-called
blue bump, a blend of a broad \heii line at \mbox{4686 \AA}, nitrogen and carbon
emission lines formed in the expanding atmospheres of the WR stars, and several
high-excitation emission lines (sometimes including a narrow  \heii line)
covering the spectral range \mbox{4600-4700 \AA{}}~\citep{Schaerer98}.
The blue WR bump, however, is generally very faint, its flux being usually only
a few percent of the flux of the \hb emission line. The red WR bump centred
around \mbox{5808 \AA,} is usually even fainter than the blue WR
bump. Therefore, a systematic search for signatures of WR emission is
intrinsically challenging.

Early studies relied on the use of imaging with
a narrow-band $\lambda$4686 filter to detect, with great effort, the
spatially-resolved emission of several WRs in very nearby low-mass
galaxies~\citep{Drissen90,Sargent91,Drissen93a,Drissen93b}. Very detailed
studies, in most cases even for individual stars, have been accomplished using
the catalogues reported in these studies \citep[e.g.][]{Abbott04,Drissen08,Ubeda09}. 
%cita Mau
At farther distances, faint WR signatures can be detected via the analysis of
integrated spectra obtained using optical fibres or long slits
\citep[e.g.][]{kunth85,Vacca92,Izotov97}. These studies
allowed the compilation of a few catalogues of 40--140 WR galaxies
\citep{Conti91,Schaerer99,Guseva00}. The most recent surveys of WR galaxies have
used single-fibre spectra from the Sloan Digital Sky Survey
(SDSS\footnote{http://www.sdss.org/}). The catalogue compiled by
B08a reported 570 new WR galaxies using SDSS,
and \cite{Shirazi12} added later 189 new WR galaxies. However, these catalogues
are limited to the size of the fibre (3\arcsec), and hence they are not able to
spatially resolve the WR population. Besides, galaxies observed by SDSS show a
very broad distribution of distances, sampling very heterogeneous sizes and
apertures. For this study, we first focused the effort on the detection of the
blue bump in our search for WR spectral signatures. Once we developed the
technique we did use it to detect the red bump, as explained below.

\subsection{Identification of regions with the blue WR bump}

Since only very massive stars ($M_{\mathrm{ini}} > 25\,$\msun for \zsun)
undergo the WR phase, typically $2-3$ Myr after their
birth~\citep{Meynet95}, the first natural place to search for the WR emission is
where the \hii regions are located. However, given that most of the stars in a
cluster will not undergo this phase, the location of the WR stars can be
restricted to a fraction of the total area of the \hii region,
%(i.e., the spatial extent of these stars can be  different than that of the
%rest of the ionising population). 
this effect was noticed by \cite{Kehrig13} (see Fig. 5 and 6). 
Thus the reason why our approach relies on performing a pixel-by-pixel analysis
of the spectra\footnote{Note that the spatial element with an
independent spectrum is usually called \emph{spaxel} in IFS, in this work we
will use pixel and spaxel indistinctly.}.
% does not correspond to that of a pixel due to the re-sampling of the data} 
The procedure followed to detect WR signatures in our galaxy sample is
described as follows:

\begin{enumerate}

 \item The first step consists in subtracting the underlying stellar population
   using the \starlight code \citep{Cid-Fernandes04b,Cid-Fernandes05} in order
   to better characterize the continuum and therefore better observe the faint
   WR features. It is important to note that by construction, the substraction
   of the underlying continuum refers to the large scale galaxy stellar content,
   not including young stars cluster populations and their likely emission.
   A reduced spectral library set of 18 populations (3,
   63, 400, 750, 2000 and 12000 Myr, combined with \zsun, \zsun/2.5 and \zsun/5
   metallicities) is used, so as to deal with the large amount of pixels that
   had to be analysed in a reasonable amount of time. These models were selected
   from the compilation by~\cite{Gonzalez-Delgado05} and the MILES library
   (\citealt{Vazdekis10}, as updated by~\citealt{Falcon-Barroso11}).

 \item As only massive, young ionising stars are able to undergo the WR phase, we
   focused only in those pixels whose approximate \ha
   equivalent width, EW\,(\ha\onespace), is of the order of 6 \AA\ or
   higher~\citep{Cid-Fernandes11,Sanchez14}. Given that the widths of the
   nebular lines are rather similar for the whole sample (\mbox{$\sigma \sim
     2.8$ \AA{}}, with no presence of emission from an AGN or other broader
   components), and assuming  a Gaussian fit, we can roughly estimate
   EW\,(\ha\onespace) just by measuring the observed peak of the \ha emission
   over the continuum (\mbox{$\equiv$ P/C}). Under these assumptions,
   $\mathrm{P/C}  > 0.85$ practically ensures that we are dealing with a young
   ionising population.

 \item Next, using a similar approach to that discussed in B08a, which simulates
   early studies that used narrow band images
   \citep[e.g.][]{Drissen90,Sargent91,Drissen93a,Drissen93b}, we define a
   pseudo-filter spanning the \mbox{4600-4700 \AA} rest-frame range. For each
   pixel, we integrate the density flux,  $F_\mathrm{bump}$, within this
   spectral range and compare it with the $rms$ in two spectral windows,
   \mbox{4500-4600 \AA{}} and  \mbox{4750-4825 \AA{}}. We then define the
   detection significance, $\varepsilon$, as (\citealt{Tresse99}):

\begin{equation}
\varepsilon = \frac{F_\mathrm{bump}}{\sigma} = \frac{F_\mathrm{bump}}{\sigma_\mathrm{c} D\sqrt{2N + \frac{\mathrm{EW}}{D}}} \approx \frac{F_\mathrm{bump}}{\sigma_\mathrm{c}D\sqrt{2N}}
\end{equation}

where $\sigma_\mathrm{c}$ is the mean standard deviation per spectral point on
the continuum on each side of the bump feature, $D$ denotes the spectral
dispersion in \AA{} per spectral point, $N$ corresponds to the number of
spectral points used in the integration of the flux density and EW refers to the
equivalent width of the bump. Since EWs as low as just a few \AA{} are
expected, we neglect its contribution in the equation. After performing several
visual inspections we set the procedure to select only those spaxels with
\mbox{$\varepsilon > 4$}. 

 %% The broad features varies significantly in appearance, given that the
 %% relative emission line strength and velocity broadening of individual WR
 %% stars can be very different.

\item WR features are broad and have different shapes, as the velocity structure of
  the WR winds changes between different WR stars, varying significantly the
  appearance of the features. The best approach to build up a
  reliable technique to detect WR features therefore is to construct a training
  set where WR features are clearly visible and the numerical values of detection
  parameters can be tested. We constructed such a training set by visual
  inspection of a set of spectra with tentative detections. 
  %% However, the structure of the emission has to be
  %% somehow similar to a bump, instead of a flat over-density flux
  %% feature. Sometimes this happens (i.e., $\varepsilon > 4$, but no bump
  %% structure was observed, just a rather flat weak detection). 
  We used the parameter P$_\mathrm{WR}$, which is the peak value within the
  WR filter range normalized to the rms. We select
  those pixels with \mbox{P$_\mathrm{WR} \equiv
    \frac{\mathrm{Peak}}{\sigma_\mathrm{c}} > 3.5$.} In a few cases,
  practically a pure narrow emission (i.e., with a width similar to that of
  \hb) is observed, identified as a narrow  \heii$\lambda$4686  emission
  line. It is still not clear if this narrow feature is intimately linked with
  the appearance of hot WR stars~\citep{Schaerer98,Crowther06a} or to O stars
  at low metallicities \citep{Kudritzki02}. We thus additionally impose
  $\frac{\varepsilon}{\mathrm{P}_\mathrm{WR}} > 1.1$ to avoid such cases.

\item Finally, a plausible detection should occupy an area similar or larger
  than the CALIFA PSF. Therefore, we imposed a minimum of 9 adjacent spatial
  elements satisfying the previous criteria to declare a positive
  detection. This corresponds roughly to the size of the PPAK PSF according to
  \cite{Garcia-Benito14}.

\end{enumerate}

Fig.~\ref{fig:detection_plots} (top) shows two examples to illustrate how our
procedure works. As can be observed, it is not difficult to obtain positive
detections with low significance levels when using the adopted
criteria. Based on our training sample, whenever there is WR emission,
$\varepsilon$ peaks in a pixel and then decreases radially down to
the levels of the cuts we have used. The criterion of gathering at least 9
grouped pixels helps to reject individual pixels with low significance levels.

\begin{figure*}[t!]
  \centering
  \includegraphics[angle=90,trim = 5cm 0cm 5cm -0.5cm,clip=true,width=0.95\textwidth]{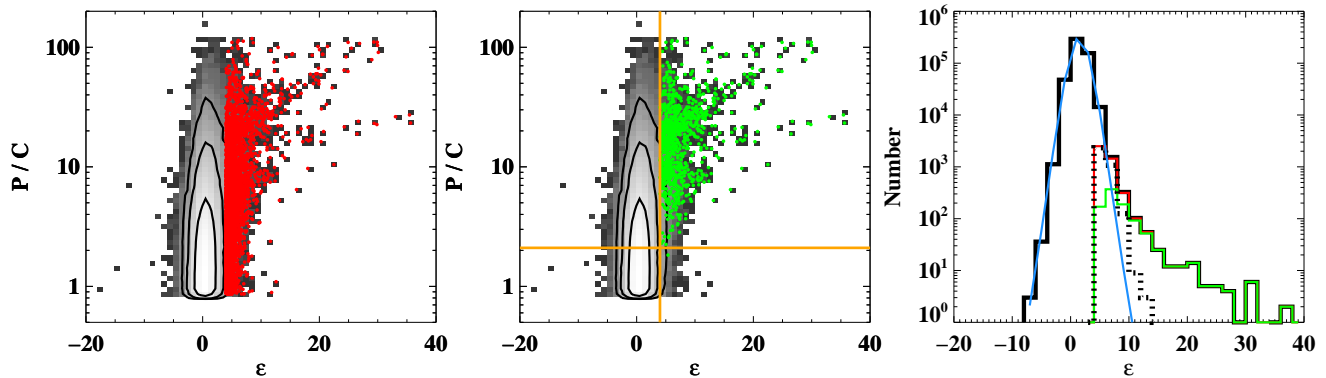}
  \caption{\textbf{Left:} Density plot (grey scale) of P/C vs. $\varepsilon$
    parameters for all the pixels in the sample for which \mbox{P/C $>$
      0.85}. Contours that contain 68\% ($1\sigma$), 95\% ($2\sigma$) and 99\%
    ($3\sigma$) of the points are overplotted. Red dots correspond to the values
    of the pre-selected regions, i.e., before imposing the grouping of at least
    9 pixels. \textbf{Middle:} Same as before, but this time the overplotted
    green dots correspond to selected regions, obeying also the grouping
    criterion. The orange vertical line indicates the cut used for $\varepsilon$
    (4), while the horizontal line denotes \mbox{P/C $\sim$ 2.1 parameter},
    having practically all the selected pixels a higher value than this
    limit. \textbf{Right:} Histogram of $\varepsilon$ for pixels with \mbox{P/C
      $>$ 0.85} (black), pre-selected pixels (red), selected pixels (green) and
    for those pre-selected pixels not satisfying the grouping criterion (dotted
    black). The curve resulting from a Gaussian fit of the black 
    histogram is overplotted in blue.}
  \label{fig:statistics1}
\end{figure*}

\subsection{Identification of regions with red WR bump}

The detection of the red WR bump was made using a similar procedure than the one
explained for detecting the blue WR bump. In this case, the continuum windows at
each side of the bump are 5600-5700~\AA{} and 5920-6000~\AA{}, the latter window
was chosen to avoid the bright \hei 5876 \AA{} emission line. Since the red WR bump
is generally fainter than the blue WR bump and very close to the \hei line,
their detection is more challenging that searching for the blue WR
bump,
%Furthermore, the continuum subtraction in this spectral range is worse
%than that obtained around the blue WR bump. 
so we increased the significance level to $\varepsilon_\mathrm{red}>5$ to select
those pixels with positive detections.

%% We also encountered some problems with a badly subtracted sky line at
%% \mbox{5770 \AA{}} (observed frame). However, given the broadness of the red bump
%% and that these spectral lines are as narrow as the instrumental width (just a few \AA{}),
%% we could correct for this by just interpolating the emission in a
%% three-spectral-point gap centred at this wavelength in the rest-frame. This
%% correction was only necessary for regions within \mbox{NGC~3991}.

Fig.~\ref{fig:detection_plots} (bottom) shows two examples to illustrate how the
procedure of searching for the red WR bump works. Now the continuum is noisier
than that for the blue WR bump, and even some structure is also
observed. Besides, the tail of the red WR bump coincides with the \hei line at
5876 \AA.

\subsection{Constraining the search of WR features}% of the parameters}

%% As mentioned before, WR features are generally very faint and thus they are very
%% hard to detect in the integrated spectrum of a kpc-sized extragalactic
%% region. After running the code presented in the previous section, 

%% After applying the method introduced above, we detected
%% thousands of pixels with a significant detection level of WR features. With a
%% sample of hundreds of galaxies and hundreds of thousands of pixels, we have
%% enough statistics to assess how these parameters behave and can help us
%% constraining the cuts for future searches.

Using the selection criteria described above on every individual pixel, a
positive detection of WR features in several thousands of pixels was obtained,
for all values of the considered P/C range (see Fig.~\ref{fig:statistics1},
left). We define this as our ``pre-selected sample''. However, after applying
criterion 5, the number of pixels with a positive detection reduced
dramatically, to several hundreds. We define this as our ``selected sample''. It
is interesting to note that the majority of the pixels in this restricted sample
have P/C values higher than 2.1 (Fig.~\ref{fig:statistics1}, middle), which
corresponds to \mbox{EW\,(\ha\onespace)} $\sim$ \mbox{15 \AA{}}. This is not a
very high value, since for simple (not binary) population models the WR phase
normally ends \mbox{5--6 Myr} after the stars are born and this EW value clearly
indicates a dominant stellar population older than that. For reference,
according to \popstar models~\citep{Molla09,Martin-Manjon10} at solar
metallicity, \mbox{EW\,(\ha\onespace) is of the order of 15 \AA{}} for a
starburst with an age \mbox{$\tau \sim 8.3$ Myr}. But we have to take into
account that the measured EWs may well be reduced by the continuum of
non-ionising stellar populations. All in all, at the depth and spatial
resolution of the data used in this study, a lower limit for the \mbox{EW
  (\ha\onespace)} can be considered as a safe value to look for WR emission
features in galaxies in the Local Universe. Had we placed the cut of the P/C
parameter at 2.1, we would have analysed less than half the initial pixel
sample.

By inspection of Fig.~\ref{fig:statistics1} (right) the histogram of $\varepsilon$ for the different samples of pixels
(black histogram, initial sample with \mbox{P/C $> 0.85$}; red histogram, the
pre-selected sample; and finally, green histogram, the selected sample) we
observe that the initial sample behaves well following a Gaussian function,
centred at zero and with a width of 2.5. 
This is the expected behaviour of a random distribution centred at zero,
i.e., with no detection at all. However, a tail appears in the distribution for
$\varepsilon > 6$. Therefore, the pixels in the selected sample are
statistically discriminated against those following a random distribution,
strengthening the validity of the detection. It is also worth noting that below
$\varepsilon = 8$ the dotted distribution (pixels from the pre-selected sample
excluding those from the selected sample) exceeds significantly the green
distribution, which means that below this value it is hard to distinguish
between real positive detections and noise. That is the main reason why the
additional grouping criterion was introduced. In contrast, when $\varepsilon >
10$ the green distribution clearly dominates over the dotted one, indicating
that when a detection is made in a pixel at this significance level it will very
likely represent a position with a resolved region with real WR features. 

\begin{figure}
\centering
\includegraphics[angle=90,trim = 0cm 1cm 0cm 0.5cm,clip=true,width=0.95\columnwidth]{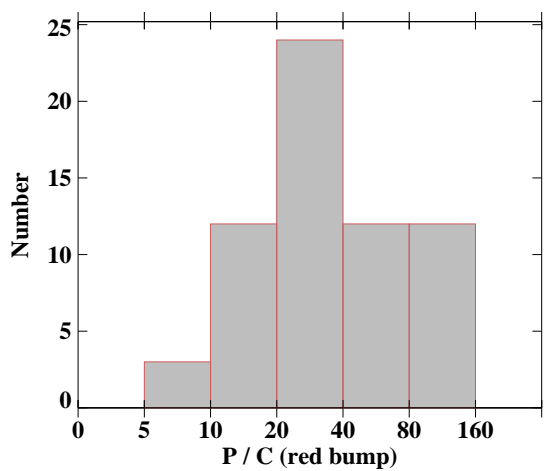}
  \caption{Histogram of the parameter P/C for pixels with positive detection of
    the red bump. The scale of the $x$ axis starts at a value of 5 and then
    every bin doubles the previous one.}
  \label{fig:statistics2}
\end{figure}

Regarding the red WR bump, it was only detected at high P/C
(Fig.~\ref{fig:statistics2}). The lowest value of this parameter for pixels with
positive detection of this bump is \mbox{P/C $\sim$ 9}, which corresponds to
\mbox{EW\,(\ha\onespace) $\sim$ 60 \AA{}}. This gives us an explanation of why
it is generally more difficult to observe the red WR bump than the blue
bump. The former is weaker than the latter and the presence of underlying
non-ionising stellar population dilutes it more easily. For this reason, we can
only observe the red bump when this contamination is not very high, and then the
observed EW (\ha\onespace), not corrected by the presence of other non-ionising
populations, approaches the EW predicted by the models for populations younger
than \mbox{6 Myr}. For instance, according to \popstar models at solar
metallicity, \mbox{EW (\ha\onespace) $\sim$ 60 \AA{}} for a starburst with age
\mbox{$\tau \sim 5.8$ Myr}.

Finally, we did not find candidates with red bump emission in regions other than
those where the blue bump was detected. This is expected since, at the distance
where our target galaxies are located (from \mbox{$\sim$ 10 Mpc} to \mbox{$\sim$
  300 Mpc}) we can only resolve sizes larger than $\sim$100\,pc. Actually, the
red WR bump has been observed in the absence of blue bump detection in very few
occasions, and always at high spatial resolution
\citep[\mbox{$\sim$10 pc};][]{Westmoquette13}.

\section{Properties of the WR population in the CALIFA survey}
\label{sec:properties}

\subsection{The catalogue of WR-rich regions}
\label{sec:catalogue}

\input{table1}

The selection procedure presented in Sec.~\ref{sec:identification} allow us to
build a mask per galaxy pinpointing the location of
those pixels satisfying criteria 1--5. We report the detection of regions with
positive WR emission in a total of 25 galaxies. The main properties of these WR
galaxies are listed in Table~\ref{table:gal_catalogue}. Our catalogue of WR
galaxies represents somewhat over 4\% of the initial galaxy sample, which we
recall is representative of the galaxy population in the Local Universe. 
As the CALIFA sample was not purposely selected to contain only
galaxies with strong star formation episodes, a small percentage of
galaxies with regions showing strong WR signatures is expected.

The WR host galaxies found are typically spirals and nearby blue dwarf galaxies
with low inclination angles.
%(i.e., statistically less underlying
%non-ionising stellar populations able to dilute the emission bumps agn present
%along the line-of-sight). 
Interestingly, 13 out of the 25 WR galaxies (52\%) are
related to galaxy interaction processes (galaxy pairs, signs of recent past
interaction, or galaxies undergoing a merger). Also the majority of the WR
galaxies analysed by \cite{Lopez-Sanchez10c} are also experiencing some kind of
interaction.
% although sometimes only deep images and/or neutral gas maps are able to identify them. 

It is surprising to find WR emission in an isolated Sa galaxy (NGC 1056), 
with low disc star formation activity than later type
galaxies~\citep{Giuricin94,Plauchu-Frayn12}, and in a dwarf elliptical (IC 0225;
also detected in B08a, WR346). In fact, this dwarf galaxy is a peculiar object
showing a blue core, something which is quite rare in this kind of
objects~\citep{Gu06,Miller08}.

\begin{figure}
\centering
 \includegraphics[trim = 0.5cm 0.5cm 11cm 21.2cm,clip=true,width=0.95\columnwidth]{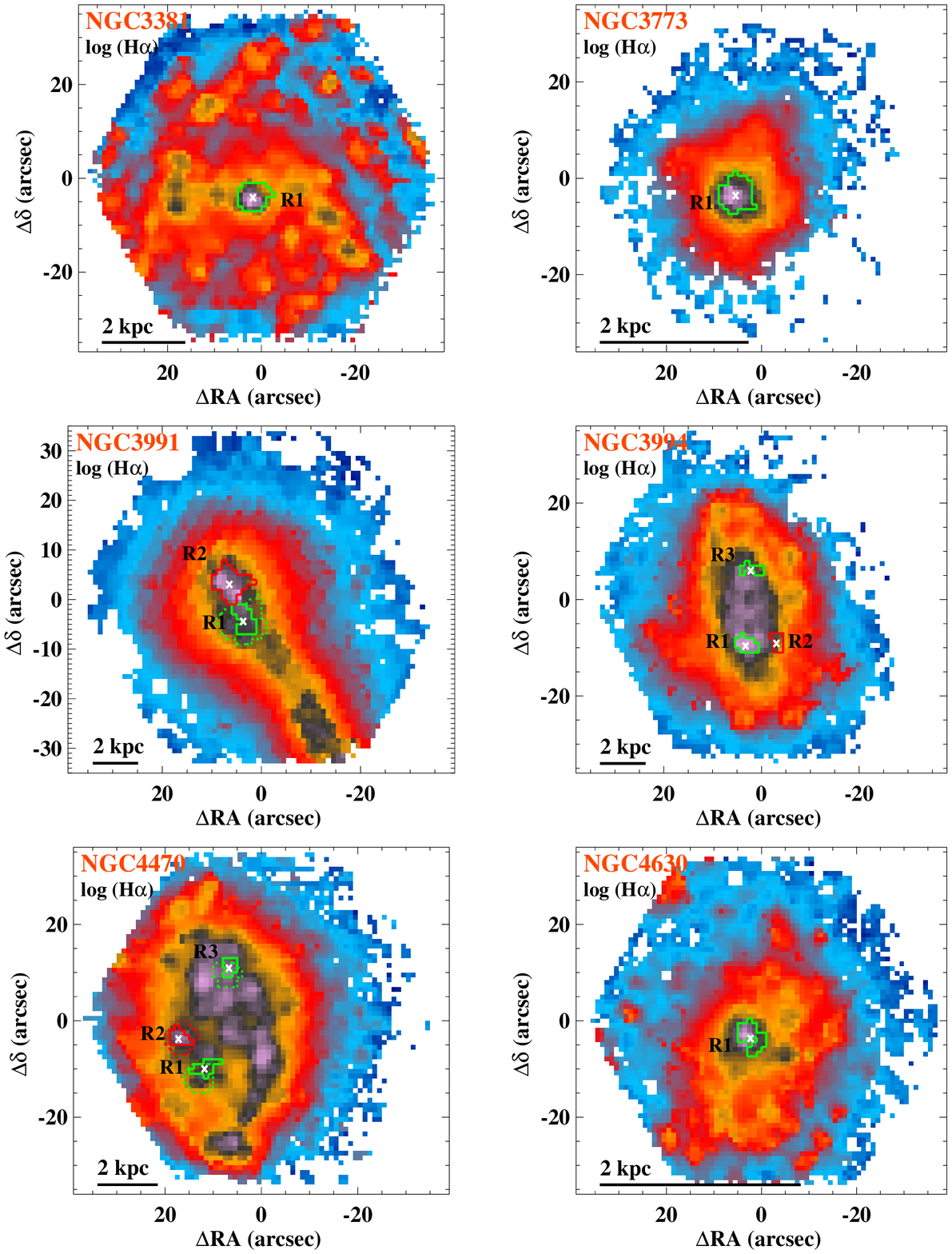} 
    \caption{\ha maps with logarithmic intensity scale of the galaxies with
      detected WR emission within the labelled regions enclosed by the green and
      red continuum contours. The pointted contours correspond to the associated
      \ha clump indentified using \hiiexplorer~\citep{Sanchez12b}, whenever the
      emission of the \ha clump is more extended than that of the WR region}. A
    cross indicates the barycenter of the region. The
      scale corresponding to \mbox{2 kpc} is drawn at the bottom-left
      corner. North points up and East to the left. \label{fig:ha_example}
\end{figure}

\begin{figure*}
\centering
\includegraphics[angle=90,trim = 4.5cm 0cm 5cm 0cm,clip=true,width=0.95\textwidth]{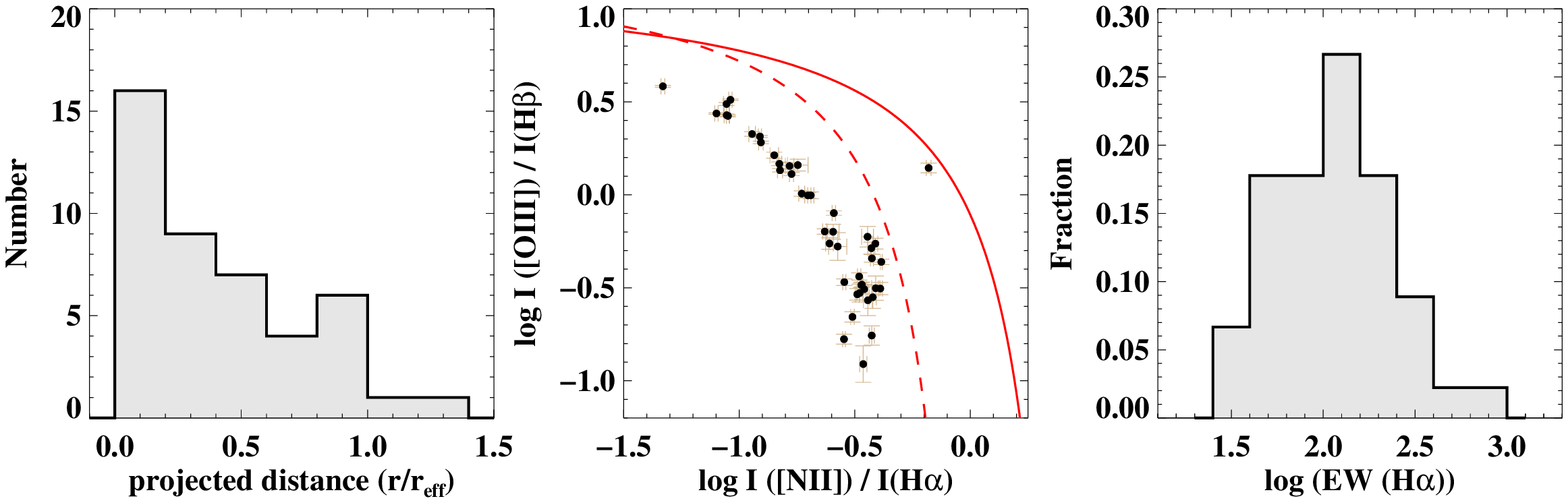}
  \caption{\textbf{Left:} distribution of the WR region projected distance to
    the galaxy centre, normalized to the effective radius of the
    galaxy. \textbf{Middle:} \oiii $\lambda$5007/\hb vs. \nii
    $\lambda$6583/\ha diagnostic diagram \citep[usually called the BPT diagram
      as described by][]{Baldwin81} for the selected regions. The solid and
    dashed lines indicate the~\cite{Kewley01a} and ~\cite{Kauffmann03}
    demarcation curves, respectively. These lines are usually used to
    distinguish between classical star-forming objects (below the dashed-line)
    and AGN powered sources (above the solid-line). Regions between both lines
    are considered to be of composite ionising source. \textbf{Right}:
    Distribution of the observed equivalent width of \ha in logarithmic units
    for the selected WR regions.}
  \label{fig:r_reff_bpt}
\end{figure*}

We identified a total of 44 WR-rich regions within the 25 galaxies listed
in Table \ref{table:gal_catalogue}, individual WR regions per galaxy are
numbered R1, R2, etc. Not all the WR regions are located in the
centre of the galaxies, but rather they are distributed in the circumnuclear
regions or found in external \ha clumps
(e.g. \mbox{NGC~5665}). Fig.~\ref{fig:ha_example} shows the \ha map of the
galaxy NGC~4470, where 3 WR regions have been identified. Similar maps for the rest of
the galaxies are shown in Fig.~\ref{fig:ha_maps}. A third of the WR regions are
located at projected distances further than half an effective radius, and the
majority of them lies within one effective radius (see Fig.~\ref{fig:r_reff_bpt}, left). 
The WR regions sometimes cover a fraction of the total area that is normally
considered a \hii region (e.g., UGC 9663). In some cases, the WR regions are not
apparently associated with an \ha clump or a \hii\onespace-like region, but
rather located within the \ha diffuse emission areas (e.g., R4 in \mbox{NGC 5665}).
This suggests that some \hii regions are not resolved with our IFS data. 
For example, the \hii regions in the circumnuclear starburst in \mbox{NGC 5953}~\citep{Casasola10} are not resolved
here but their WR emission has been detected. The main ionising source of these
regions usually is young, massive stars, as inferred from the diagnostic BPT
diagram \citep{Baldwin81} shown in Fig.~\ref{fig:r_reff_bpt} (middle
panel). Only the line ratios of one WR region is very close to the location of
the AGN domain. This object is the nucleus of \mbox{NGC 7469}, a well known
\mbox{Seyfert 1} galaxy with a circumnuclear
starburst~\citep{Cutri84,Heckman86,Wilson86}, interacting with its companion
\mbox{IC 5283}.
Note that the sistematics of this effect could have produced a bias in searches
of WR features with SDSS or other surveys of nearby galaxies \citep[see e.g.][]{Kehrig13,Shirazi12}.

\input{table2}

The main properties of the catalogue of the selected WR regions are listed in
Table~\ref{table:reg_catalogue}. Given that WR stars reside in \hii regions we
tried, if possible, to associate the WR region with an \ha clump in order to
define their radius. As expected, the \ha equivalent widths of these regions are
close to or higher than 100 \AA{}. It is interesting to note that the
distribution of EW~(\ha\onespace) does not peak at the highest values, but
within the range \mbox{125--160 \AA{}}, and then it decreases (see
Fig.~\ref{fig:r_reff_bpt}, right). This is consistent with the fraction of
star-forming galaxies that contain WR features as a function of
\mbox{EW~(\hb\onespace)} reported in B08a, where the curve reaches a maximum
value and then turns over. This turnover is expected to happen when either
\mbox{EW (\hb\onespace)} or \mbox{EW (\ha\onespace)} (good indicators of the age
of young populations) sample burst ages that are short relative to the starting
time of the WR phase (after 2--3 Myr). 
Obviously, our measured equivalent widths actually represent lower-limits to the
real values due to the presence of underlying continuum of non-ionising
populations. That is the reason why, according to \popstar models, our peak
represents ages of  \mbox{$\tau \sim$ 5.5 Myr}, practically when the WR phase is
about to end. Yet, the real shape of the curve or the distribution of equivalent
widths (i.e., in this study) should be similar to the shapes reported. As
already seen in the previous section with the spectra of the pixels, we have
detected the red bump in regions with higher EWs than average (\mbox{150--300
  \AA{}}). Taking into account that both bumps are originated in populations
with similar age, this suggests that in order to better observe the red bump a
lower contamination of the continuum by underlying non-ionising populations
seems to be necessary.

 \begin{figure*}
\centering
\includegraphics[angle=90,trim = 1cm 0cm 1cm -1.5cm,clip=true,width=0.9\textwidth]{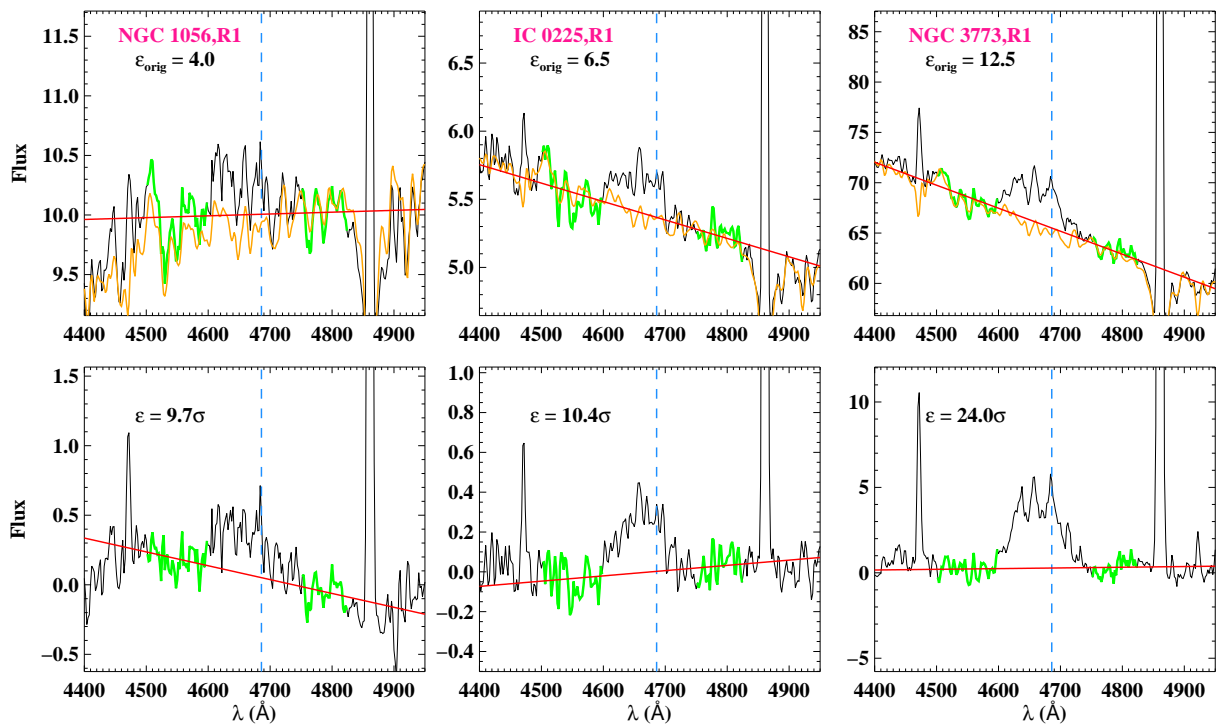}
  \caption{Examples of the integrated observed spectra around the blue WR bump
    for three selected regions: R1 in NGC~1056 (left), R1 in IC~225 (middle) and
    R1 in NGC~3773 (right). The blue vertical line shows the location of the
    \heii$\lambda$4686  line. The significance level
    ($\varepsilon_\mathrm{orig}$) is given on the left-top corner. The red solid
    line shows a linear fit to the continuum. The green sections of the spectrum
    correspond to those spectral ranges used to compute the $rms$. The fitted
    continuum spectrum with \starlight is overplotted in
    orange. \textbf{Bottom:} The corresponding continuum-subtracted spectra
    derived for each of the three selected regions. The value of $\epsilon$ is
    provided in each case. The flux is given in units of \mbox{$10^{-16}$ erg
      s$^{-1}$ cm$^{-2}$ \AA{}$^{-1}$}.}
  \label{fig:SL_fits}
\end{figure*}

For each WR region detected we added the spectra of all the corresponding
spaxels in order to increase the S/N and perform a detailed
analysis of its properties. We derived the star formation
history (SFH) of the WR region using the \starlight code in order to subtract
the underlying stellar continuum.
%\footnote{Note that the stellar templates used in
%  \starlight do not include emission from WR stars.}. 
%Since only a few dozen regions are available, 
At this time we used a full set of several hundreds of stellar templates to
perform a better modelling of their SFH.
%% As done in \cite{Miralles-Caballero14}, prior to applying the
%% \starlight code, the nebular continuum emission was also modelled and subtracted
%% from the observed spectrum using the derived \ha luminosity for each region,
%% following the procedure explained in~\cite{Molla09} and~\cite{Martin-Manjon10}. 

We divided our WR region sample in two classes depending on the significance level
of the observed integrated spectra,
$\varepsilon_\mathrm{orig}$. As shown in Fig.~\ref{fig:SL_fits} 
%justifies the reason of this subdivision. As this figure shows, 
the continuum is not only composed by
noise but also by real absorption and emission stellar features. This fact is
very notorious around 4500 \AA{}, where the fitted continuum (solid orange line)
follows well the observed continuum (solid green line) in the case of R1 in
\mbox{NGC 1056}. Hence, the $rms$ computed using this continuum is higher than
the one obtained using the subtracted fitted spectrum 
%(now the continuum basically corresponds to noise), 
and hence it induces an increase of the
significance level. However, when $\varepsilon_\mathrm{orig}$ is very low (e.g.,
region R1 in \mbox{NGC 1056}), $\varepsilon$ is also higher because the WR
feature is highly enhanced. We can also see that in this situation the continuum subtraction at the level of the emission feature is not very well achieved. Although slight deviations from a zero-valued horizontal line for the subtracted continuum is expected due to fluctuations of the noise, in cases such as  R1 in \mbox{NGC 1056} the resulting subtracted continuum has a considerable slope.  Therefore, the emission and shape of the WR feature
extremely depends on the model subtraction when its detection level is close to the noise level in the observed spectrum.
%For weak emission lines (e.g., sometimes the \sii~$\lambda\lambda$6717,6731
%emission lines), the uncertainty introduced by this subtraction can be even
%larger than 60\% (Rosales-Ortega, in prep.).
The higher the significance level in the observed spectrum, the lower the
dependence on the model subtraction. For instance, for the spectrum of R1 in
\mbox{NGC 3773}, $\varepsilon$ is higher than $\varepsilon_\mathrm{orig}$ by a
factor of 2, but this increase practically comes from the improvement of the
continuum emission. The shape and area of the WR emission feature is very
similar in the observed and subtracted spectrum.

Following this reasoning, we define as class-0 regions those with
$\varepsilon_\mathrm{orig} < 5$, and class-1 regions as those with higher
significance level on the observed spectrum. Hence, with this subdivision we
can distinguish between regions whose WR emission features highly depend on the
continuum subtraction and those whose dependence is either not very significant
or even negligible. Under this classification scheme, 14 regions are class-0
regions and the rest (30) are class-1 regions. Table~\ref{table:reg_catalogue}
also lists this subdivision.

Finally, Table~\ref{table:reg_catalogue} indicates those regions with a positive
detection of the red bump. In this case, the significance level of the detection
is generally higher than 20, but only five WR regions with red bump have been found
in our sample.

\subsection{Multiple line-fitting of the WR features}
\label{sec:line_fit}

%As mentioned in the previous subsection, when the continuum subtraction is
%performed, the WR features in the spectra are enhanced. Therefore, sometimes the

The exercise explained in the previous section showed that in some cases 
WR emission features show up after stellar continuum subtraction in regions 
where there is a barely significant detection in the originally observed spectrum. 
In those cases, the emission line fluxes can be
misleading since we are dealing with faint emission features at a continuum
level where the uncertainties and systematics are not well understood. We
therefore analysed WR features found {\em only} in class-1 regions, avoiding extremely
model-dependent results (although class-0 regions still make a sample of
promising candidates).

The blue bump is quite a complex emission structure formed from the blend of
broad stellar lines of helium, nitrogen and carbon: \heii~$\lambda$4686, \nv
$\lambda\lambda$4605,4620, \niii $\lambda\lambda\lambda$4628,4634,4640, and
\ciii/\civ $\lambda\lambda$4650,4658
\citep{Conti89,Guseva00,Crowther07}. Furthermore, a number of nebular emission
lines, such as the narrow \heii~$\lambda$4686 emission line, \feiii
$\lambda\lambda\lambda$4658,4665,4703,  \ariv $\lambda\lambda$4711,4740, \hei
$\lambda$4713 and \neiv $\lambda$4713 are often superimposed on the broad
features~\citep{Izotov98,Guseva00}. This makes the disentangling of the
individual fluxes rather challenging. These nebular emission lines within the
bump should be properly removed and not included in the flux of the broad
stellar lines.

In the last years, multiple-Gaussian line fitting procedures have been performed
in order to fit this complex feature (e.g., B08a;~\citealt{Lopez-Sanchez10a};
MC14b). Following similar prescriptions used in those studies, we performed an
automatic procedure to achieve a satisfactory fit in each case:

\begin{enumerate}

 \item The code fits a linear plus a Gaussian function to the broad
   \heii~$\lambda$4686, to a broad feature centred at \mbox{$\sim$ 4645 \AA{}},
   and to the two brightest nebular lines within this spectral range,
   \feiii~$\lambda$4658 and the narrow  \heii$\lambda$4686. The chosen continuum
   windows are the spectral ranges 4500$-$4550  and 4750$-$4820 \AA{}. The
   central wavelength of the Gaussian located at \mbox{4645~\AA{}} is set free
   within \mbox{15~\AA{}}. In that way, the code selects which blend is better
   observed, either the component centred around the nitrogen lines
   (\mbox{$\sim$ 4634 \AA{}}) or the one centred around the carbon lines
   (\mbox{$\sim$ 4650 \AA{}}). The width ($\sigma$) of the narrow (i.e.,
   nebular) lines is fixed to that of \hb while the width of the broad
   components are set free with a maximum value of \mbox{22 \AA{}} \mbox{(FWHM
     $\sim$ 52\AA{})}, which corresponds to a FWHM of about 3300 km/s. This is
   appropriate taking into account the spectral resolution of the data
   (FWHM$\sim 400$ km/s) and the typical upper limit adopted to the width of
   individual WR features (\citealt{Smith82,Crowther07};~B08a). If the code
   cannot successfully fit a nebular line (due to the absence of if or low
   $SNR$), that component is discarded in the next step.

 \item If the width of the broad component centred at \mbox{4645 \AA{}} equals
   the maximum adopted value, a broad component around \mbox{4612 \AA{}} is
   included in the fit. If the added component is not successfully fitted, it is
   discarded.

 \item Next, the $rms$ of the continuum is compared to the peak of the
   residuals. If this peak is higher than 4$\times rms$ within the spectral
   range \mbox{4600$-$4750 \AA{}}, a new component is added. If the central
   wavelength of the new component is shorter than \mbox{4650 \AA{}}, it will
   represent a broad emission line (in a few cases two distinct components are
   fitted, once centred at around 4634 \AA{} and another at around 4650
   \AA{}). If the central wavelength of the new component is longer than
   \mbox{4650 \AA{}}, it will be considered as a narrow line (i.e., a nebular
   line). 

 \item Finally, the previous step is done iteratively, adding each time a new
   component, until the peak is lower than 4$\times rms$.

\end{enumerate}

\begin{figure}
\centering
\includegraphics[trim = 10cm 0.5cm 1.5cm 21cm,clip=true,width=0.95\columnwidth]{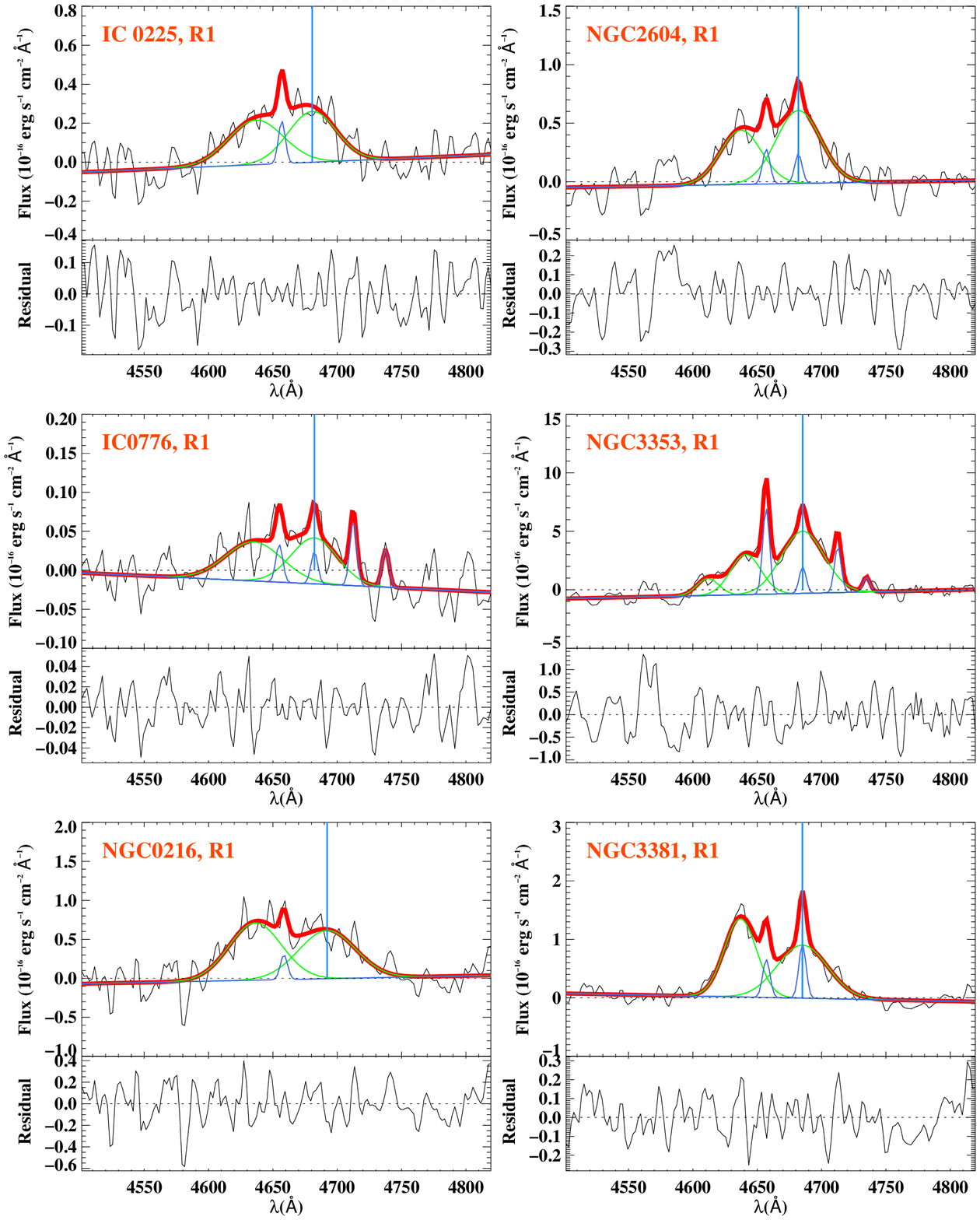}
 \caption{Example of the multiple-line fit of WR features for the region found
   in NGC 3381. An almost horizontal blue line denotes the resulting continuum
   of the fit. The total fitted continuum plus emission lines on the blue bump
   is drawn by a thick-red line. The nebular (blue) and broad stellar (green)
   components of the fit are also drawn. The vertical blue line indicate the
   position of the \mbox{\heii~$\lambda$4686} line. An auxiliary plot shows in
   black the residuals (in flux units) after modelling all the stellar and
   nebular features.}
 \label{fig:example_bb_fits}
\end{figure}

This procedure was run 50 times for each WR region, which is the total number of
times we applied the \starlight fit to subtract the stellar continuum. In that
way, we added to the error budget the uncertainty of the continuum
subtraction. Fig.~\ref{fig:example_bb_fits} shows an example of our procedure
using the fit to R1 in NGC~3381. The fits obtained to the rest of WR regions are
shown in  Fig.~\ref{fig:bb_fits}. In general, between three and six components
are needed to properly fit the WR features. The derived uncertainties of the
fits range between 10 and 50\%. The intensities derived for the broad features
and the nebular \heii$\lambda$4686 emission line are compiled in
Table~\ref{table:wr_fluxes}. In general, the flux of the broad blue bump is of
the order of 10\% or lower of that of H$\beta$.

In the case of R1 in NGC 7469, which is the nucleus of a \mbox{Seyfert~1}, we
did not apply any continuum subtraction and just performed the fitting to the
observed spectrum. Most Seyfert nuclei present a featureless ultraviolet and
optical continuum following a power-law, \mbox{$f_\nu \propto \nu^{-\alpha}$},
attributed to a non-thermal source~\citep{Osterbrock78}. Although within a
diameter \mbox{$\gtrsim 2$ kpc} the optical continuum is probably dominated by
the stellar emission, its modelling at the level of the weak emission of the WR
features is quite poor due to the presence of a power-law component, not
included in the stellar libraries used in \starlight. In any case, the
significance level of the WR feature in the observed spectrum of this source is
high enough ($\varepsilon_\mathrm{orig} = 8.9$; see
Table~\ref{table:reg_catalogue}) so that a proper correction due to the
continuum subtraction would not probably alter significantly the emission shape
of the WR feature.

Regarding the red bump, basically two broad stellar lines form this broad
emission feature, \mbox{\civ $\lambda\lambda$5801,5812}
~\citep{Vanderhucht01,Ercolano04}. Given the broadness of these stellar lines,
the auroral \mbox{\nii~$\lambda$5755} emission line is usually superposed on the
red bump. Therefore, we fitted this narrow line together with a broad component
centred at \mbox{5808 \AA{}}, setting the central wavelength of this broad
component free within \mbox{20 \AA{}}. The fits are presented in
Fig.~\ref{fig:rb_fits}. We note that assuming a single broad component provided
better results than fitting two close broad components. As we did for fitting
the blue bump, an upper limit to the width of the broad component was
needed. Hence, we adopted an upper limit of \mbox{$\sigma =$ 30 \AA{}}
(\mbox{FWHM $\sim$ 70 \AA{}}). Changing this value by 5~\AA{} induces a
variation of $\sim$10\% in the integrated flux. We tried to include a fit to the
\ciii $\lambda$5696 emission line, which is mainly originated in carbon WR
subtypes, but given the noisy residual spectrum, even in the case of 
\mbox{NGC 3773} (see Fig.~\ref{fig:rb_fits}), no successful fit could be achieved with
an uncertainty of less than 50\%. Finally, this analysis does not include a fit
to the nebular \hei~$\lambda$5876 emission line since, after the continuum
subtraction, some residual emission up to \mbox{$\sim $5920 \AA{}} was still
found. The intensities derived for the red bump are listed in
Table~\ref{table:wr_fluxes}, being only a few percent of the \hb emission.

\section{Discussion}
\label{sec:discussion}

\subsection{Nature and number of WR stars}

Learning about the nature of the WR population from the measured lines that form
the blue and red bumps is not an easy task. 
%The strength of the stellar broad
%emission lines depends on the nature of the WR stars located within each region. 
For instance, the broad
\heii$\lambda$4686 and the blue bump emission are mainly linked to WN
stars. However, some emission from WC stars may also be expected in the blue WR
bump (e.g.,~\citealt{Schaerer98}), so all WR types contribute to this broad
emission. The broad \civ$\lambda\lambda$5801,5812 emission feature essentially
originates in WC stars (mainly in early-types, WCE). However, given that this
feature is harder to detect than the blue bump, its non-detection does not
necessarily imply the absence of WC stars. Other lines that are directly linked
with other WR subtypes (e.g., \ovi$\lambda\lambda$3811,3834, which are related
to WO types) are not detected either. Finally, both WN and WC stars contribute
to the emission of the broad \niii$\lambda\lambda\lambda$4628,4634,4640 (WN) and
\ciii/\civ$\lambda\lambda$4650,4658 (WC) blends. Following MC14b, we consider
these steps to identify the different sub-types of WRs that cause the WR bump
emission for a given region and estimate their approximate number:

\begin{itemize}

 \item Although \heii lines are also produced by O stars with ages \mbox{$\tau
   \lesssim$ 3 Myr} ~\citep{Schaerer98,Massey04,Brinchmann08b}, we assume that
   all the emission of the \heii$\lambda$4686 line comes from WRs.\\

 \item The luminosity of WN late-type stars (WNL), the main contributors to the
   emission of the  broad \heii$\lambda$4686  emission, is not constant and can
   vary within factors of a few. Given the reported metallicity dependence of
   this broad emission, we use the approach proposed by~\cite{Lopez-Sanchez10a}
   to estimate the luminosity of a single WNL, $L_{\mathrm{WNL}} (\textrm{\heii}
   \lambda4686)$, as a function of metallicity:

   \vspace{-10pt}
   \begin{equation}
\label{eq:wnl_4686}
     L_{\mathrm{WNL}} (\textrm{\heii} \lambda4686) = (-5.430 + 0.812 x) \times 10^{36}~\mathrm{erg s}^{-1}
   \end{equation}
   with $x$ = 12 + log(O/H). \\

  \item The \nv$\lambda\lambda$4605,4620 emission originates in early-type
    nitrogen stars (WNE). The mean luminosity of a single Large Magellanic Cloud
    WNE star is, with an dispersion of a factor of 2, about
    $1.6\times10^{35}~\mathrm{erg\, s}^{-1}$ \citep{Crowther06a}. These stars
    also contribute to the \mbox{\heii$\lambda$4686} emission in about
    $8.4\times10^{35}~\mathrm{erg\, s}^{-1}$.

  \item Given that we do not observe the \ciii $\lambda$5696 line, mainly
    originated in late-type WC stars, we assume that the red  bump emission is
    mainly produced due to the presence of WCEs. We also use the approach
    proposed by~\cite{Lopez-Sanchez10a} to estimate the luminosity of this time
    a single WCE ($L_{\mathrm{WCE}}, \textrm{\civ}\lambda5808)$, as a function
    of metallicity:

   \vspace{-10pt}
    \begin{equation}
    \label{eq:wce}
      L_{\mathrm{WCE}} (\textrm{\heii}\lambda5808) = (-8.198 + 1.235 x) \times 10^{36}~\mathrm{erg s}^{-1}
    \end{equation}
    with $x$ = 12 + log(O/H). These stars also contribute to the emission of the
    broad  \heii$\lambda$4686 line. The estimation of the contribution of a single
    WCE to this emission feature is explained in the following item. \\

  \item As mentioned before, the non-detection of the red bump does not guarantee that there are not WCs. MC14b estimated an upper-limit to the number of WC stars in \mbox{NGC 3310} to be in the range 5--20\% of the number of WN stars. There, in general \mbox{EW($\lambda$4650) $<$ EW($\lambda$4686)}. However, as can be seen in Table~\ref{table:wr_fluxes}, in the current study \mbox{EW($\lambda$4650)} is of the order of (or in some cases even higher than) \mbox{EW($\lambda$4686)}. This points towards a number of WC stars quite higher than just the 5\%. Under these circumstances, we have made a rough estimation of the number of WCEs even if no red bump was detected. To do that, a system of two equations with two unknowns has to be solved:
\begin{equation}
\label{eq:system}
\systeme{
x \times L_{\mathrm{WNL}} (\textrm{\heii} \onespace) + y \times  L_{\mathrm{WCE}} (\textrm{\heii} \onespace) = L(4686),\\ x \times L_{\mathrm{WNL}} (\textrm{\niii}\onespace) + y \times  L_{\mathrm{WCE}} (\textrm{\ciii /\civ}\onespace) = L (4650)
} 
\end{equation}

where $L (4686)$ and $L (4650)$ refer to the measured flux of the \heii $\lambda4686$ and the $\lambda4650$ (including \niii $\lambda\lambda\lambda$4628-34-41 and \ciii\onespace/\civ $\lambda\lambda$4650-58) broad features, respectively; $L_{\mathrm{WNL}} (\textrm{\heii} \onespace)$ and $ L_{\mathrm{WCE}} (\textrm{\heii} \onespace)$ correspond to the luminosity of a single WNL (from Eq.~\ref{eq:wnl_4686}) and a single WCE star in the $\lambda4686$ broad feature, respectively; $L_{\mathrm{WNL}} (\textrm{\niii}\onespace)$ and $L_{\mathrm{WCE}} (\textrm{\ciii /\civ}\onespace)$ indicate the luminosity of a single WNL (3.8$\times10^{35}$ erg s$^{-1}$;~\citealt{Crowther06a}) and a single WCE star in the $\lambda4650$ broad feature, respectively; and the unkowns $x$ and $y$ are the number of WNL and WCE stars.

As reported in~\cite{Crowther06a}, the combined \ciii $\lambda4650$ +
\hii $\lambda4686$ flux of LMC WC4 stars averages around 49$\times10^{35}$ erg
s$^{-1}$. In our study, 12\% of this value is assigned to a contribution to
the $\lambda4686$ feature whereas the remaining 88\% is assumed for the
contribution to the $\lambda4650$ emission. We split it that way because
several works support that the \mbox{\heii $\lambda4686$} emission in WC stars
contributes on average by 12\% to the combined 4650/4686 blend
(\citealt{Schaerer98,Smith90}), ranging from 8 to 30\%. Finally, if the
\nv$\lambda\lambda$4605,4620 emission is present in the spectra, the number of
WNE stars is first obtained and then their contribution to $L (4686)$ subtracted
before solving the equation system (Eq.~\ref{eq:system}).

\begin{figure}
  \hspace{0.5cm}
  \includegraphics[trim = 0.5cm 0cm 1cm -1cm,clip=true,angle=90,width=0.9\columnwidth]{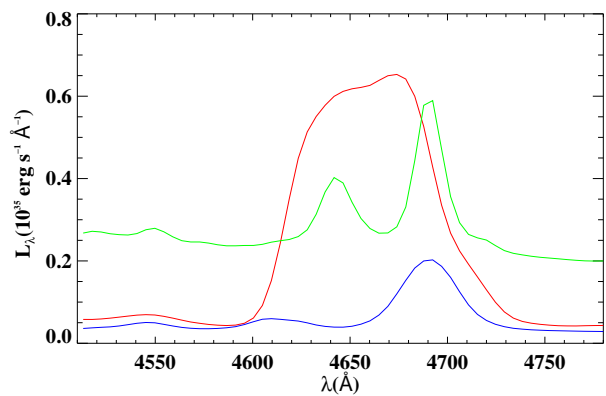}
   \caption{Template of LMC WC (red), early WN (blue) and late WN (green) from~\cite{Crowther06a}.}
  \label{fig:templates}
\end{figure}

\item Finally, we must note that the fits performed on these features are not
  necessarily physical, since WC lines can be much broader than the upper limits
  adopted, as shown in Fig.~\ref{fig:templates} (as broad as almost the entire
  emission bump). Depending on the contribution of WC stars to the blue bump the
  results might be misleading. We chose not to include another component to the
  fit much wider than the others and with no clue on how to constrain it (in
  most cases the red bump is not detected). Instead, we tried to minimize this
  by, instead of fitting directly this feature, assigning different
  contributions of this emission to the \ciii $\lambda4650$ and \hii
  $\lambda4686$ fitted features, as mentioned in the previous item.

\end{itemize}

\begin{table*}
\begin{footnotesize}
  \begin{sideways}
  \begin{minipage}{1.01\textheight}
  \vspace{1.5cm}
  \caption{Fluxes derived from our fitting of the WR bumps, and derived WR
    numbers for the class-1 WR regions identified in this study}
  \label{table:wr_fluxes}
  \begin{tabular}
{@{\hspace{0.1cm}}l@{\hspace{0.1cm}}c@{\hspace{0.15cm}}c@{\hspace{0.15cm}}c@{\hspace{0.15cm}}c@{\hspace{0.15cm}}c@{\hspace{0.15cm}}c@{\hspace{0.15cm}}c@{\hspace{0.15cm}}c@{\hspace{0.15cm}}c@{\hspace{0.15cm}}c@{\hspace{0.15cm}}c@{\hspace{0.15cm}}c@{\hspace{0.1cm}}@{\hspace{0.1cm}}c@{\hspace{0.05cm}}}
\hline \hline
   \noalign{\smallskip}
\multicolumn{1}{l}{Galaxy} &
\multicolumn{1}{c}{Region}  &
\multicolumn{1}{c}{H$\beta$ (WR)}  &
\multicolumn{1}{c}{c (H$\beta$)} \vline &
\multicolumn{1}{c}{\nv $\lambda$4612} &
\multicolumn{1}{c}{N/C  $\lambda$4650} &
\multicolumn{1}{c}{Narrow \heii $\lambda$4686} &
\multicolumn{1}{c}{Blue bump}  &
\multicolumn{1}{c}{Red bump} \vline &
\multicolumn{1}{c}{12 + log(O/H)}  &
\multicolumn{1}{c}{N$_{\mathrm{WNE}}$}  &
\multicolumn{1}{c}{N$_{\mathrm{WNL}}$}  &
\multicolumn{1}{c}{ N$_{\mathrm{WCE}}$}  &
\multicolumn{1}{c}{ N$_{\mathrm{WCE}}$}  \\

\multicolumn{1}{l}{ } &
\multicolumn{1}{c}{ID} &
\multicolumn{1}{c}{($\times 10^{-14}$ cgs) } &
\multicolumn{1}{c}{} \vline &
\multicolumn{5}{c}{I / I(H$\beta$) (WR)} \vline &
\multicolumn{1}{c}{(O3N2)} &
\multicolumn{1}{c}{ } &
\multicolumn{1}{c}{ } &
\multicolumn{1}{c}{(RB) } &
\multicolumn{1}{c}{(4650) } \\

\multicolumn{1}{l}{(1)} &
\multicolumn{1}{c}{(2)} &
\multicolumn{1}{c}{(3)} &
\multicolumn{1}{c}{(4)} \vline &
\multicolumn{5}{c}{(5--9)} \vline &
\multicolumn{1}{c}{(10)} &
\multicolumn{1}{c}{(11)} &
\multicolumn{1}{c}{(12)} &
\multicolumn{1}{c}{(13)} &
\multicolumn{1}{c}{(14)} \\
\noalign{\smallskip}
 \hline
   \noalign{\smallskip}
IC 0225 & R1 & 1.62 $\pm$ 0.03 & 0.09 $\pm$ 0.01 & $-$ & 9.3 $\pm$ 4.7 & $-$ & 9.9 $\pm$ 4.6 & $-$ & 8.40 $\pm$ 0.04 &  $-$ & 61 $\pm$29 &  $-$ & 15 $\pm$9 \\
IC 0776 & R1 & 0.67 $\pm$ 0.00 & 0.00 $\pm$ 0.01 & $-$ & 4.1 $\pm$ 3.1 & 0.44 $\pm$ 0.20 & 3.9 $\pm$ 0.8 & $-$ & 8.21 $\pm$ 0.02 &  $-$ & 28 $\pm$6 &  $-$ & 7 $\pm$5 \\
NGC 0216 & R1 & 3.94 $\pm$ 0.09 & 0.12 $\pm$ 0.01 & $-$ & 10.7 $\pm$ 2.4 & 0.60 : & 11.7 $\pm$ 1.3 & $-$ & 8.33 $\pm$ 0.06 &  $-$ & 181 $\pm$18 &  $-$ & 40 $\pm$11 \\
NGC 2604 & R1 & 4.92 $\pm$ 0.08 & 0.25 $\pm$ 0.01 & $-$ & 7.1 $\pm$ 1.1 & 0.67 $\pm$ 0.32 & 11.0 $\pm$ 1.3 & $-$ & 8.46 $\pm$ 0.04 &  $-$ & 393 $\pm$43 &  $-$ & 53 $\pm$13 \\
NGC 3353 & R1 & 81.51 $\pm$ 0.54 & 0.19 $\pm$ 0.01 & 0.45 $\pm$ 0.09 & 2.3 $\pm$ 0.3 & 0.31 $\pm$ 0.06 & 4.5 $\pm$ 0.3 & 0.71 $\pm$ 0.10 & 8.22 $\pm$ 0.02 & 481 $\pm$92 & 268 $\pm$122 & 62 $\pm$8 & 68 $\pm$12 \\
NGC 3381 & R1 & 6.61 $\pm$ 0.12 & 0.15 $\pm$ 0.01 & $-$ & 9.8 $\pm$ 0.6 & 1.54 $\pm$ 0.17 & 10.5 $\pm$ 0.7 & $-$ & 8.58 $\pm$ 0.07 &  $-$ & 269 $\pm$16 &  $-$ & 75 $\pm$4 \\
NGC 3773 & R1 & 22.07 $\pm$ 0.31 & 0.07 $\pm$ 0.01 & $-$ & 9.7 $\pm$ 1.7 & 0.71 $\pm$ 0.15 & 8.2 $\pm$ 0.9 & 6.82 $\pm$ 1.07 & 8.35 $\pm$ 0.04 &  $-$ & 216 $\pm$62 & 151 $\pm$22 & 83 $\pm$17 \\
NGC 3991 & R1 & 4.73 $\pm$ 0.07 & 0.02 $\pm$ 0.01 & $-$ & 5.7 $\pm$ 0.5 & 1.32 $\pm$ 0.19 & 3.3 $\pm$ 0.7 & 4.36 $\pm$ 1.00 & 8.27 $\pm$ 0.04 &  $-$ & 187 $\pm$80 & 261 $\pm$56 & 137 $\pm$14 \\
 & R2 & 15.10 $\pm$ 0.16 & 0.07 $\pm$ 0.01 & $-$ & 4.5 $\pm$ 0.3 & 0.70 $\pm$ 0.14 & 2.4 $\pm$ 0.3 & 4.13 $\pm$ 0.80 & 8.24 $\pm$ 0.03 &  $-$ & 348 $\pm$128 & 808 $\pm$149 & 353 $\pm$24 \\
NGC 3994 & R1 & 8.20 $\pm$ 0.18 & 0.50 $\pm$ 0.01 & $-$ & 10.7 $\pm$ 1.3 & $-$ & 5.9 $\pm$ 0.8 & $-$ & 8.53 $\pm$ 0.05 &  $-$ & 592 $\pm$89 &  $-$ & 428 $\pm$47 \\
NGC 4470 & R1 & 0.44 $\pm$ 0.03 & 0.09 $\pm$ 0.03 & $-$ & 32.5 $\pm$ 4.5 & $-$ & 23.8 $\pm$ 3.7 & $-$ & 8.50 $\pm$ 0.17 &  $-$ & 81 $\pm$9 &  $-$ & 37 $\pm$3 \\
 & R2 & 0.86 $\pm$ 0.03 & 0.19 $\pm$ 0.02 & $-$ & 17.8 $\pm$ 2.7 & $-$ & 14.2 $\pm$ 3.6 & $-$ & 8.46 $\pm$ 0.09 &  $-$ & 97 $\pm$26 &  $-$ & 39 $\pm$6 \\
NGC 4630 & R1 & 11.01 $\pm$ 0.14 & 0.28 $\pm$ 0.01 & $-$ & 5.0 $\pm$ 0.6 & 0.65 $\pm$ 0.16 & 6.5 $\pm$ 0.8 & $-$ & 8.60 $\pm$ 0.06 &  $-$ & 49 $\pm$5 &  $-$ & 10 $\pm$1 \\
NGC 5630 & R1 & 1.74 $\pm$ 0.05 & 0.28 $\pm$ 0.02 & $-$ & 8.8 $\pm$ 1.5 & $-$ & 9.5 $\pm$ 1.7 & $-$ & 8.37 $\pm$ 0.06 &  $-$ & 194 $\pm$34 &  $-$ & 44 $\pm$9 \\
NGC 5665 & R2 & 7.44 $\pm$ 0.13 & 0.30 $\pm$ 0.01 & $-$ & 5.7 $\pm$ 0.6 & $-$ & 4.9 $\pm$ 0.5 & $-$ & 8.56 $\pm$ 0.05 &  $-$ & 260 $\pm$25 &  $-$ & 97 $\pm$11 \\
 & R3 & 1.97 $\pm$ 0.02 & 0.23 $\pm$ 0.01 & $-$ & 8.6 $\pm$ 0.9 & 1.13 $\pm$ 0.21 & 7.2 $\pm$ 0.9 & $-$ & 8.53 $\pm$ 0.03 &  $-$ & 100 $\pm$13 &  $-$ & 39 $\pm$4 \\
 & R6 & 12.98 $\pm$ 0.14 & 0.36 $\pm$ 0.01 & $-$ & 5.8 $\pm$ 0.5 & 0.40 $\pm$ 0.15 & 2.9 $\pm$ 0.4 & $-$ & 8.56 $\pm$ 0.03 &  $-$ & 226 $\pm$42 &  $-$ & 192 $\pm$18 \\
NGC 5954 & R2 & 5.28 $\pm$ 0.08 & 0.42 $\pm$ 0.01 & $-$ & 8.6 $\pm$ 1.0 & 0.57 $\pm$ 0.27 & 5.1 $\pm$ 1.0 & $-$ & 8.55 $\pm$ 0.04 &  $-$ & 129 $\pm$31 &  $-$ & 83 $\pm$10 \\
 & R3 & 7.67 $\pm$ 0.06 & 0.37 $\pm$ 0.01 & $-$ & 5.5 $\pm$ 0.6 & $-$ & 3.8 $\pm$ 0.4 & $-$ & 8.55 $\pm$ 0.02 &  $-$ & 147 $\pm$19 &  $-$ & 76 $\pm$9 \\
NGC 6090 & R1 & 51.64 $\pm$ 0.76 & 0.55 $\pm$ 0.01 & $-$ & 3.1 $\pm$ 0.3 & 0.72 $\pm$ 0.14 & 4.9 $\pm$ 0.5 & $-$ & 8.55 $\pm$ 0.03 &  $-$ & 31023 $\pm$2851 &  $-$ & 4538 $\pm$623 \\
NGC 7469 & R1 & 98.29 $\pm$ 3.48 & 0.35 $\pm$ 0.02 & $-$ & 12.2 $\pm$ 1.7 & $-$ & 7.5 $\pm$ 2.1 & $-$ & 8.48 $\pm$ 0.07 &  $-$ & 21772 $\pm$7053 &  $-$ & 12893 $\pm$1660 \\
UGC 00312 & R1 & 1.33 $\pm$ 0.01 & 0.11 $\pm$ 0.01 & $-$ & 5.8 $\pm$ 2.1 & 0.75 $\pm$ 0.22 & 2.2 $\pm$ 0.9 & $-$ & 8.23 $\pm$ 0.02 &  $-$ & 72 $\pm$41 &  $-$ & 75 $\pm$29 \\
 & R2 & 1.55 $\pm$ 0.03 & 0.17 $\pm$ 0.01 & $-$ & 8.1 $\pm$ 1.2 & $-$ & 3.4 $\pm$ 0.4 & $-$ & 8.39 $\pm$ 0.04 &  $-$ & 122 $\pm$18 &  $-$ & 120 $\pm$16 \\
UGC 10297 & R1 & 1.21 $\pm$ 0.01 & 0.32 $\pm$ 0.01 & $-$ & 8.0 $\pm$ 1.2 & 1.57 $\pm$ 0.45 & 10.6 $\pm$ 1.7 & $-$ & 8.22 $\pm$ 0.03 &  $-$ & 130 $\pm$21 &  $-$ & 18 $\pm$4 \\
UGC 10331 & R1 & 4.08 $\pm$ 0.05 & 0.45 $\pm$ 0.01 & $-$ & 6.2 $\pm$ 0.7 & 0.56 $\pm$ 0.26 & 3.8 $\pm$ 0.7 & $-$ & 8.47 $\pm$ 0.03 &  $-$ & 438 $\pm$98 &  $-$ & 256 $\pm$30 \\
 & R2 & 2.92 $\pm$ 0.04 & 0.32 $\pm$ 0.01 & $-$ & 7.7 $\pm$ 0.7 & 0.63 $\pm$ 0.23 & 5.3 $\pm$ 0.8 & $-$ & 8.36 $\pm$ 0.03 &  $-$ & 476 $\pm$77 &  $-$ & 220 $\pm$21 \\
UGC 10650 & R1 & 3.29 $\pm$ 0.04 & 0.15 $\pm$ 0.01 & $-$ & 5.9 $\pm$ 1.0 & $-$ & 6.5 $\pm$ 0.7 & $-$ & 8.35 $\pm$ 0.03 &  $-$ & 319 $\pm$34 &  $-$ & 70 $\pm$16 \\
UGC 6320 & R1 & 13.90 $\pm$ 0.12 & 0.13 $\pm$ 0.01 & 0.66 $\pm$ 0.15 & 4.4 $\pm$ 0.4 & 0.29 $\pm$ 0.11 & 6.6 $\pm$ 0.4 & 1.05 $\pm$ 0.15 & 8.29 $\pm$ 0.02 & 162 $\pm$35 & 86 $\pm$42 & 20 $\pm$3 & 33 $\pm$4 \\
 & R2 & 2.82 $\pm$ 0.03 & 0.13 $\pm$ 0.01 & $-$ & 3.2 $\pm$ 0.8 & 0.55 $\pm$ 0.19 & 6.8 $\pm$ 0.8 & $-$ & 8.29 $\pm$ 0.02 &  $-$ & 41 $\pm$4 &  $-$ & 2 $\pm$1 \\
UGC 9663 & R1 & 0.98 $\pm$ 0.01 & 0.10 $\pm$ 0.01 & $-$ & 4.9 $\pm$ 1.2 & 0.95 $\pm$ 0.30 & 7.9 $\pm$ 1.4 & $-$ & 8.14 $\pm$ 0.02 &  $-$ & 90 $\pm$17 &  $-$ & 7 $\pm$3 \\
\hline \noalign{\smallskip}
\multicolumn{14}{@{} p{\textwidth} @{}}{\textbf{Notes.} Column (1): name of the
  galaxy. Column (2): region identification number. Column (3): the extinction
  corrected \hb line intensity of the region showing WR features. Column (4):
  extinction coefficient. Columns (5--9): line intensity ratios of the broad
  band components of the blue bump plus the nebular \heii$\lambda$4686  emission
  line and the broad component of the red bump with respect to \hb, normalized
  to \mbox{\hb = 100}. We used the law by \cite{Cardelli89}, assuming $R_V =
  3.1$  and case B recombination~\citep{Osterbrock89}, in order to correct the
  emission-line fluxes for internal extinction. Column (10): gaseous oxygen
  abundance of the region showing WR features, obtained using the
  $O3N2$-parameter (see text). Column (11): derived number of early-type WN
  stars. Column (12): derived number of late-type WN stars. Column (13): derived
  number of early-type WC stars using the flux measurement of the red bump. Column(14): estimated number of early-type WC stars using the flux measurement of the broad $\lambda$4650 feature.}
\end{tabular}
\end{minipage}
\end{sideways}
\end{footnotesize}
\end{table*}

All things considered, to find the number of each type of WR, we need first to estimate the metallicity
of the regions. Considering that we did not detect electron temperature
sensitive lines such as the \oiii $\lambda$4363 for the direct determination of
metallicity, we used strong line calibrations based on easily observable,
optical lines. 
%In the last decades several abundance calibrators involving different
%emission-line ratios have been proposed.
These empirical techniques have been applied to estimate oxygen
abundances in objects as different as individual \hii regions in spiral
galaxies, dwarf irregular galaxies, nuclear starbursts and emission-line
galaxies. However, caution must be taken when using the strong-line methods to
derive the oxygen abundance, as some of them differ by up to 0.6~dex
\citep[see][for a recent review]{Lopez-Sanchez12}. Given that some of the
galaxies in our sample of WR regions are relatively close (\mbox{$D_\mathrm{L} <
  30$ Mpc}; see Table~\ref{table:gal_catalogue}), the
\oii~$\lambda\lambda$3727,3729 lines do not lie in our observed
spectrum. Therefore, we used a calibration based on the $O3N2$ parameter, first
introduced by \cite{Alloin79}:

\begin{equation}
  O3N2 = \textrm{log} \left (\frac{\textrm{H}\alpha}{\textrm{\nii}\lambda6584} \times \frac{\textrm{\oiii}\lambda5007}{\textrm{H}\beta} \right ).
\end{equation}

We consider the calibration proposed by~\cite{Marino13}, 
%which is based on direct determinations of the oxygen abundance using the
%electron temperature of the ionised gas. This calibration 
which is valid for \mbox{12 + log(O/H) $> 8.1$} and
has a dispersion somewhat lower than 0.2~dex. In any case, we compared the
derived oxygen abundances with those obtained using the ``counterpart'' C-method,
introduced by~\cite{Pilyugin12b}, which 
%uses \oii~$\lambda\lambda$3727,3729 
is valid for all the metallicity range, 
%When the comparison was possible, 
obtaining consistent values within 0.1~dex.
%although the dispersion is somewhat higher, $\sim$0.2~dex, in the
%extreme cases, 12 + log(O/H) $\sim 8.0-8.2$ and $\sim 8.6-8.8$. 
The metallicities derived using the $O3N2$ calibration are listed in
Table~\ref{table:wr_fluxes}. The oxygen abundances range between 12 + log(O/H) =
8.2 and 8.6, with the typical value being about half of the solar value.

%% \begin{figure}
%% \centering
%% \includegraphics[angle=90,trim = 0cm 0cm 0cm 0cm,clip=true,width=0.95\columnwidth]{figs/dillution.eps}
%%   \caption{Flux of the \mbox{\heii~$\lambda$4686} broad component in 10$^{-16}$
%%     cgs units for WR regions and the associated \ha clumps (\hii\onespace) when
%%     the aperture of the WR region is clearly smaller. The solid red lines
%%     indicates the 1:1 relation, the dashed lines $\pm$ 20\% and the dotted lines
%%     $\pm$ a factor of 2. Dots filled in green denote when
%%     \mbox{$\varepsilon_\mathrm{orig}$ (\hii\onespace) $<$ 5}, circled in blue
%%     those cases with a broader component in the \hii spectra, and with a red dot
%%     in the centre those with fewer components are fitted in the \hii spectra.}
%%   \label{fig:dillution}
%% \end{figure}

\begin{figure*}
  \hspace{0.5cm}
  \includegraphics[trim = 0cm 8.5cm 11.2cm 13cm,clip=true,width=0.9\columnwidth]{figs/bb_fits_1.eps}
  \hspace{0.5cm}
  \includegraphics[trim = -1cm 7.5cm 10.2cm 8.5cm,clip=true,width=0.9\columnwidth]{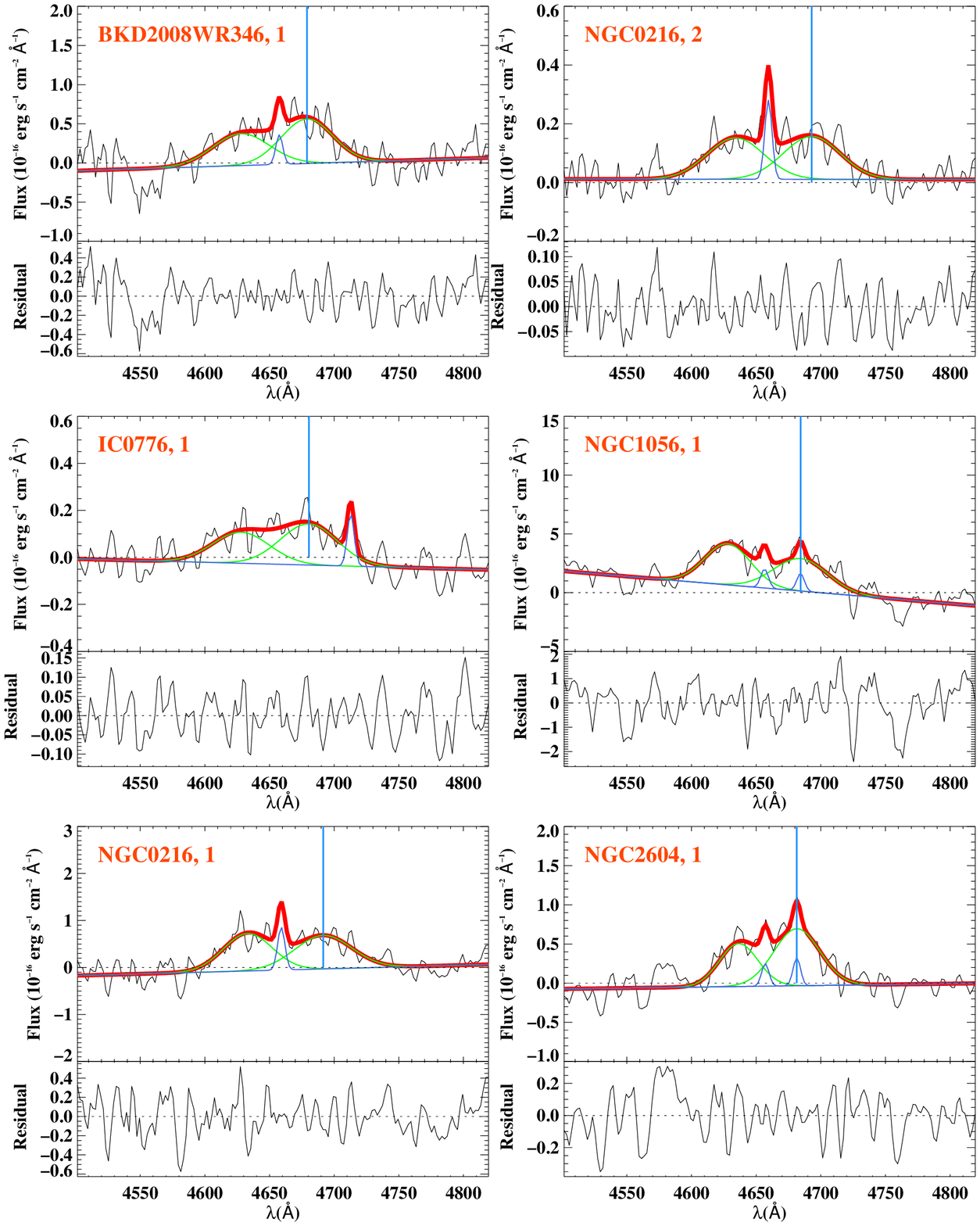}
  \caption{R1 in IC~0776 as typical example showing the effects of the dilution,
    for the spectrum of the WR region (left) and for the spectrum of the
    associated \ha clump (right). Lines and colours are the same as in
    fig.~\ref{fig:example_bb_fits}. Some structure is lost and fewer nebular
    lines are fitted when the WR feature is diluted. It is also noticeable that
    the broad lines widen.}
  \label{fig:lines_dilluted}
\end{figure*}

%%of the fluxes of the broad line emission features and the
%% oxygen abundance of the regions that show WR emission we estimate, when
%% possible, . 

With this information we derived the number of WR stars for each sub-type, when
possible. The number of WNE, WNL and WCE stars are compiled in the last columns of
Table~\ref{table:wr_fluxes}. We estimated the number of WNL and WCE stars by solving Eq.~\ref{eq:system} in all cases,
of WNE stars in just 2 cases (those showing a clear detection of the
\nv$\lambda\lambda$4605,4620 blend) and the number of WCE stars using Eq.~\ref{eq:wce} in the five
regions where the red WR bump is present in our spectra. In those cases where the red bump is detected we considered the number of WCE stars derived via the flux measurement of such bump to compute their contribution to the broad \mbox{\heii $\lambda4686$} emission. It is also convenient to mention that if we did not consider to solve the system and estimated instead the number of WNL stars by using directly Eq.~\ref{eq:wnl_4686}, the resulting values would be higher by $10-20$\%.

The derived values span a wide
range, from a few dozens in regions within the closest galaxies (\mbox{NGC 4630}
and \mbox{UGC 6320}) to more than 30 thousands in regions belonging to the most
distant objects (\mbox{NGC 6090} and \mbox{NGC 7469}). It is worth noting the
detection of more than 30 thousand WRs in the nuclear region of a Seyfert 1
galaxy. It is not the first time that the coexistence of the AGN and a
circumnuclear  compact starburst located several hundreds of pc away has been
reported (e.g.,in \mbox{Mrk 477}~\citealt{Heckman97}). In fact, the estimated
number of WRs in that compact source amounts to about 30 thousands. 

We can compare the estimation of the number of WCE stars in those cases where the red bump is detected, N$_{\mathrm{WCE}}$ (RB) and N$_{\mathrm{WCE}}$ (4650) in Table~\ref{table:wr_fluxes}. Only in one case (NGC 3353) can we consider both values compatible within the uncertainties. A difference within a factor of two is encountered between both estimates in the remaining WR regions. This is not surprising given the uncertainties involved in these estimates, like the possible systematics when using average luminosities that range enormously in each methodology (for instance, in~\citealt{Crowther06a} luminosities of single WC4s in LMCs range from less than 2 to more than \mbox{4$\times 10^{36}$ erg s$^{-1}$}). Finally, we remark that the number of WCEs that we obtain by solving the equation system (Eq.~\ref{eq:system}) represents generally between 15 and 25\% the number of WNLs\footnote{Should we assume the \mbox{\heii $\lambda4686$} broad emission comes entirely from WNLs.} as we 
expected, given 
that \mbox{EW($\lambda$ 4650) $\gtrsim$ EW($\lambda$ 4686)}. We even find some unexpected cases (e.g., R1 in NGC 3994, R6 in NGC 5665, R1 and R2 in UGC 00312) 
where \mbox{EW($\lambda$ 4650) $>$ $2\times$EW($\lambda$ 4686)}, 
and where the percentage is higher than 40\%. Again, the systematics may play an important role in these estimates. All in all, we conclude that the estimates in the number of WCEs are probably correct within a factor of two.

\subsection{Dilution of the WR features}

One of the main biases when interpreting observational data is the limitation
imposed by the spatial resolution. As mentioned before, WRs are usually very
localized, sometimes the emission of the associated \hii region is significantly
more extended, i.e., the spatial extent of WRs can be considerably smaller than
that of the rest of the ionising population~\citep{Kehrig13}. This is one of the
main reasons why the emission of WRs easily dilutes in non-resolved studies. In
this section, we discuss the effects of this dilution on the detectability and
determination of the fluxes of the broad emission features when using the
techniques described in this paper. Indeed, this directly affects the
determination of the number of WR stars and constitutes a crucial step when
interpreting the observed flux and when comparing them with stellar population
models.

%It is quite clear that 
In several cases the spatial extent of the \ha clumps
(i.e., \hii regions or aggregates) associated to our WR regions is larger than
the area showing WR emission (see Fig.~\ref{fig:ha_maps}). If we integrated the
spectrum of these clumps we would expect the WR bump to either disappear or
dilute significantly. Then, if we performed the multiple-fit of the WR emission
features we would expect to obtain lower equivalent widths or detection
significance. 
Several factors affect the final measured fluxes when the dilution effect is
present:

%% We did this exercise with those
%% regions for which the associated \ha clump is more extended. It is
%% clear, from Fig.~\ref{fig:dillution} that our expectations were not confirmed,
%% since in most cases the fluxes were found to be enhanced by more than a factor
%% of two. Several reasons can explain this:

\begin{itemize}

 \item Once the bump is diluted, its detection significance level decreases;
   and, if it is too much diluted it can actually disappear in the observed
   spectrum. As mentioned in Sec.~\ref{sec:line_fit}, when
   \mbox{$\varepsilon_\mathrm{orig} < 5$}, the shape of the WR feature is not
   very well constrained, since it is extremely model dependent on the continuum
   subtraction, which leads to unrealistic features.
%   Besides, when performing this subtraction, the
%   feature is normally enhanced and, if the continuum increases in the spectrum
%   of the \ha clump (we have added up the spectra of more pixels), then it is
%   not uncommon to recover a unrealistic feature, more intense than that in the
%   spectrum of the original region.
   This happens, for instance, in R1 in
   \mbox{IC 0225}. When we integrate only the pixels showing WR features,
   \mbox{$\varepsilon_\mathrm{orig} = 6.3$}, and when we integrate the whole \ha
   clump \mbox{$\varepsilon_\mathrm{orig} = 4.2$}. In \mbox{NGC 1056}, the bump
   actually disappears in the observed spectrum
   (\mbox{$\varepsilon_\mathrm{orig} = 2.6$}). 

 \item If the bump dilutes, barely resolved lines (i.e., weak nebular lines,
   weak broad components) dissolve, thus fewer and broader components are
   fitted. In order to properly interpret the measurement of the fluxes of the
   broad bump features, we have to correct for the nebular emission lines; and,
   specifically for the case of the broad \mbox{\heii ~$\lambda$4686} line, try
   to isolate it from the blend at \mbox{4640--4650  \AA{}}. This effect is
   illustrated in Fig.~\ref{fig:lines_dilluted}, for R1 in \mbox{IC 0776}.

 \item In some cases, even if the same number of components are fitted, they
   become broader. If the lines are broader and the peak is similar then more
   flux is measured.

\end{itemize}

%% As Fig.~\ref{fig:dillution} illustrates, when we integrate the spectra of the
%% \ha clumps, in most cases one, two or even the three effects just described
%% above occur. 
These effects justify why we did not consider in our analysis the
0-class regions. Even if none of these effects is observed the recovered fluxes
for the 0-class  WR regions can vary for large factors, leading to very uncertain WR
properties. Note that this does not affect the WR regions with the highest
detection significance since their aperture is similar to that of the associated
\ha clump (e.g., in \mbox{NGC 3353}, \mbox{NGC 3773}, \mbox{NGC 3991}).
It is worth noting that dilution effects should be taken into account when
deriving properties of WR populations, specially in observations with poor spatial
resolution and if the sampled area is larger than the area showing WR
emission.

\subsection{Comparison with stellar population models} 

\subsubsection{Definition of the sub-sample}

\begin{table*}
\fontsize{8.2}{10}\selectfont
\begin{minipage}{\textwidth}
\renewcommand{\footnoterule}{}  % to avoid a line before footnotes
%\begin{footnotesize}
\caption{Derived properties of the WR population and WR number ratios}
\label{table:wr_pop}
\begin{center}
\begin{tabular}{@{\hspace{0.05cm}}l@{\hspace{0.20cm}}c@{\hspace{0.20cm}}c@{\hspace{0.20cm}}c@{\hspace{0.20cm}}c@{\hspace{0.20cm}}c@{\hspace{0.20cm}}c@{\hspace{0.20cm}}c@{\hspace{0.20cm}}c@{\hspace{0.20cm}}c@{\hspace{0.20cm}}c@{\hspace{0.20cm}}c@{\hspace{0.10cm}}c@{\hspace{0.05cm}}}
\hline \hline
   \noalign{\smallskip}
Galaxy & Region & H$\beta$ (clump) & c (H$\beta$)  & Blue WR bump &  Red WR bump  & EW (\hb) &  $\tau$ & $\eta_{0}$ &  N$_{\mathrm{0}}$ & N$_{\mathrm{WR}}$ / N$_{\mathrm{0}}$ & $Q_{0}^{\mathrm{Total}}$ & $Q_{0}^{\mathrm{WR}}$   \\
 	& ID & ($\times 10^{-14}$ cgs) & &  EW (\AA{}) & EW (\AA{}) & (\AA{})  & (Myr) &  &  &   &log (cgs)  & log (cgs)\\
(1)	& (2) &   (3) &  (4) & (5) & (6) & (7) & (8) & (9) & (10)  & (11)  & (12) & (13) \\
 \hline
    \noalign{\smallskip}
IC 0225 & R1 & 3.23 $\pm$ 0.07 & 0.06 & 1.4 $\pm$ 0.8 & $-$ & 31  $^{+3}_{-1}$  & 5.5 $\pm$ 0.3 & 0.30 $ \pm $0.05  & 551 $^{+276}_{-251}$  & 0.14 $^{+0.22}_{-0.08}$  & 51.56  & 51.31 $\pm$0.16  \\
   \noalign{\smallskip}
NGC 3353 & R1 & 81.51 $\pm$ 0.54 & 0.19 & 3.8 $\pm$ 0.4 & 0.9 $\pm$ 0.2 & 90  $^{+5}_{-3}$  & 4.5 $\pm$ 0.3 & 0.25 $^{+0.10}_{-0.05}$  & 4919 $^{+2430}_{-2095}$  & 0.17 $^{+0.21}_{-0.09}$  & 52.52  & 52.32 $\pm$0.11  \\
   \noalign{\smallskip}
NGC 3381 & R1 & 6.61 $\pm$ 0.12 & 0.15 & 4.5 $\pm$ 0.8 & $-$ & 45  $^{+5}_{-6}$  & 4.3 $\pm$ 0.2 & 0.35 $^{+0.15}_{-0.10}$  & < 172   & > 1.75  & 51.92  & > 51.89   \\
   \noalign{\smallskip}
NGC 3773 & R1 & 22.07 $\pm$ 0.31 & 0.07 & 3.5 $\pm$ 1.1 & 4.5 $\pm$ 1.6 & 46  $^{+18}_{-3}$  & 5.2 $\pm$ 0.2 & 0.25 $ \pm $0.05  & < 1139   & > 0.19  & 51.95  & > 51.91   \\
   \noalign{\smallskip}
NGC 4630 & R1 & 11.01 $\pm$ 0.14 & 0.28 & 3.0 $\pm$ 0.7 & $-$ & 49  $^{+7}_{-3}$  & 4.3 $\pm$ 0.5 & 0.35 $^{+0.15}_{-0.13}$  & 252 $^{+138}_{-80}$  & 0.24 $^{+0.13}_{-0.09}$  & 51.38  & 51.19 $\pm$0.04  \\
   \noalign{\smallskip}
NGC 5954 & R2 & 12.03 $\pm$ 0.20 & 0.42 & 0.9 $\pm$ 0.2 & $-$ & 41  $^{+4}_{-3}$  & 5.5 $\pm$ 0.3 & 0.30 $ \pm $0.05  & 5123 $^{+1018}_{-783}$  & 0.04 $^{+0.01}_{-0.01}$  & 52.32  & 51.77 $\pm$0.06  \\
   \noalign{\smallskip}
UGC 6320 & R1 & 13.90 $\pm$ 0.12 & 0.13 & 4.5 $\pm$ 0.6 & 1.1 $\pm$ 0.2 & 74  $^{+9}_{-3}$  & 4.8 $\pm$ 0.2 & 0.22 $ \pm $0.05  & 744 $^{+759}_{-506}$  & 0.31 $^{+0.85}_{-0.20}$  & 51.88  & 51.78 $\pm$0.08  \\
   \noalign{\smallskip}
 & R2 & 6.64 $\pm$ 0.07 & 0.09 & 1.9 $\pm$ 0.5 & $-$ & 79  $^{+23}_{-8}$  & 4.8 $\pm$ 0.2 & 0.22 $^{+0.13}_{-0.05}$  & 1124 $^{+249}_{-285}$  & 0.04 $^{+0.01}_{-0.01}$  & 51.56  & 51.04 $\pm$0.05  \\
   \noalign{\smallskip}  
\hline \noalign{\smallskip} 
\multicolumn{13}{@{} p{\textwidth} @{}}{{\footnotesize \textbf{Notes.} Column (1): name of the galaxy. Column (2): region identification number. Column (3): the extinction
  corrected \hb line intensity of the corresponding \ha clump. Column (4):
  extinction coefficient. Column (5): equivalent width of the
  \mbox{\heii~$\lambda$4686} line measured in the WR region, corrected by the
  continuum emission of non-ionising populations. Column (6): same as (5) but
  for the broad \mbox{\civ~$\lambda$5808} line. Column (7): equivalent width of
  \hb measured in the \ha clump and corrected by the continuum emission of
  non-ionising populations. Column (8): derived age of the ionising population
  using the EWs reported in (7) and the \popstar models. Column (9): adopted
  $\eta_0$  parameter. Column (10): derived number O stars. Column (11): number
  ratio of WR stars with respect to the total numbers of O stars. Column
    (12): total Lyman Continuum flux, as derived with the \ha luminosity (the
    random uncertainty is in all cases it is below 2\%).  Column (13): estimated
    Lyman Continuum flux emerging from WR stars. }}

\end{tabular}
\end{center}
%\end{footnotesize}
\end{minipage}
\end{table*}

%% As discussed in the previous subsection, if we integrate the whole associated
%% \ha clump, the structure of the feature can be diluted and, when subtracting the
%% continuum, misleading results can be obtained. On the other hand, 

The \ha emission in \hh regions 
%is produced by ionising photons of O stars, this emission 
is typically more extended than the localized WR stellar emission in a given
cluster, giving rise to the dilution effect discussed above, a fact that has to
be taken into account when comparing spectroscopic observables with stellar
population models.
For instance, to make a proper comparison of the flux ratio of
the blue bump to \hb, we have to obtain the flux of \hb from the whole \hii spectrum,
%constrained to the spatial region of the WR emission, 
as done for instance in~B08a and discussed in \cite{Kehrig13} for \mbox{Mrk 178}.
% The former used a 3\arcsec~fibre ($\sim$ 60 pc) while the size of the radius of
% the associated giant \hii region is of the order of 150 pc.

Aperture effects can also deceive the comparison with models the other
way round. Non-resolved \ha clumps with sizes of several hundreds of pc or
larger can harbour several \hii regions. For instance, within an aperture of
\mbox{1 kpc} several regions in NGC 4630 are observed, but only the central
region shows significant WR emission features. This can happen even at smaller
apertures, such as in the giant \hii region \mbox{Tol 89} in \mbox{NGC 5398}
\citep{Sidoli06}. Within an aperture of about 300 pc there are at least
2  complexes (A and B) of massive (M $> 10^5$ \msun) clusters with ages within
the range $\tau \sim 2-5$ Myr, the most massive and youngest (B) not showing
signs of WR emission. Thus, if we compare the integrated WR and \hb emission in
this case, the interpretation of the properties of the population might be
wrong. This gives us an idea that \mbox{100--200 pc} can represent an
appropriate aperture to perform this kind of studies.

In this study, we have performed this comparison for regions with a radius
smaller than 400 pc (see Table~\ref{table:reg_catalogue}), in order to have at
least several regions at our disposal. Thus, in a few cases (
\mbox{NGC 3353}, \mbox{NGC 3381}, \mbox{NGC 5954}) our results may still be somehow misleading
provided that there is a cluster young enough so that WRs are still not
present.

\subsubsection{WR ratios and predictions from stellar population models}

Empirical results such as the ratio of WR to O stars provide sensitive tests of
evolutionary models which involve complex processes (i.e., rotation,
binarity, feedback, IMF, etc.) that remain not sufficiently constrained. 
% Although is out of the main scope of this work to make a deep analysis of...
This ratio can be roughly derived by first estimating the number of O
stars using the \hb luminosity. ~We remind the reader that the Balmer emission in \hii regions is typically more extended than the localized WR stellar emission. Therefore, in order to properly compute the numer of O stars we have to use the \hb luminosity of the whole \ha clump where the WR emission is detected. Then, assuming a contribution to the \hb luminosity by an O7V star of \mbox{$L_{\mathrm{O7V}} = 4.76\times10^{36}$ erg s$^{-1}$}, a first estimation of the number of such stars is obtained by \mbox{N$_{\mathrm{O7V}} = L(\mathrm{H}\beta)/L_{\mathrm{O7V}}$}. Finally, the contribution of the WRs and other O subtypes to the ionising flux is corrected as explained below:

\begin{itemize}

 \item Following~\cite{Crowther06a}, the average numbers of ionising photons of
   a WN and a WC star are assumed to be log $Q_{0}^{\mathrm{WN}} = 49.4$ and
   $Q_{0}^{\mathrm{WC}} = 49.5$, respectively.   

 \item The total number of O stars (N$_\mathrm{O}$) can be derived from the
   number of O7V (N$_\mathrm{O7V}$) stars by correcting for other O stars
   subtypes, using the parameter $\eta_0$ introduced by~\cite{Vacca92}
   and~\cite{Vacca94}. This parameter depends on the initial mass function for
   massive stars and is a function of time because of their secular
   evolution~\citep{Schaerer98}. We made a rough estimation of the age of the
   population using the EW (\hb\onespace) measured on the \hii region
   spectra. This value was corrected by the emission of non-stellar ionising
   population, since the \starlight code provides us with the star formation
   history of the region (see~\citealt{Miralles-Caballero14}). With the
   approximate knowledge of the age of the population we estimated $\eta_0$
   using the SV98 models. The strongly non-linear temporal evolution of this
   parameter during some time intervals (see Fig.~21 in~\citealt{Schaerer98})
   causes some asymmetries in the determination of its
   uncertainty. Table~\ref{table:wr_pop} lists the derived age, corrected EW
   (\hb\onespace) and estimated $\eta_0$ for our sub-sample of regions.

\end{itemize}

\noindent
With this we determined the number of O stars as:

\begin{equation}
\label{eq:no}
 \mathrm{N}_\mathrm{O} = \frac{Q_0^{\mathrm{Total}} - N_\mathrm{WN}Q_0^{\mathrm{WN}} - N_\mathrm{WC}Q_0^{\mathrm{WC}}}{\eta_0 Q_0^{\mathrm{O7V}}},
\end{equation}

\noindent
where $Q_0^{\mathrm{Total}} = \mathrm{N}_{\mathrm{O7V}} Q_0^{\mathrm{O7V}} $ and
$Q_0^{\mathrm{O7V}}$ are the total and O7V number of ionising photons,
respectively. We have adopted an average Lyman continuum flux per O7V star of
\mbox{log $Q_0^{\mathrm{O7V}} = 49.0$}
(\citealt{Vacca92,Schaerer98,Schaerer99}). The number of O stars and the WR to O
ratio are reported in Table~\ref{table:wr_pop}. The errors are generally large,
making evident the high uncertainties involved when trying to derive these
properties. In two cases (R1 in \mbox{NGC 3353} and \mbox{NGC 3381}), the computed number of O stars is negative, a non-physical result. Nevertheless, given the large uncertainties involved in the determination of $\eta_O$ and the number of each sub-type of WR, we have been able to provide an upper-limit within a confidence level of 90\%. This allowed us to obtain a lower limit to the WR to O ratio. We remark that given the large uncertainties involved a Montecarlo simulation was run in order to obtain all the uncertainties. The median values in each simulation of N$_\mathrm{O}$ and N$_{\mathrm{WR}}/$N$_{\mathrm{O}}$ were the same to those obtained applying Eq.~\ref{eq:no} directly to the derived average values of its elements, except for R1 in \mbox{UGC 6320}. Taking advantage of the Montecarlo simulation, we realized that in this case a non-negligible percentage of estimates for N$_\mathrm{O}$ were negative, which indicates that we were close to the noise level to obtain a physical 
result. Using the Bayessian approach we restricted the distribution considering only positive N$_\mathrm{O}$ and then renormalized this new distribution with only physical values. We thereby obtained a more representative value for N$_\mathrm{O}$ by taking the median of the renormalized distribution, and which is reported in the table. The same accounts for N$_{\mathrm{WR}}/$N$_{\mathrm{O}}$ for R1 in \mbox{UGC 6320}.

Table~\ref{table:wr_pop} also provides the total estimated Lyman
Continuum flux and that emerging from WR stars. It is interesting to note that
the contribution from the WR population is generally higher than the flux
emerging from the other populations altogether. Even in those cases where the
number of WR stars is almost negligible compared to that of O stars (around
4\%, in \mbox{NGC 5954} and in R2 in \mbox{UGC 6320}), the ionising flux
emerging from WR populations accounts for 30\% of the total Lyman Continuum
budget.

We have compared our observational data with the predictions of two different
theoretical model sets: (i) \popstar~\citep{Molla09,Martin-Manjon10}, a
self-consistent set of models including the chemical and the spectro-photometric
evolution, for spiral and irregular galaxies, where star formation and dust
effects are important; and (ii) \bpass~\citep{Eldridge08,Eldridge09}, which includes the
binary evolution in modelling the stellar populations, that can extend the WR
phase up to longer than 10 Myr. Both models use the photoionization code
\cloudy~\citep{Ferland98} to predict the nebular emission.

The disagreement between models and data at moderate and low metallicities have
been known for some time (e.g.,~\citealt{Guseva00,Crowther06a,Perez-Montero10,Lopez-Sanchez10a}),
especially when trying to explain the derived large WR to O ratios from
observations. As discussed in~\cite{Crowther07}, the production of more WRs than
currently favoured by the models can be achieved by including binarity in the
evolutionary codes or when rotation is included in the stellar tracks. They have
become promising sources for an increased WR population. In fact, several
studies of samples of Galactic massive O stars support that binary interaction
dominates the evolution of massive stars 
(e.g.~\citealt{Kobulnicky07,Sana12,Kiminki12}). \cite{Eldridge08} developed a
synthesis population code, \bpass\footnote{http://www.bpass.org.uk/}, where
binarity is included. They found out that a third of the population evolves as
single stars, while the remaining two thirds correspond to interacting
binaries. The inclusion of binaries led to a prolonged WR phase (up to
\mbox{$\tau \sim$ 15 Myr}), consistent with earlier predictions
by~\cite{vanBever03}.

\begin{figure*}
\centering
\includegraphics[angle=90,width=0.95\textwidth]{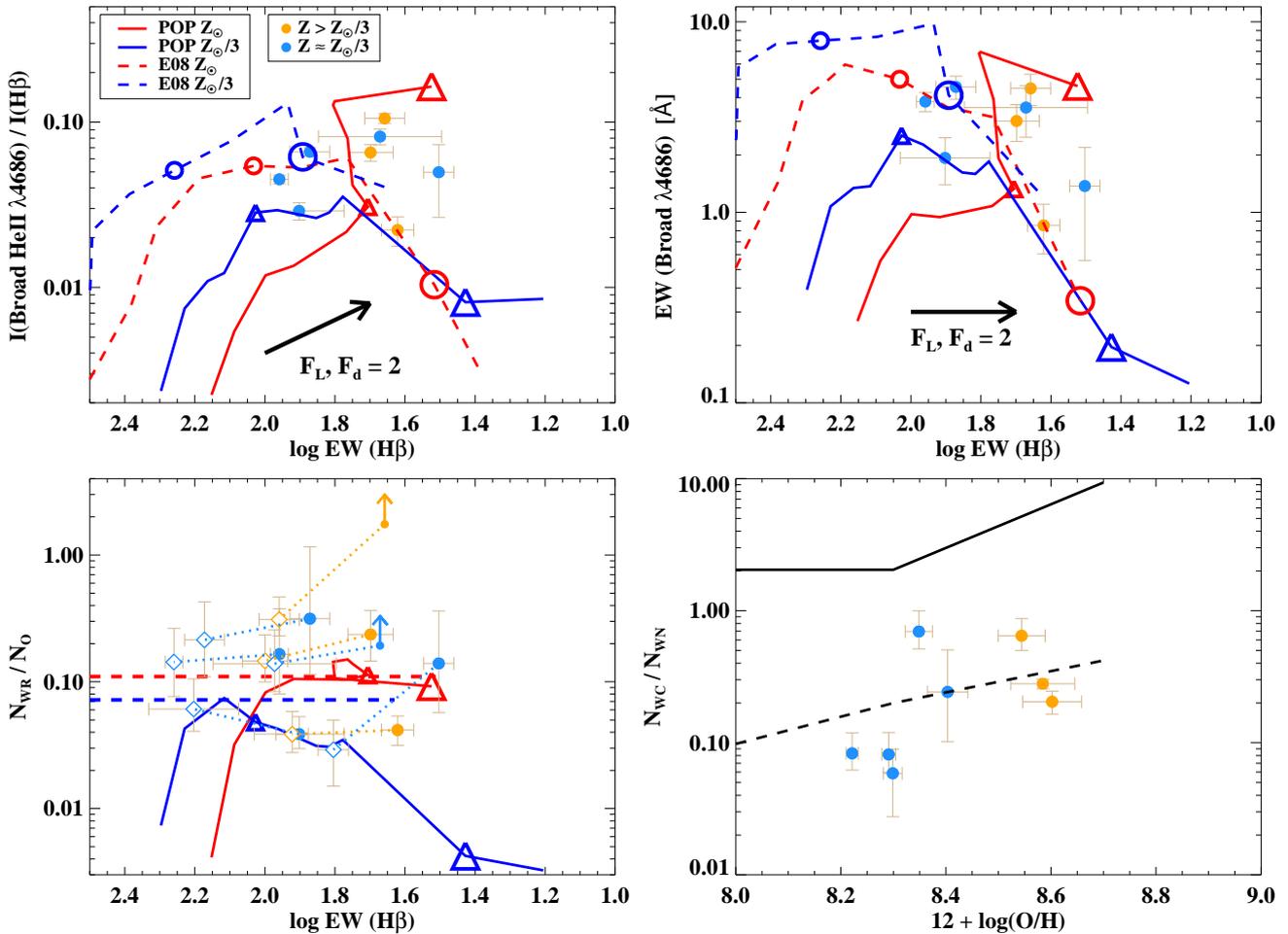}
  \caption{\textbf{Top}: Intensity ratio of the \mbox{\heii 4686 \AA{}} broad
    line to \hb (left) and the EW of the broad line (right)  vs. EW
    (\hb\onespace), a good indicator of the age of young ionising
    populations. The data are coloured depending on their metallicity. Tracks
    from \popstar (solid line; simple population) and \bpass (dashed line;
    binary population) are displayed in different colours also depending on the
    metallicity. Open triangles mark the values for 4.5 (small triangles) and
    5.5 Myr  (large triangles), respectively, on the \popstar tracks. Open
    circles mark the values for 6 (small circles) and 12.5 Myr  (large circles),
    respectively, on the \bpass tracks. The arrows labelled as
    \mbox{F$_\mathrm{L}$ = 2} and \mbox{F$_\mathrm{d}$ = 2} illustrate how the
    tracks would move on the plot if half of the ionising photons actually
    escape the ionised regions or are absorbed by dust grains within the \hii
    region. \textbf{Bottom-left:} Derived ratio of the number of WRs over the
    number of O stars  vs. EW (\hb\onespace). Lower-limits are provided for R1 in \mbox{NGC 3381} and \mbox{NGC 3773}. Open diamonds show how the data
    would move in the plot if half of the photons escaped or were absorbed (the
    dotted lines connect each case); here, the upper-limits dissapear because having been the \hb flux increased, the derived number of O stars is not negative any more. The tracks from the \bpass models are
    completely horizontal because the only average values over the time of the
    WR phase are provided. \textbf{Bottom-right:} Ratio of the derived number of
    WC over WN stars vs. metallicity. Solid (dashed) line marks the track for
    the POPSTAR (BPASS) model. 
}
  \label{fig:pop}
\end{figure*}

As observed in Fig.~\ref{fig:pop} (top) simple stellar
population models (\popstar\onespace) are not able to reproduce for the low-metallicity
regions (Z $\sim$ Z$_{\sun}$/3) either the flux ratio \mbox{I (broad \heii
  $\lambda$4686) / I(\hb\onespace)} or the EW of the blue bump corrected by the
continuum emission of the non-ionising population. For the sake of completeness,
we also show how the modelled tracks would move if half of the ionising photons
escape from the \hii regions (i.e., $F_\mathrm{L} \equiv 2.0$) or if they are
absorbed by dust grains within the nebula (i.e., $F_\mathrm{d} \equiv 2.0$).
Thus, if we allow for some Lyman photon leakage or some absorption by dust
grains within the nebula, the \popstar models could reproduce the flux
ratio. However, in any case, for three regions only the binary models could
explain the relatively high blue bump equivalent widths (Fig.~\ref{fig:pop}, top-right).

Reproducing the WR to O ratio, namely N$_\mathrm{WR}$/N$_\mathrm{O}$, is hard
for the stellar population models, especially for the low-metallicity regions
(Fig.~\ref{fig:pop}, bottom-left). Besides, as mentioned in MC14b, this ratio
can vary dramatically if we allow for some Lyman photon leakage, some
absorption by dust grains within the nebula, or even a different IMF. 
Note that this variation is not the same for all regions (like in the top panel), since if EW (\hb\onespace)
increases, the age decreases and $\eta_0$ (which is practically constant in some
age bins and practically highly non-linear in others) may vary almost nothing or
dramatically, thus the variation of N$_\mathrm{WR}$/N$_\mathrm{O}$ also ranges
from almost nothing to a dramatic change. Those regions in which the ratio is
higher than 0.2 cannot be reproduced at all with any kind of model if at least
half of the ionising photons are missing by either process. Since the ratios in
binary models are in general higher than those in simple models, the former are
generally close to the observed ratios.

We could also compare the ratio of carbon to nitrogen WRs, namely
N$_\mathrm{WC}$/N$_\mathrm{WN}$, with the predictions of models (Fig.~\ref{fig:pop}, bottom-right), taking into account the high uncertainties involved specially when deriving N$_\mathrm{WC}$. While simple models
predict too high ratios, our derived values are closer to the predictions made
by binary models. Yet, discrepancies with factors of $\sim$2 can be
encountered. A similar agreement is generally found at moderate and low
metallicities when effects of rotation and binarity are included in the
modelling (e.g.,~\citealt{Neugent11,Neugent12}).

As a conclusion, we find that only in a few cases the observables
can be reproduced by simple star models. In general, processes such as binarity,
fast rotation, or even more complex processes (e.g. photon leakage, absorption by dust grains, IMF
variations, etc.) are need to be included in the theoretical modelling of the
stellar populations in order to better understand how massive stars evolve. 
This follows the recent line of discussion in observational studies of WR populations
(e.g.,~\citealt{Crowther07};~B08a;~\citealt{Lopez-Sanchez10a,Bibby10,Bibby12,Kehrig13}).

\subsection{WRs and the environment: the GRB-WR connection}

\begin{figure*}
\centering
\includegraphics[angle=90,trim = 0cm 0cm 0cm 0cm,clip=true,width=0.95\columnwidth]{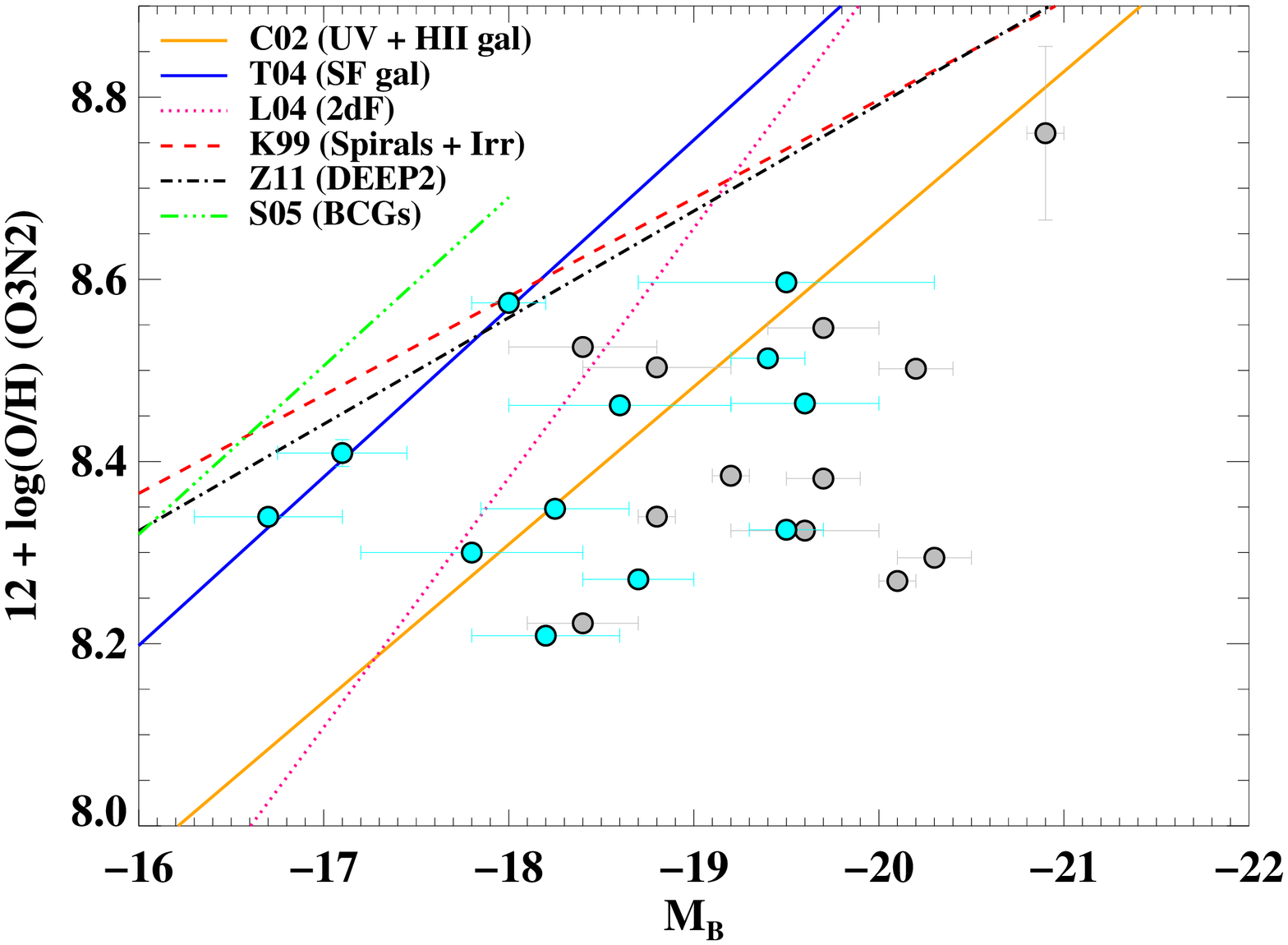}
\includegraphics[angle=90,trim = 0cm 0cm 0cm 0cm,clip=true,width=0.95\columnwidth]{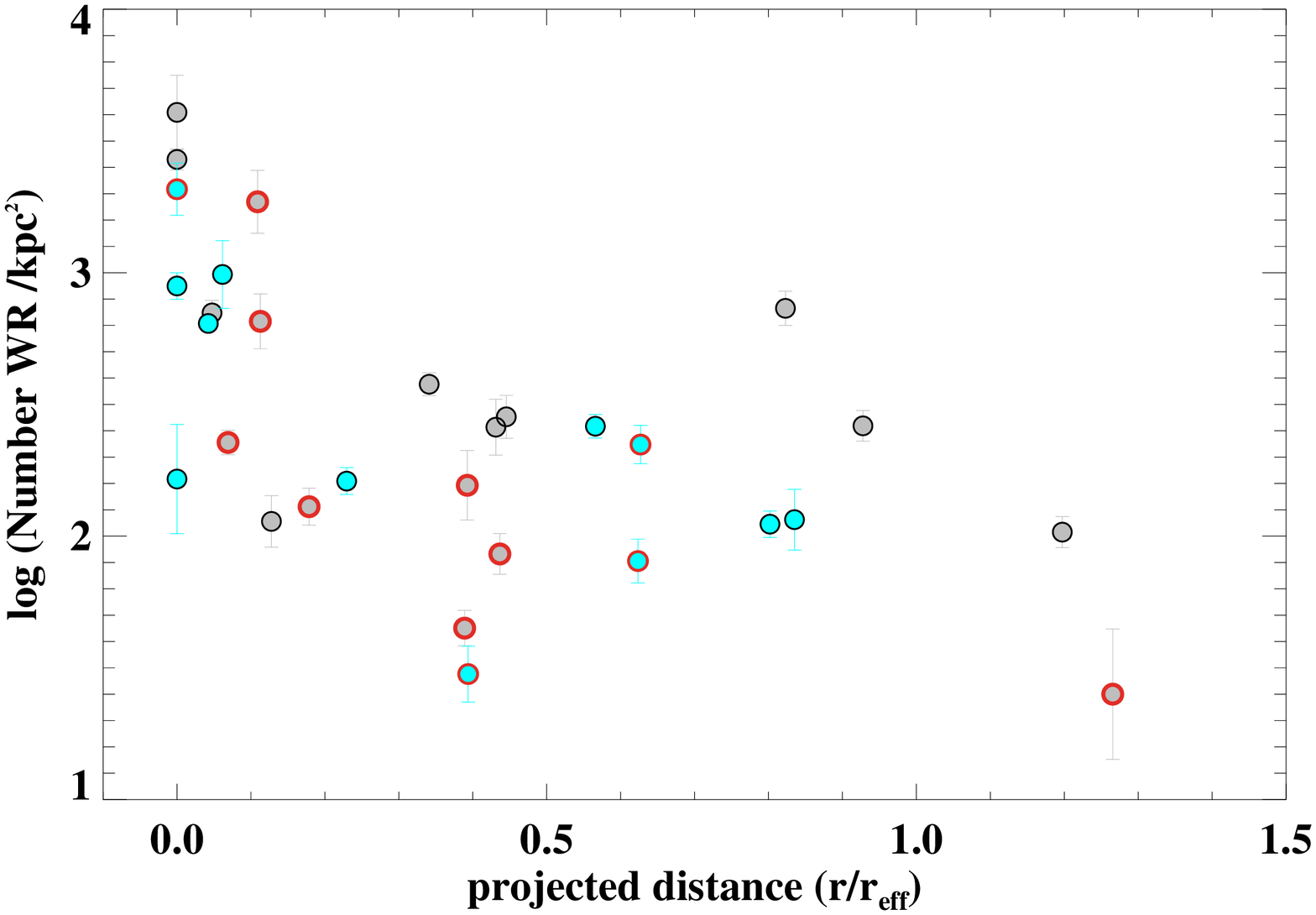}
 \caption{\textbf{Left:} The luminosity-metallicity ($L - Z$) relation of the
   galaxies in the CALIFA survey with detected WR stars, in grey those showing
   some signs of undergoing or recent interaction (M), and in blue those
   classified as isolated (I). $L - Z$ relations for various galaxy samples from
   the literature are also drawn: UV-selected galaxies (solid orange; Contini et
   al.~\citeyear{Contini02}), SDSS star-forming galaxies (solid blue; Tremonti
   et al.~\citeyear{Tremonti04}), a large magnitude-limited sample (dotted
   pink;~Lamareille et al.\citeyear{Lamareille04}), irregular and spiral galaxies
   (dashed red; Kobulnicky et al.~\citeyear{Kobulnicky99a}), emission-line
   galaxies at \mbox{z $\sim$ 0.8} from the Deep Extragalactic Evolutionary
   Probe 2 survey~(dotted-dashed;~Zahid et al.~\citeyear{Zahid11}) and
   star-forming BCGs (three dotted-dashed green; Shi et
   al.~(\citeyear{Shi05}). The large error bars in the absolute magnitudes refer
   to the uncertainties in the galaxy distance moduli, taken from
   NED. \textbf{Right:} number density of WRs vs. their projected
   galactocentric distance. The colours of the regions denote if they belong to
   a merging (M) or an isolated (I) galaxy. Regions circled in red are those
   whose host galaxy does not follow the luminosity-metallicity relation and
   have simultaneously a metallicity \mbox{12 + log (O/H) $<$ 8.4}.}
  \label{fig:mb_Z}
\end{figure*}

As mentioned in Sec.~\ref{sec:catalogue}, 13 out of the 25 galaxies hosting WR
regions show signs of interaction. Taking into account that young 
populations are also expected in normal spirals, irregulars and 
dwarf blue galaxies, this fact suggests that the number of WRs 
is particularly high in galaxy interactions. This is not entirely surprising, 
since merging processes are known to enhance the star formation activity in the
galaxies involved
\citep[e.g.][]{Kennicutt87,Barton03,Geller06,diMatteo08,Rodriguez-Zaurin10,Barrera-Ballesteros15}.
This is also in agreement with studies of supernova radial
distributions that found SNe Ibc (those associated with more massive stars and
GRBs) more centrally concentrated in disturbed/interacting galaxies
\citep{Habergham12}.

Very few studies have focused on the issue of connecting 
the environment of the galaxy, the presence of WRs and the 
production of GRBs (e.g.,\citealt{Hammer06,Han10}). MC14b investigated 
the environment of a single galaxy with strong presence WRs 
and tried to connect it with the properties of the hosts galaxies 
in \cite{Han10}, finding similarities in the integrated properties of 
GRB and WR hosts. A detailed study of the environment of local
galaxies with similar properties to the GRB hosts is essential to better
understand the physical properties of GRBs observed at moderate and high
redshifts and the nature of their progenitors. The 25 galaxies with positive
detection in this work allow us to dig deeper into this connection.

H10 performed a spectral analysis of 8 LGRB hosts in order to study the
environment in which such energetic events can take place. Since, according to
the core-collapse model, WRs are considered as the most favoured candidates to
being the progenitors of LGRBs, the presence of WRs in their hosts provided
evidence in favour. They identified some characteristics that support this
connection. One of them is that these galaxies do not follow the luminosity- and
mass-metallicity relations obeyed typically by irregular, spiral and
star-forming galaxies. The scenario of LGRBs occurring in host galaxies with
\textit{lower metallicities} than the general population is also supported by a
few observational studies \citep[e.g.][]{Modjaz11,Levesque10a,Graham13}. 
This under-abundance is naturally explained in galaxy mergers, since they induce
radial gas mixing processes, such as inflows of external gas on to the central
regions (e.g.,~\citealt{Barnes96,Rupke10a}).

We have explored in Fig.~\ref{fig:mb_Z} (left) if our sample of host galaxies
follow the luminosity-metallicity relation observed in the general
population. We obtained the metallicity of the galaxies by integrating the whole
spectrum within the CALIFA FoV and using the $O3N2$ calibration, as we did for
the \hii regions. For metallicities higher than about \mbox{12 + log(O/H) = 8.4}
the galaxies tend to follow the relation, although there are a couple of
outliers. In contrast, galaxies with lower metallicities show a clear dichotomy:
those with no clear signs of recent or current interaction processes (classified
as isolated; I in Table~\ref{table:gal_catalogue}) tend to follow the relation
while those identified as merging (M in Table~\ref{table:gal_catalogue}) are
clearly outliers. Actually, even the most diverted object of the sample of
isolated galaxies, \mbox{UGC 10297}, may present a disc truncation which could
indicate a past minor merger~\citep{Comeron12}.

We have at least a sub-sample of about 10 galaxies, most of them undergoing an
interaction, with similar characteristics to GRB hosts, namely moderate to low
metallicity \citep{Modjaz11} and showing a significant offset in
the luminosity-metallicity relation followed by the general galaxy
population \citep{Levesque10b}.
But, are those hosts the ones that have larger number of WRs? Where
are they located? Fig.~\ref{fig:mb_Z} (right) shows a correlation between the
number density of WRs (number of WRs/kpc$^2$) and their projected distance to
the centre of the galaxy. The Spearman's correlation coefficient is very similar
($\rho \sim$ -0.6) in regions from both the isolated and merging galaxy sample,
being somewhat stronger in the latter. While at distances shorter than about 0.3
\reff several hundreds and even thousands of WR are observed per kpc$^2$, in
general an order of magnitude or less is observed at longer galactocentric
distances.

The two regions with the highest number density are rather metal rich, and 2
regions out of 12 whose galaxies do not follow the luminosity-metallicity
correlation have densities higher than 1000 WRs/kpc$^2$. We could assume that in
those 12 regions the probability for a GRB to occur is higher. Therefore, the
existence of more WRs would not necessarily guarantee more GRB events and they
would not necessarily be in the very central regions (where larger number of WR
are found). In fact, there are some studies that do not associate the presence
of GRBs to that of WRs, at least not in the exact location of the WRs but at
spatial shifts of about hundreds of pc
away~\citep{Hammer06,Christensen08,Levesque11,Thone14}. Yet, the presence of WRs
is found in the host galaxies. In those studies, the location of the GRBs does
not lie necessarily in the central regions of the galaxy either (e.g.,
in~\citealt{Thone14} it is found about 7--8 kpc away). Although we have found
galaxies with similar properties to the GRB hosts, more resolved studies of
nearby GRB hosts are needed to better understand how these energetic phenomena
occur and what is their relation with the WR population.

\section[]{Conclusions}
\label{sec:conclusions}

In this exploratory work, 
we have developed a technique to perform an automated search of Wolf-Rayet (WR)
signatures through a pixel-by-pixel analysis of integral field spectroscopy
(IFS) data of local galaxies. This procedure has been applied to a sample of 558
galaxies from the CALIFA survey plus the extended projects, spanning a wide
range of physical properties. This represents the first systematic search of
these peculiar populations in a large sample of galaxies observed with IFS,
where we can have simultaneous spectral and spatial information. We present a
catalogue of 44 regions with clear signatures of WR emission in the blue bump (five
of them also showing the red bump), in 25 star-forming galaxies presenting
a variety of morphologies (and environments): irregulars, blue dwarfs, spirals
and interacting galaxies. We have performed a detailed analysis of the main
properties of these regions, drawing the following conclusions:

\begin{enumerate}

 \item We are able to detect WR emission in the spectrum of a pixel (with
   spatial scales spanning from 60 to about 600 pc) with equivalent widths (EW)
   of \ha typically higher than about 15 \AA{}. The detection of the red bump
   implies less contamination by underlying non-ionising populations, i.e.,
   \mbox{EW (\ha) $>$ 60 \AA{}}. For the candidate regions showing WR features,
   namely, WR regions, the distribution of EWs (ranging from 25 to 700 \AA{})
   peaks at 125--160 \AA{}, decreasing at larger values (i.e., younger
   ages). The presence of a turnover in this distribution is consistent with the
   WR phase starting a few Myr after the population is born.

 \item While most of the regions are distributed within one effective radius,
   only a third of them are found at the centre of the galaxy, within
   \mbox{$\sim$ 1 kpc} or less from the nucleus. While most WR regions are found
   in \ha clumps (i.e., \hii regions or aggregates), several seem to be
   associated with the diffuse \ha emission in the central regions of the
   galaxy. The latter are probably associated to \hii regions that are not
   resolved and/or not distinguished due to the high \ha surface brightness in
   the centre of the galaxies.

 \item We have performed a detailed fitting of the spectra of these regions
   considering the broad stellar and narrow nebular emission lines in the WR
   blue bump. Although large part of the emission is likely to be
   originated from late-type nitrogen WRs (WNL), direct positive detection from
   early-types (WNE) and carbon WRs (WC) has been obtained in 2 and 5 regions,
   respectively. Plausible evidence of a non-negligible percentage of early-type carbon WRs (WCEs) has been found in all regions.
  The WR regions host from dozens to more than 30,000 WRs, the latter
   in the two most distant galaxies. Actually, in one of the two (NGC 7469) a
   direct evidence has been found of the existence of strong star formation with
   large number of WRs and Seyfert activity within the central 2 kpc.

 \item There is a strong presence of WRs in
   galaxy interactions, irregulars and blue dwarf galaxies, where strong star
   formation is observed. Galaxies showing signatures of recent past or
   undergoing merging processes have in general similar integrated properties
   %(i.e., they do not follow the luminosity-metallicity correlation normally
   %found in other star-forming galaxies) 
   to the hosts of gamma-ray burst events
   that also show an important presence of WR populations. However, the
   existence of a large number of WRs does not necessarily guarantee more GRB
   events or that they have to occur necessarily in the host very central
   regions. This is consistent with recent studies on GRB hosts.

 \item We have discussed the effect of the dilution of the WR features when
   integrating the spectra at larger apertures than those where the feature is
   originally detected.
   %Due to the technique used to subtract the stellar
   %continuum, when the feature is diluted in the observed spectrum, a higher
   %flux is generally later recovered (up to or even larger than a factor of 2)
   %when fitting the features in the subtracted spectrum. S
   %Several reasons have been discussed in order to explain such behaviour: 
   %the loss of detection in the
   %original observed spectrum, the loss of structure of the feature
   %(i.e., loss of some nebular lines) 
   %and an increase of the width of the broad features due to
   %the dilution. 
   These effects have to be taken into account when applying WR searching
   techniques that are currently widely used.

 \item We have found clear evidence that the effects of binary
   stars and other processes (i.e., UV photon leakage, absorption by UV photons
   within the \hii nebula) need to be taken into account in the stellar population
   modelling in order to better reproduce the observed fluxes of the WR stellar
   emission lines, equivalent widths and the ratios of WR to O stars as well as
   between WR sub-types, especially at moderate to low metallicities. If
   binarity is important, the WR phase could last up to 10 Myr or so. Although
   very low metallicity regimes are not explored in this study, our result is
   consistent with previous studies that claim that simple star models fail
   dramatically to reproduce many of the observed properties of these stars in
   the Local Universe in low metallicity environments.

\end{enumerate}

\section*{Acknowledgments}

DMC,\,AID and FFRO would like to acknowledge financial support provided by the
project AYA2010-21887-C04-03
(former \emph{Ministerio de Ciencia e Innovaci\'on}, Spain) as well as the
exchange programme ``Study of Emission-Line Galaxies with Integral-Field Spectroscopy''
(SELGIFS, FP7-PEOPLE-2013-IRSES-612701), funded by the EU through the IRSES
scheme.
AMI acknowledges support from Agence Nationale de la Recherche through the
STILISM project (ANR-12-BS05-0016-02).
RGD acknowledges support through the project AYA2014-57490-P.
CJW acknowledges support through the Marie Curie Career Integration Grant 303912.
This research has made use of the NASA/IPAC Extragalactic Database (NED) which
is operated by the Jet Propulsion Laboratory, California Institute of
Technology, under contract with the National Aeronautics and Space
Administration.
Support for LG is provided by the Ministry of Economy, Development, and
Tourism's Millennium Science Initiative through grant IC120009, awarded to The
Millennium Institute of Astrophysics, MAS. LG acknowledges support by CONICYT
through FONDECYT grant 3140566.

\bibliographystyle{aa}

\bibliography{my_bib.bib}{}

\clearpage

\appendix

\section{Additional figures}

%\begin{figure*}
%\hspace{-9cm} %odd page
\begin{minipage}{\textwidth}
%\centering
\includegraphics[trim = 0cm 0cm 0cm 4cm,clip=true,width=0.95\textwidth]{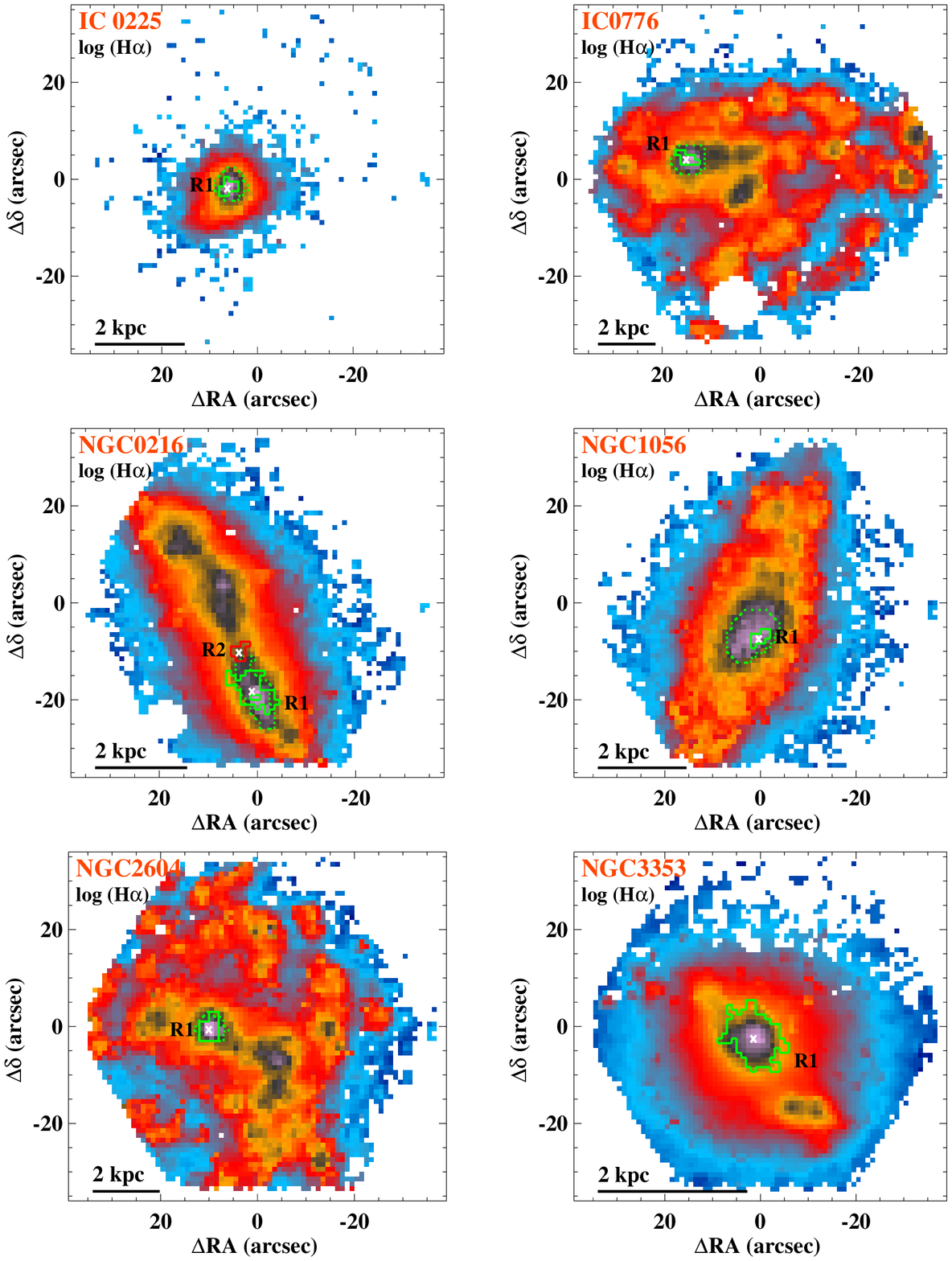}
  \captionof{figure}{\ha maps with logarithmic intensity scale of the galaxies
    with detected WR emission within the labelled regions enclosed by the green
    and red continuum contours. The pointted contours correspond to the
      associated \ha clump indentified using \hiiexplorer, whenever the emission
      of the \ha clump is more extended than that of the WR region. A cross
    indicates the barycenter of the region. The
    scale corresponding to \mbox{2 kpc} is drawn at the bottom-left
    corner. North points up and East to the left.}

 \label{fig:ha_maps}
\end{minipage}
%\end{figure*}

\begin{figure*}
\hspace{1cm}
 \includegraphics[trim = 0cm 0cm 0cm 4cm,clip=true,width=0.95\textwidth]{figs/ha_wr_maps_2.eps}
\addtocounter{figure}{-1}   
    \caption{-- Continued}
\end{figure*}

\begin{figure*}
\hspace{1cm}
 \includegraphics[trim = 0cm 0cm 0cm 4cm,clip=true,width=0.95\textwidth]{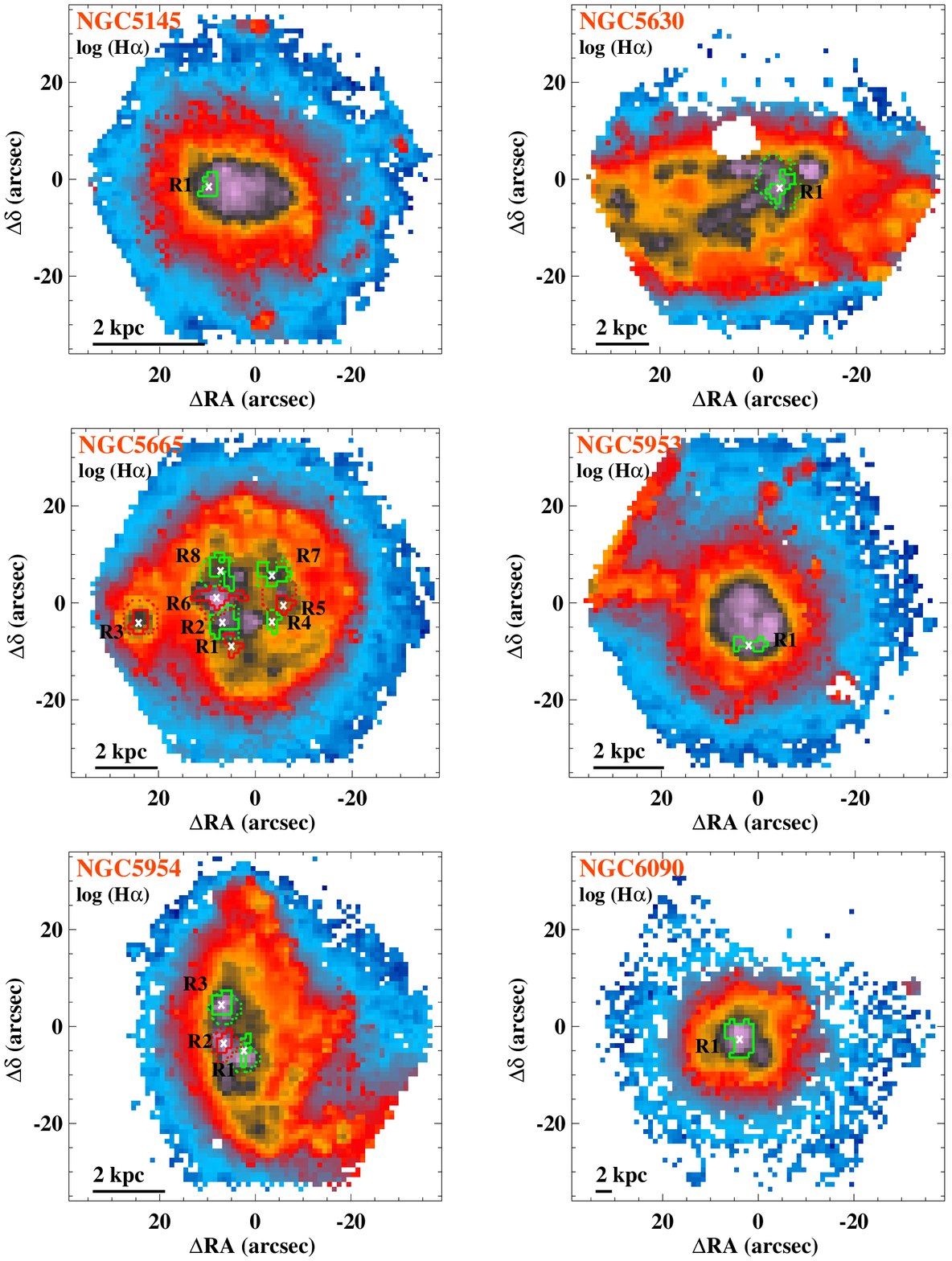}
\addtocounter{figure}{-1}   
    \caption{-- Continued}
\end{figure*}

\begin{figure*}
\hspace{1cm}
 \includegraphics[trim = 0cm 0cm 0cm 4cm,clip=true,width=0.95\textwidth]{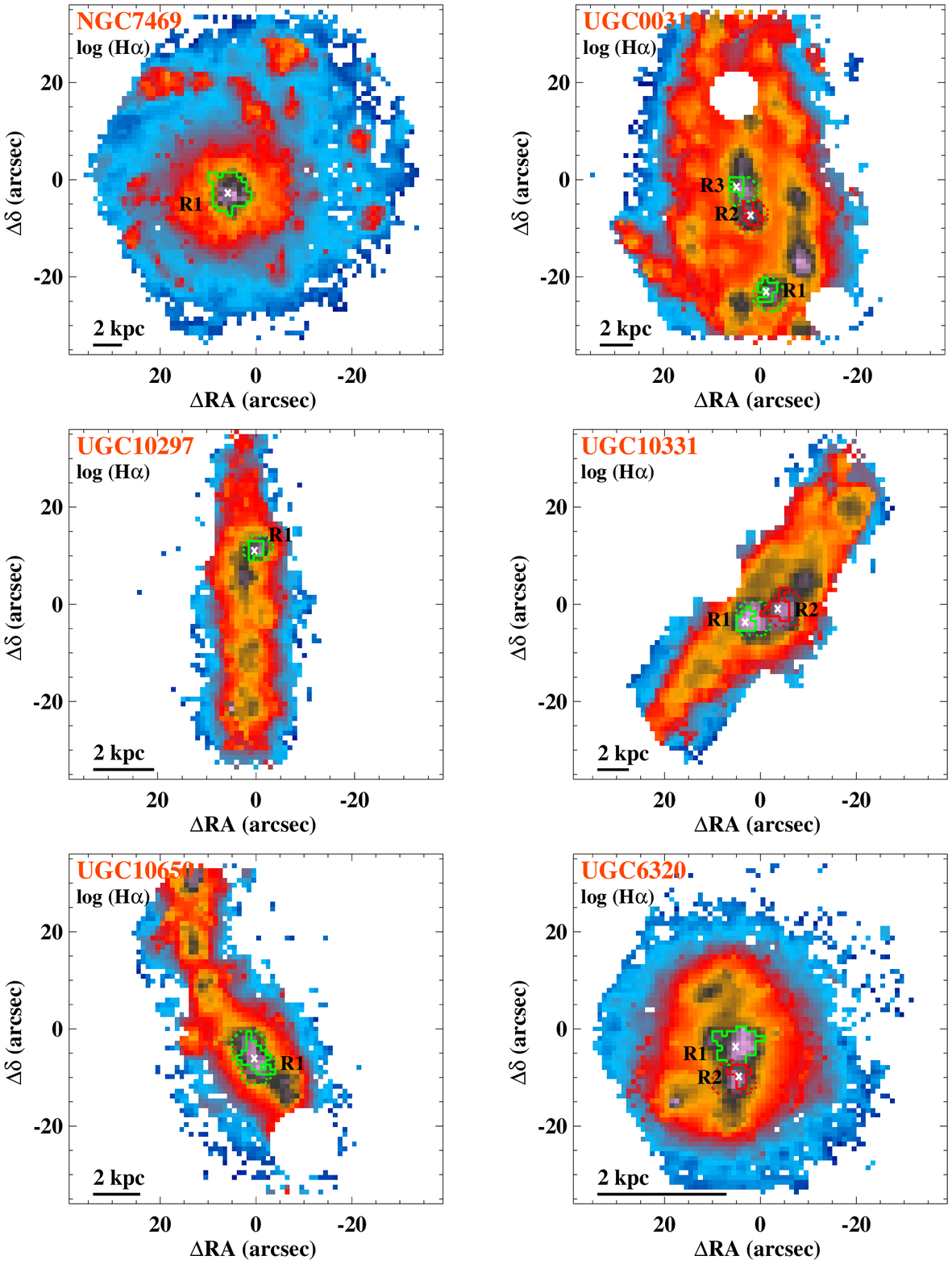}
\addtocounter{figure}{-1}   
    \caption{-- Continued}
\end{figure*}

\begin{figure*}
\hspace{1cm}
 \includegraphics[trim = 0cm 17cm 0cm 4cm,clip=true,width=0.95\textwidth]{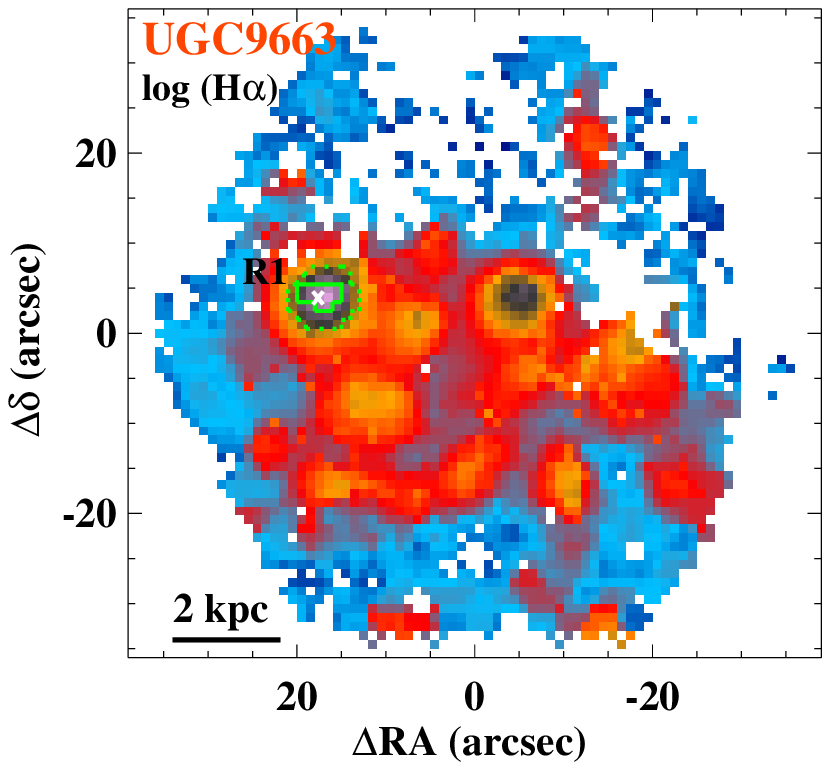}
\addtocounter{figure}{-1}   
    \caption{-- Continued}
\end{figure*}

\clearpage

\hspace{-9cm}
\begin{figure*}
\begin{minipage}{\textwidth}
\centering
\includegraphics[trim = 0cm 0cm 0cm 4cm,clip=true,width=0.95\textwidth]{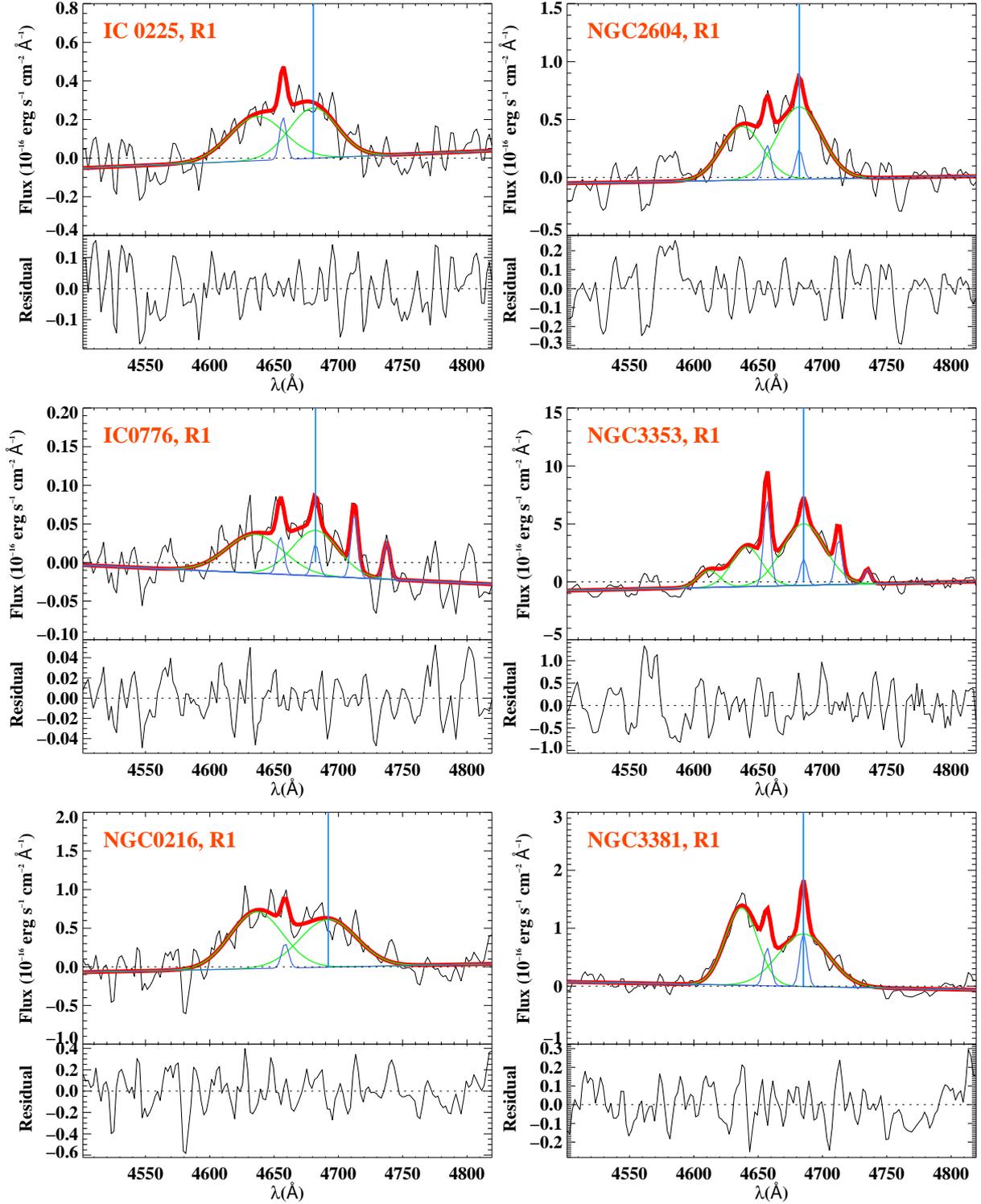}
  \captionsetup{width=0.8\textwidth}
 \captionof{figure}{Multiple-line fit of WR features within the blue bump. An almost horizontal blue line denotes the resulting continuum of the fit. The total fitted continuum plus emission lines to the blue bump is drawn by a  thick-red line. The nebular (blue) and broad stellar (green) components of the fit are also drawn. The vertical blue line indicate the position of the \mbox{\heii 4686 \AA{}} line. In each case, an auxiliary plot shows in black the residuals (in flux units) after modelling all the stellar and nebular features. For the case of NGC 7469, the fitting has been done on the observed spectrum, as explained in the text.}
 \label{fig:bb_fits}
\end{minipage}
\end{figure*}

\begin{figure*}
\hspace{1cm}
 \includegraphics[trim = 0cm 0cm 0cm 4cm,clip=true,width=0.99\textwidth]{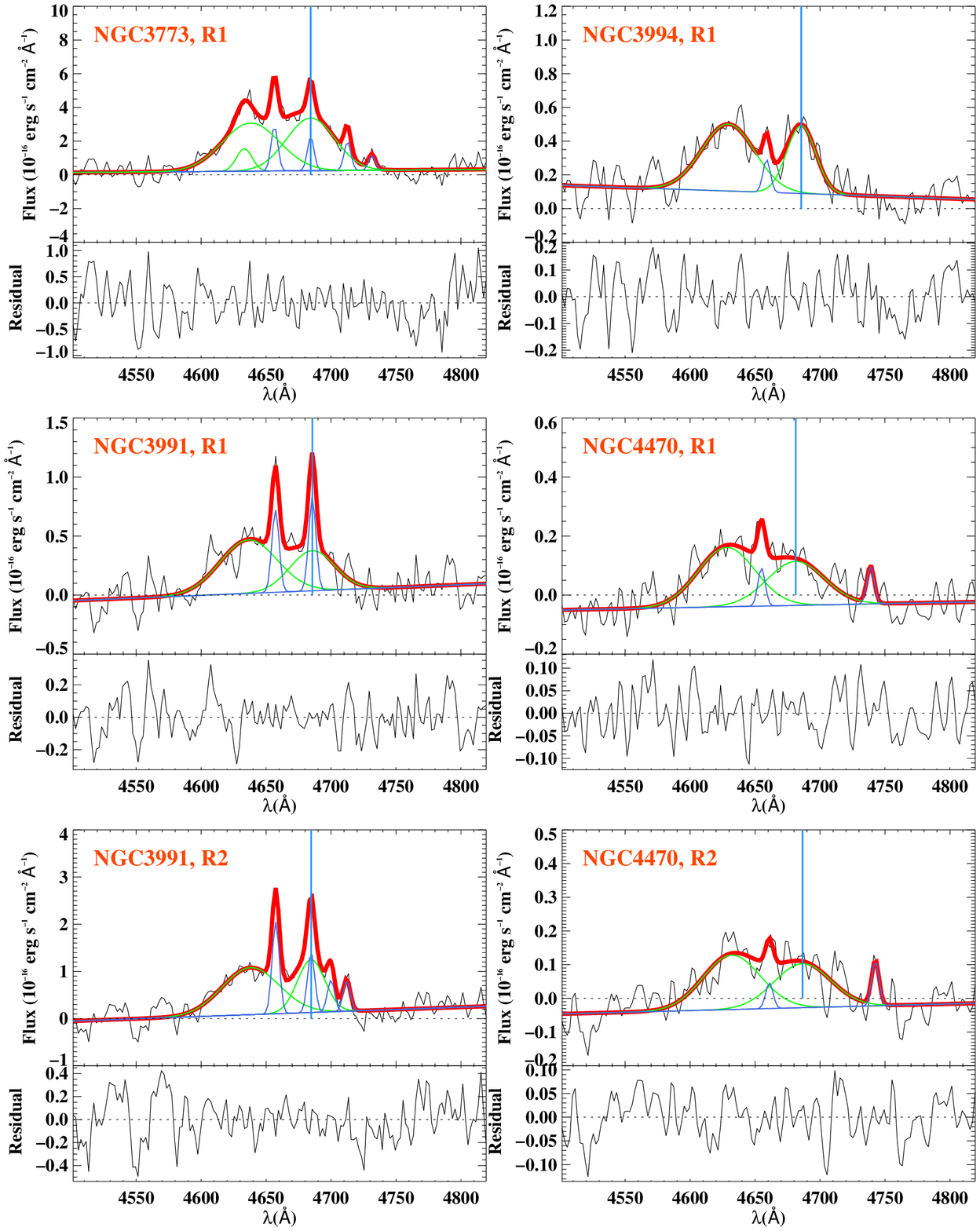}
\addtocounter{figure}{-1}   
    \caption{-- Continued}
\end{figure*}

\begin{figure*}
\hspace{1cm}
 \includegraphics[trim = 0cm 0cm 0cm 4cm,clip=true,width=0.99\textwidth]{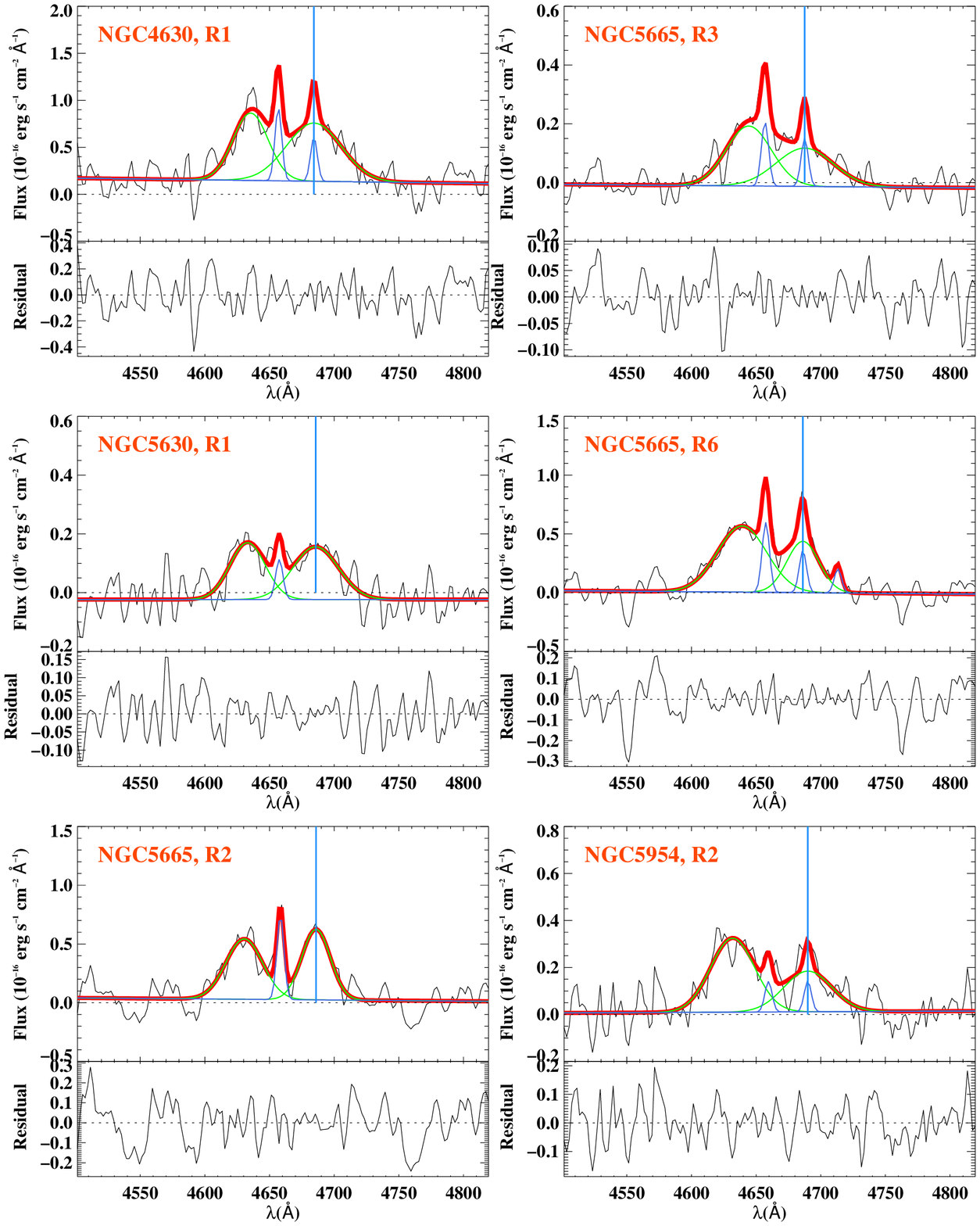}
\addtocounter{figure}{-1}   
    \caption{-- Continued}
\end{figure*}

\begin{figure*}
\hspace{1cm}
 \includegraphics[trim = 0cm 0cm 0cm 4cm,clip=true,width=0.99\textwidth]{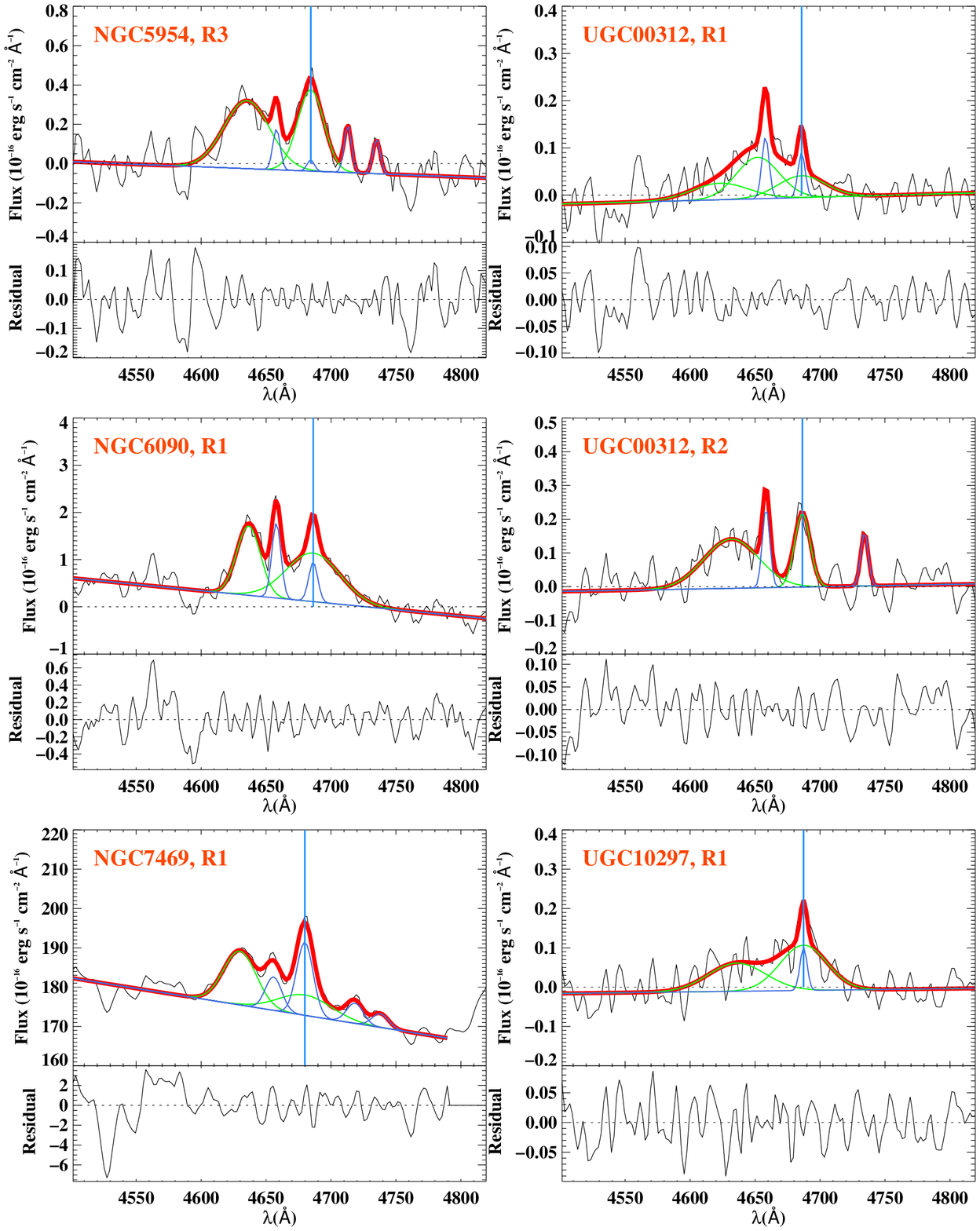}
\addtocounter{figure}{-1}   
    \caption{-- Continued}
\end{figure*}

\begin{figure*}
\hspace{1cm}
 \includegraphics[trim = 0cm 0cm 0cm 4cm,clip=true,width=0.99\textwidth]{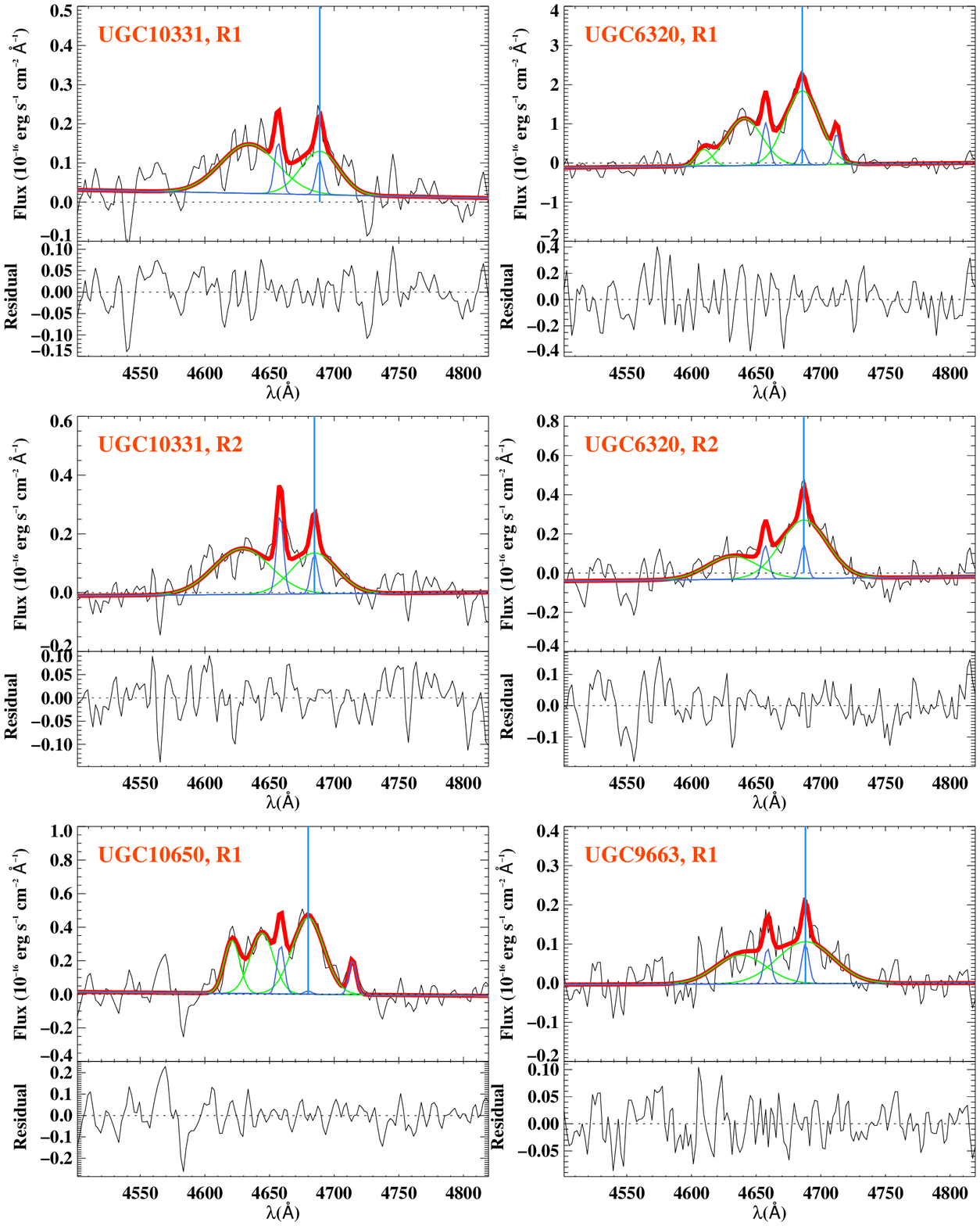}
\addtocounter{figure}{-1}   
    \caption{-- Continued}
\end{figure*}

\clearpage

%\hspace{-9.6cm}
%\begin{figure*}
\begin{minipage}{\textwidth}
\centering
\includegraphics[trim = -1cm -0.5cm 0cm -1cm,clip=true,width=0.9\textwidth]{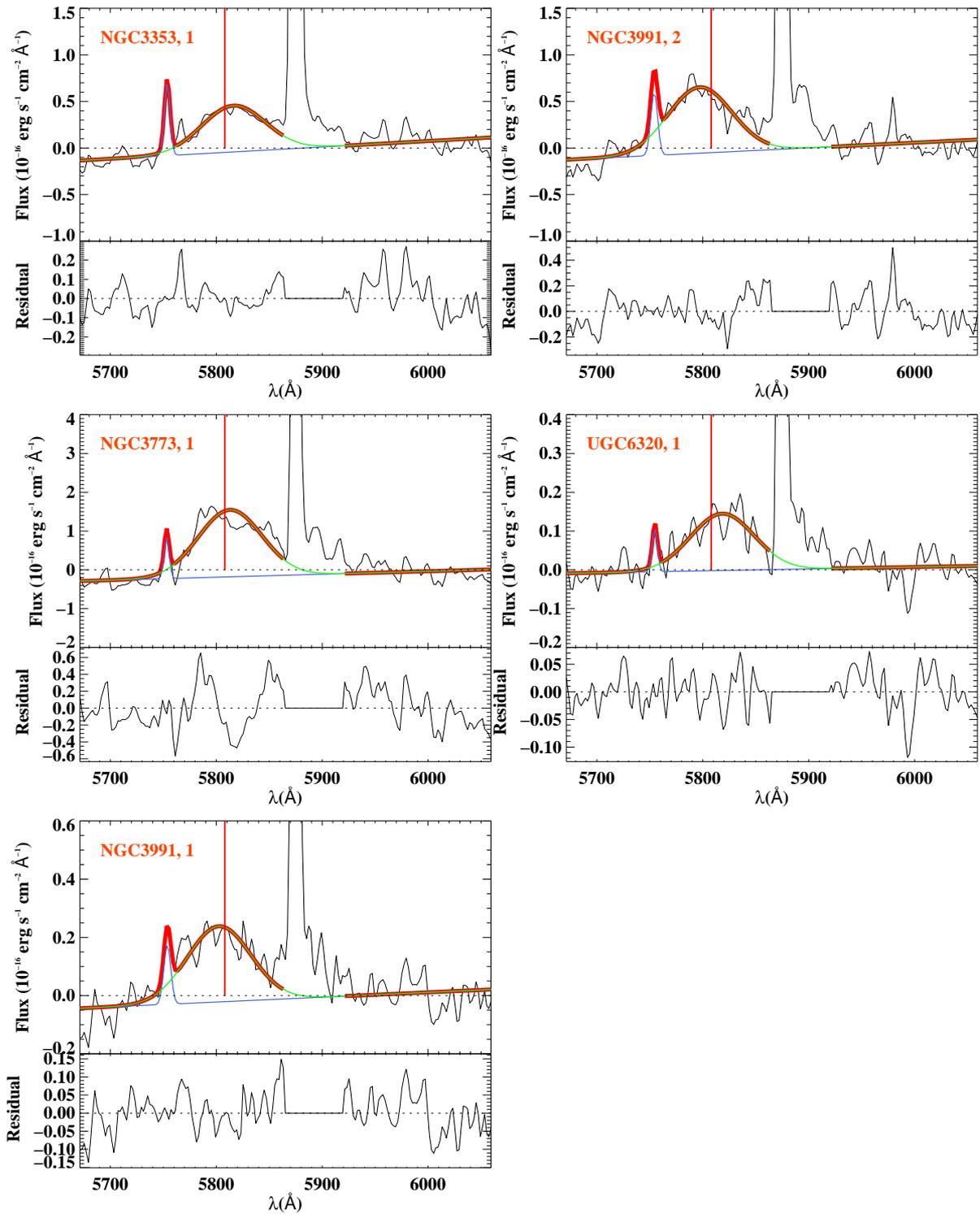}
  \captionsetup{width=0.85\textwidth}
 \captionof{figure}{Multiple-line fit of WR features within the red bump. In each figure the \textit{residual} spectrum is shown in black. This corresponds to the emission gaseous and stellar line spectrum minus the modelled feature, in flux units. An almost horizontal blue line denotes the resulting continuum of the fit. The total fitted continuum plus emission lines to the blue bump is drawn by a thick-red line. The fit to the auroral \mbox{\nii~5755 \AA{}} line (blue) and broad stellar \mbox{\civ 5808 \AA{}} feature (green) components of the fit are also drawn. The vertical red line indicate the position of the \mbox{\civ 5808 \AA{}} feature. }
 \label{fig:rb_fits}
\end{minipage}
%\end{figure*}

\label{lastpage}

\end{document}

%% file: journals.tex
\def\apjl{ApJ}%
\def\apjs{ApJS}%
\def\ao{Appl.~Opt.}%
\def\aap{A\&A}%

%% file: table1.tex
\begin{table*}
\hspace{1.5cm}
\begin{minipage}{0.85\textwidth}
\renewcommand{\footnoterule}{}  % to avoid a line before footnotes
\begin{normalsize}
\caption{Catalogue of galaxies showing WR features using the CALIFA survey data}
\label{table:gal_catalogue}
\begin{center}
\begin{tabular}{lccccccc}
\hline \hline
   \noalign{\smallskip}
Galaxy & $D_{\mathrm{L}}$ & AR & DEC  & Morph. & Stage & i = cos$^{-1}$ (b/a) & $M_{\mathrm{B}}$  \\
 	& (Mpc) & & & type & &  (deg) & (mag)  \\
 (1)	& (2) & (3) & (4) & (5) & (6) &  (7) & (8)  \\
 \hline
   \noalign{\smallskip}
IC 0225 & 21.3 & 02h 26m 28.2s & +01$^{\circ}$ 09m 39.2s & dE & I & 30 & -17.1 \\
IC 0776 & 33.9 & 12h 19m 3.1s & +08$^{\circ}$ 51m 23.5s & Sdm & I & 53 & -18.7 \\
NGC 0216 & 21.1 & 00h 41m 27.2s & $-$21$^{\circ}$ 02m 40.4s & Sd & I & 70 & -16.7 \\
NGC 1056 & 21.6 & 02h 42m 48.2s & +28$^{\circ}$ 34m 30.3s & Sa & I & 61 & -19.6 \\
NGC 2604 & 29.0 & 08h 33m 23.0s & +29$^{\circ}$ 32m 21.1s & Sd & M & 47 & -19.7 \\
NGC 3353 & 12.7 & 10h 45m 22.1s & +55$^{\circ}$ 57m 39.4s & BCD/Irr & M & 47 & -18.4 \\
NGC 3381 & 22.5 & 10h 48m 24.9s & +34$^{\circ}$ 42m 43.1s & SBd & I & 47 & -19.4 \\
NGC 3773 & 12.7 & 11h 38m 12.8s & +12$^{\circ}$ 06m 44.4s & dE/dS0 & I & 29 & -18.2 \\
NGC 3991 & 44.3 & 11h 57m 31.6s & +32$^{\circ}$ 20m 29.4s & Sm & M & 74 & -20.1 \\
NGC 3994 & 42.7 & 11h 57m 36.9s & +32$^{\circ}$ 16m 41.3s & SABbc & M & 61 & -20.2 \\
NGC 4470 & 32.2 & 12h 29m 37.8s & +07$^{\circ}$ 49m 27.9s & Sc & I & 47 & -18.6 \\
NGC 4630 & 9.4 & 12h 42m 31.1s & +03$^{\circ}$ 57m 37.7s & IBm & I & 39 & -18.0 \\
NGC 5145 & 17.0 & 13h 25m 13.8s & +43$^{\circ}$ 16m 03.2s & S? & I & 52 & -19.5 \\
NGC 5630 & 36.7 & 14h 27m 36.9s & +41$^{\circ}$ 15m 28.1s & SBdm & M & 67 & -19.6 \\
NGC 5665 & 30.8 & 14h 32m 25.7s & +08$^{\circ}$ 04m 45.3s & SABc & M & 46 & -19.7 \\
NGC 5953 & 27.6 & 15h 34m 32.4s & +15$^{\circ}$ 11m 38.8s & Sa & M & 28 & -18.8 \\
NGC 5954 & 26.5 & 15h 34m 35.0s & +15$^{\circ}$ 12m 01.5s & SABc & M & 41 & -18.4 \\
NGC 6090 & 123.7 & 16h 11m 40.6s & +52$^{\circ}$ 27m 24.9s & Sd pec & M & 65 & -20.9 \\
NGC 7469 & 67.7 & 23h 03m 15.6s & +08$^{\circ}$ 52m 27.6s & SAB(s)dm & M & 45 & -21.2 \\
UGC 00312 & 59.3 & 00h 31m 23.9s & +08$^{\circ}$ 28m 02.2s & SBd & M & 63 & -20.3 \\
UGC 10297 & 32.1 & 16h 15m 28.9s & +18$^{\circ}$ 54m 16.6s & Sc & I & 83 & -19.5 \\
UGC 10331 & 62.2 & 16h 17m 21.0s & +59$^{\circ}$ 19m 14.0s & SABc & M & 75 & -19.2 \\
UGC 10650 & 41.1 & 17h 00m 14.6s & +23$^{\circ}$ 06m 24.8s & Scd & M & 80 & -18.8 \\
UGC 6320 & 14.8 & 11h 18m 17.2s & +18$^{\circ}$ 50m 50.5s & S? & I & 36 & -17.8 \\
UGC 9663 & 33.2 & 15h 01m 13.6s & +52$^{\circ}$ 35m 44.9s & Im & I & 41 & -18.2 \\
\hline \noalign{\smallskip}
\multicolumn{8}{@{} p{1.0\textwidth} @{}}{\textbf{Notes.} Column (1): name of galaxy. Column (2): luminosity distance, taken from NED\footnote{http://ned.ipac.caltech.edu/}. Columns (3) and (4): Right Ascension and declination of the centre of the galaxy.  Columns (5) and (6): morphological type and stage  of the galaxy (isolated, I, or merging, M). They were generally inferred by combining the independent visual classifications of several members of the CALIFA collaboration (see~\citealt{Walcher14}). For some galaxies references from the literature were taken (NED; IC~0225,~\citealt{Gu06}; IC~0776,~\citealt{Garcia-Lorenzo14}; NGC 3353,~\citealt{Sanchez-Portal00}; NGC~3373,~\citealt{Dellenbusch08}).  Columns (7) and (8) inclination angle (i) and B absolute magnitude, taken from NED.}
\end{tabular}
\end{center}
\end{normalsize}
\end{minipage}
\end{table*}

%% file: table2.tex
\begin{table*}
\hspace{0.6cm}
\begin{minipage}{0.93\textwidth}
\renewcommand{\footnoterule}{}  % to avoid a line before footnotes
\begin{normalsize}
\caption{Sample of regions found with positive detection of WR features}
\label{table:reg_catalogue}
\begin{center}
\begin{tabular}{lccccccccccc}
\hline \hline
   \noalign{\smallskip}
Galaxy & Region & $\Delta$AR & $\Delta$DEC  & d / $r_{\mathrm{eff}}$  & r & $\varepsilon$  & $\varepsilon~_{\mathrm{orig}}$ & EW (H$\alpha)_{\mathrm{obs}}$  & \ha & Class & Red \\
 	& ID & (arcsec) & (arcsec) &   & (pc) & &  & (\AA{}) & clump &  & bump\\
  (1)	& (2) & (3) & (4) & (5) & (6) & (7) & (8) &  (9) & (10) & (11) & (12) \\
 \hline
   \noalign{\smallskip}
IC 0225 & R1 & 1.7 & 0.0 & 0.0 & 346 & 10.4 & 6.3 & 121& 1 &  1 &  0 \\
IC 0776 & R1 & 12.6 & 6.5 & 0.4 & 547 & 9.0 & 6.6 & 421& 1 &  1 &  0 \\
NGC 0216 & R1 & -4.7 & -16.9 & 0.6 & 470 & 11.7 & 6.3 & 86& 1 &  1 &  0 \\
 & R2 & -2.0 & -8.9 & 0.3 & 180 & 11.0 & 4.6 & 25& 0 &  0 &  0 \\
NGC 1056 & R1 & -1.7 & -3.6 & 0.3 & 554 & 9.7 & 3.8 & 53& 1 &  0 &  0 \\
NGC 2604 & R1 & 1.8 & -0.3 & 0.0 & 421 & 13.9 & 9.0 & 139& 1 &  1 &  0 \\
NGC 3353 & R1 & -0.5 & 0.7 & 0.1 & 372 & 29.1 & 11.7 & 317& 1 &  1 & 1 \\
NGC 3381 & R1 & -0.8 & -0.4 & 0.0 & 365 & 31.2 & 10.5 & 115& 1 &  1 &  0 \\
NGC 3773 & R1 & 2.3 & -0.4 & 0.0 & 237 & 24.0 & 11.8 & 148& 1 &  1 & 1 \\
NGC 3991 & R1 & -1.0 & -6.7 & 0.4 & 956 & 13.0 & 5.9 & 154& 1 &  1 & 1 \\
 & R2 & 1.8 & 0.8 & 0.1 & 750 & 17.0 & 8.4 & 229& 1 &  1 & 1 \\
NGC 3994 & R1 & 1.0 & -7.8 & 0.8 & 428 & 10.2 & 5.5 & 109& 1 &  1 &  0 \\
 & R2 & -5.4 & -7.2 & 1.0 & 343 & 8.6 & 4.4 & 57& 1 &  0 &  0 \\
 & R3 & -0.1 & 7.9 & 0.8 & 343 & 7.8 & 4.2 & 66& 1 &  0 &  0 \\
NGC 4470 & R1 & 6.4 & -10.0 & 0.8 & 482 & 11.3 & 5.0 & 26& 1 &  1 &  0 \\
 & R2 & 11.8 & -3.8 & 0.8 & 519 & 9.8 & 5.0 & 58& 1 &  1 &  0 \\
 & R3 & 1.3 & 10.9 & 0.7 & 432 & 9.5 & 4.2 & 68& 1 &  0 &  0 \\
NGC 4630 & R1 & -1.2 & -0.2 & 0.0 & 133 & 14.1 & 6.7 & 168& 1 &  1 &  0 \\
NGC 5145 & R1 & 8.2 & -0.3 & 0.6 & 172 & 9.4 & 4.6 & 54& 0 &  0 &  0 \\
NGC 5630 & R1 & -10.0 & 3.8 & 0.4 & 849 & 8.8 & 5.4 & 72& 1 &  1 &  0 \\
NGC 5665 & R1 & 5.0 & -6.0 & 0.4 & 262 & 10.1 & 4.0 & 46& 0 &  0 &  0 \\
 & R2 & 6.9 & -1.0 & 0.3 & 468 & 11.7 & 5.2 & 125& 1 &  1 &  0 \\
 & R3 & 24.4 & -1.1 & 1.2 & 556 & 16.6 & 11.3 & 192& 1 &  1 &  0 \\
 & R4 & -3.5 & -0.9 & 0.2 & 262 & 11.1 & 4.4 & 38& 0 &  0 &  0 \\
 & R5 & -5.9 & 2.5 & 0.3 & 468 & 8.6 & 4.3 & 48& 1 &  0 &  0 \\
 & R6 & 8.1 & 4.1 & 0.4 & 504 & 15.6 & 7.1 & 219& 1 &  1 &  0 \\
 & R7 & -3.5 & 8.8 & 0.5 & 476 & 8.8 & 4.9 & 58& 1 &  0 &  0 \\
 & R8 & 7.2 & 9.8 & 0.6 & 461 & 9.9 & 4.8 & 101& 1 &  0 &  0 \\
NGC 5953 & R1 & 0.5 & -6.2 & 0.4 & 268 & 9.1 & 3.9 & 74& 0 &  0 &  0 \\
NGC 5954 & R1 & 0.7 & 0.9 & 0.1 & 372 & 9.3 & 3.9 & 74& 1 &  0 &  0 \\
 & R2 & 4.9 & 2.4 & 0.4 & 398 & 9.9 & 5.4 & 148& 1 &  1 &  0 \\
 & R3 & 5.4 & 10.5 & 0.9 & 423 & 10.7 & 6.7 & 279& 1 &  1 &  0 \\
NGC 6090 & R1 & 0.2 & -1.3 & 0.0 & 1915 & 16.2 & 8.9 & 51& 1 &  1 &  0 \\
NGC 7469 & R1 & 1.1 & -0.7 & 0.0 & 1306 & 10.0 & 8.9 & 67& 1 &  1 &  0 \\
UGC 00312 & R1 & -3.9 & -23.1 & 1.3 & 958 & 7.8 & 6.4 & 371& 1 &  1 &  0 \\
 & R2 & -0.7 & -7.2 & 0.4 & 932 & 8.5 & 7.7 & 109& 1 &  1 &  0 \\
 & R3 & 2.3 & -1.2 & 0.1 & 848 & 7.0 & 3.8 & 83& 1 &  0 &  0 \\
UGC 10297 & R1 & -1.1 & 16.2 & 0.6 & 432 & 8.7 & 6.9 & 231& 1 &  1 &  0 \\
UGC 10331 & R1 & 2.5 & -2.0 & 0.1 & 1107 & 8.8 & 5.4 & 161& 1 &  1 &  0 \\
 & R2 & -4.3 & 0.8 & 0.2 & 1082 & 10.7 & 6.4 & 137& 1 &  1 &  0 \\
UGC 10650 & R1 & 0.3 & -0.9 & 0.1 & 669 & 13.0 & 8.5 & 165& 1 &  1 &  0 \\
UGC 6320 & R1 & 1.7 & -0.3 & 0.1 & 295 & 29.1 & 12.4 & 256& 1 &  1 & 1 \\
 & R2 & 1.1 & -6.5 & 0.2 & 284 & 9.4 & 6.1 & 216& 1 &  1 &  0 \\
UGC 9663 & R1 & 15.1 & 8.4 & 0.6 & 598 & 10.0 & 7.6 & 699& 1 &  1 &  0 \\
\hline \noalign{\smallskip}
\multicolumn{12}{@{} p{\textwidth} @{}}{\textbf{Notes.} Column (1): name of the galaxy. Column (2): region identification number. Column (3): offset in Right Ascension from the centre of the galaxy. Column (4): offset in declination from the centre of the galaxy. Column (5): projected distance to the centre of the galaxy, normalised to the effective radius. Column (6): radius of the associated \ha clump (if not possible association, radius of the region), derived as $r = \sqrt{\mathrm{A}/\pi}$, where A corresponds to the number of pixels that form the region (note that each pixel represent an area of 1 arcsec$^2$). Column (7): $\varepsilon$ for the integrated spectrum of the region, after the subtracting the underlying stellar emission with \starlight (see text). Column (8): same as (7), but for the observed integrated spectrum. Column (9): \ha equivalent width of the observed integrated spectrum. Column (10): flag indicating whether a given region belongs to an \ha clump (1) or to the more extended \ha 
emission of the galaxy (0). Column (11): adopted class for the detection significance of the WR features in the regions. Column(12): flag indicating if the red bump is detected (1) or not (0).}
\end{tabular}
\end{center}
\end{normalsize}
\end{minipage}
\end{table*}

%% file: draft_WR_v4.bbl
\begin{thebibliography}{160}
\expandafter\ifx\csname natexlab\endcsname\relax\def\natexlab#1{#1}\fi

\bibitem[{{Abbott} {et~al.}(2004){Abbott}, {Crowther}, {Drissen}, {Dessart},
  {Martin}, \& {Boivin}}]{Abbott04}
{Abbott}, J.~B., {Crowther}, P.~A., {Drissen}, L., {et~al.} 2004, \mnras, 350,
  552

\bibitem[{{Allen} {et~al.}(1976){Allen}, {Wright}, \& {Goss}}]{Allen76}
{Allen}, D.~A., {Wright}, A.~E., \& {Goss}, W.~M. 1976, \mnras, 177, 91

\bibitem[{{Alloin} {et~al.}(1979){Alloin}, {Collin-Souffrin}, {Joly}, \&
  {Vigroux}}]{Alloin79}
{Alloin}, D., {Collin-Souffrin}, S., {Joly}, M., \& {Vigroux}, L. 1979, \aap,
  78, 200

\bibitem[{{Baldwin} {et~al.}(1981){Baldwin}, {Phillips}, \&
  {Terlevich}}]{Baldwin81}
{Baldwin}, J.~A., {Phillips}, M.~M., \& {Terlevich}, R. 1981, \pasp, 93, 5

\bibitem[{{Barnes} \& {Hernquist}(1996)}]{Barnes96}
{Barnes}, J.~E. \& {Hernquist}, L. 1996, \apj, 471, 115

\bibitem[{{Barrera-Ballesteros} {et~al.}(2015){Barrera-Ballesteros},
  {S{\'a}nchez}, {Garc{\'{\i}}a-Lorenzo}, {Falc{\'o}n-Barroso}, {Mast},
  {Garc{\'{\i}}a-Benito}, {Husemann}, \& {van de Ven}}]{Barrera-Ballesteros15}
{Barrera-Ballesteros}, J.~K., {S{\'a}nchez}, S.~F., {Garc{\'{\i}}a-Lorenzo},
  B., {et~al.} 2015, \aap, 579, A45

\bibitem[{{Barton Gillespie} {et~al.}(2003){Barton Gillespie}, {Geller}, \&
  {Kenyon}}]{Barton03}
{Barton Gillespie}, E., {Geller}, M.~J., \& {Kenyon}, S.~J. 2003, \apj, 582,
  668

\bibitem[{{Bibby} \& {Crowther}(2010)}]{Bibby10}
{Bibby}, J.~L. \& {Crowther}, P.~A. 2010, \mnras, 405, 2737

\bibitem[{{Bibby} \& {Crowther}(2012)}]{Bibby12}
{Bibby}, J.~L. \& {Crowther}, P.~A. 2012, \mnras, 420, 3091

\bibitem[{{Brinchmann} {et~al.}(2008{\natexlab{a}}){Brinchmann}, {Kunth}, \&
  {Durret}}]{Brinchmann08a}
{Brinchmann}, J., {Kunth}, D., \& {Durret}, F. 2008{\natexlab{a}}, \aap, 485,
  657

\bibitem[{{Brinchmann} {et~al.}(2008{\natexlab{b}}){Brinchmann}, {Pettini}, \&
  {Charlot}}]{Brinchmann08b}
{Brinchmann}, J., {Pettini}, M., \& {Charlot}, S. 2008{\natexlab{b}}, \mnras,
  385, 769

\bibitem[{{Cair{\'o}s} {et~al.}(2010){Cair{\'o}s}, {Caon}, {Zurita}, {Kehrig},
  {Roth}, \& {Weilbacher}}]{Cairos10}
{Cair{\'o}s}, L.~M., {Caon}, N., {Zurita}, C., {et~al.} 2010, \aap, 520, A90

\bibitem[{{Cardelli} {et~al.}(1989){Cardelli}, {Clayton}, \&
  {Mathis}}]{Cardelli89}
{Cardelli}, J.~A., {Clayton}, G.~C., \& {Mathis}, J.~S. 1989, \apj, 345, 245

\bibitem[{{Casasola} {et~al.}(2010){Casasola}, {Hunt}, {Combes},
  {Garc{\'{\i}}a-Burillo}, {Boone}, {Eckart}, {Neri}, \&
  {Schinnerer}}]{Casasola10}
{Casasola}, V., {Hunt}, L.~K., {Combes}, F., {et~al.} 2010, \aap, 510, A52

\bibitem[{{Castellanos} {et~al.}(2002){Castellanos}, {D{\'{\i}}az}, \&
  {Terlevich}}]{Castellanos02}
{Castellanos}, M., {D{\'{\i}}az}, A.~I., \& {Terlevich}, E. 2002, \mnras, 337,
  540

\bibitem[{{Cervi{\~n}o}(1998)}]{Cervinyo98}
{Cervi{\~n}o}, M. 1998, PhD thesis, , UMC, Spain, (1998)

\bibitem[{{Christensen} {et~al.}(2008){Christensen}, {Vreeswijk}, {Sollerman},
  {Th{\"o}ne}, {Le Floc'h}, \& {Wiersema}}]{Christensen08}
{Christensen}, L., {Vreeswijk}, P.~M., {Sollerman}, J., {et~al.} 2008, \aap,
  490, 45

\bibitem[{{Cid Fernandes} {et~al.}(2004){Cid Fernandes}, {Gu}, {Melnick},
  {Terlevich}, {Terlevich}, {Kunth}, {Rodrigues Lacerda}, \&
  {Joguet}}]{Cid-Fernandes04b}
{Cid Fernandes}, R., {Gu}, Q., {Melnick}, J., {et~al.} 2004, \mnras, 355, 273

\bibitem[{{Cid Fernandes} {et~al.}(2005){Cid Fernandes}, {Mateus}, {Sodr{\'e}},
  {Stasi{\'n}ska}, \& {Gomes}}]{Cid-Fernandes05}
{Cid Fernandes}, R., {Mateus}, A., {Sodr{\'e}}, L., {Stasi{\'n}ska}, G., \&
  {Gomes}, J.~M. 2005, \mnras, 358, 363

\bibitem[{{Cid Fernandes} {et~al.}(2011){Cid Fernandes}, {Stasi{\'n}ska},
  {Mateus}, \& {Vale Asari}}]{Cid-Fernandes11}
{Cid Fernandes}, R., {Stasi{\'n}ska}, G., {Mateus}, A., \& {Vale Asari}, N.
  2011, \mnras, 413, 1687

\bibitem[{{Comer{\'o}n} {et~al.}(2012){Comer{\'o}n}, {Elmegreen}, {Salo},
  {Laurikainen}, {Athanassoula}, {Bosma}, {Knapen}, {Gadotti}, {Sheth}, {Hinz},
  {Regan}, {Gil de Paz}, {Mu{\~n}oz-Mateos}, {Men{\'e}ndez-Delmestre},
  {Seibert}, {Kim}, {Mizusawa}, {Laine}, {Ho}, \& {Holwerda}}]{Comeron12}
{Comer{\'o}n}, S., {Elmegreen}, B.~G., {Salo}, H., {et~al.} 2012, \apj, 759, 98

\bibitem[{{Conti}(1991)}]{Conti91}
{Conti}, P.~S. 1991, \apj, 377, 115

\bibitem[{{Conti} \& {Massey}(1989)}]{Conti89}
{Conti}, P.~S. \& {Massey}, P. 1989, \apj, 337, 251

\bibitem[{{Contini} {et~al.}(2002){Contini}, {Treyer}, {Sullivan}, \&
  {Ellis}}]{Contini02}
{Contini}, T., {Treyer}, M.~A., {Sullivan}, M., \& {Ellis}, R.~S. 2002, \mnras,
  330, 75

\bibitem[{{Crowther}(2007)}]{Crowther07}
{Crowther}, P.~A. 2007, \araa, 45, 177

\bibitem[{{Crowther} \& {Hadfield}(2006)}]{Crowther06a}
{Crowther}, P.~A. \& {Hadfield}, L.~J. 2006, \aap, 449, 711

\bibitem[{{Crowther} {et~al.}(2006){Crowther}, {Hadfield}, {Clark},
  {Negueruela}, \& {Vacca}}]{Crowther06b}
{Crowther}, P.~A., {Hadfield}, L.~J., {Clark}, J.~S., {Negueruela}, I., \&
  {Vacca}, W.~D. 2006, \mnras, 372, 1407

\bibitem[{{Cutri} {et~al.}(1984){Cutri}, {Rieke}, {Tokunaga}, {Willner}, \&
  {Rudy}}]{Cutri84}
{Cutri}, R.~M., {Rieke}, G.~H., {Tokunaga}, A.~T., {Willner}, S.~P., \& {Rudy},
  R.~J. 1984, \apj, 280, 521

\bibitem[{{Dellenbusch} {et~al.}(2008){Dellenbusch}, {Gallagher}, {Knezek}, \&
  {Noble}}]{Dellenbusch08}
{Dellenbusch}, K.~E., {Gallagher}, III, J.~S., {Knezek}, P.~M., \& {Noble},
  A.~G. 2008, \aj, 135, 326

\bibitem[{{Di Matteo} {et~al.}(2008){Di Matteo}, {Bournaud}, {Martig},
  {Combes}, {Melchior}, \& {Semelin}}]{diMatteo08}
{Di Matteo}, P., {Bournaud}, F., {Martig}, M., {et~al.} 2008, \aap, 492, 31

\bibitem[{{Drissen} {et~al.}(2008){Drissen}, {Crowther}, {{\'U}beda}, \&
  {Martin}}]{Drissen08}
{Drissen}, L., {Crowther}, P.~A., {{\'U}beda}, L., \& {Martin}, P. 2008,
  \mnras, 389, 1033

\bibitem[{{Drissen} {et~al.}(1990){Drissen}, {Moffat}, \& {Shara}}]{Drissen90}
{Drissen}, L., {Moffat}, A.~F.~J., \& {Shara}, M.~M. 1990, \apj, 364, 496

\bibitem[{{Drissen} {et~al.}(1993{\natexlab{a}}){Drissen}, {Moffat}, \&
  {Shara}}]{Drissen93a}
{Drissen}, L., {Moffat}, A.~F.~J., \& {Shara}, M.~M. 1993{\natexlab{a}}, \aj,
  105, 1400

\bibitem[{{Drissen} {et~al.}(1993{\natexlab{b}}){Drissen}, {Roy}, \&
  {Moffat}}]{Drissen93b}
{Drissen}, L., {Roy}, J.-R., \& {Moffat}, A.~F.~J. 1993{\natexlab{b}}, \aj,
  106, 1460

\bibitem[{{Eldridge} {et~al.}(2008){Eldridge}, {Izzard}, \&
  {Tout}}]{Eldridge08}
{Eldridge}, J.~J., {Izzard}, R.~G., \& {Tout}, C.~A. 2008, \mnras, 384, 1109

\bibitem[{{Eldridge} \& {Stanway}(2009)}]{Eldridge09}
{Eldridge}, J.~J. \& {Stanway}, E.~R. 2009, \mnras, 400, 1019

\bibitem[{{Ercolano} {et~al.}(2004){Ercolano}, {Wesson}, {Zhang}, {Barlow}, {De
  Marco}, {Rauch}, \& {Liu}}]{Ercolano04}
{Ercolano}, B., {Wesson}, R., {Zhang}, Y., {et~al.} 2004, \mnras, 354, 558

\bibitem[{{Falc{\'o}n-Barroso} {et~al.}(2011){Falc{\'o}n-Barroso},
  {S{\'a}nchez-Bl{\'a}zquez}, {Vazdekis}, {Ricciardelli}, {Cardiel}, {Cenarro},
  {Gorgas}, \& {Peletier}}]{Falcon-Barroso11}
{Falc{\'o}n-Barroso}, J., {S{\'a}nchez-Bl{\'a}zquez}, P., {Vazdekis}, A.,
  {et~al.} 2011, \aap, 532, A95

\bibitem[{{Ferland} {et~al.}(1998){Ferland}, {Korista}, {Verner}, {Ferguson},
  {Kingdon}, \& {Verner}}]{Ferland98}
{Ferland}, G.~J., {Korista}, K.~T., {Verner}, D.~A., {et~al.} 1998, \pasp, 110,
  761

\bibitem[{{Fruchter} {et~al.}(2006){Fruchter}, {Levan}, {Strolger},
  {Vreeswijk}, {Thorsett}, {Bersier}, {Burud}, {Castro Cer{\'o}n},
  {Castro-Tirado}, {Conselice}, {Dahlen}, {Ferguson}, {Fynbo}, {Garnavich}, \&
  {Gibbons}}]{Fruchter06}
{Fruchter}, A.~S., {Levan}, A.~J., {Strolger}, L., {et~al.} 2006, \nat, 441,
  463

\bibitem[{{Galama} {et~al.}(1998){Galama}, {Vreeswijk}, {van Paradijs},
  {Kouveliotou}, {Augusteijn}, {B{\"o}hnhardt}, {Brewer}, {Doublier},
  {Gonzalez}, {Leibundgut}, {Lidman}, \& {Hainaut}}]{Galama98}
{Galama}, T.~J., {Vreeswijk}, P.~M., {van Paradijs}, J., {et~al.} 1998, \nat,
  395, 670

\bibitem[{{Galbany} {et~al.}(2014){Galbany}, {Stanishev}, {Mour{\~a}o},
  {Rodrigues}, {Flores}, {Garc{\'{\i}}a-Benito}, {Mast}, {Mendoza},
  {S{\'a}nchez}, {Badenes}, \& {Barrera-Ballesteros}}]{Galbany14}
{Galbany}, L., {Stanishev}, V., {Mour{\~a}o}, A.~M., {et~al.} 2014, \aap, 572,
  A38

\bibitem[{{Garc{\'{\i}}a-Benito} {et~al.}(2010){Garc{\'{\i}}a-Benito},
  {D{\'{\i}}az}, {H{\"a}gele}, {P{\'e}rez-Montero}, {L{\'o}pez},
  {V{\'{\i}}lchez}, {P{\'e}rez}, {Terlevich}, {Terlevich}, \&
  {Rosa-Gonz{\'a}lez}}]{Garcia-Benito10}
{Garc{\'{\i}}a-Benito}, R., {D{\'{\i}}az}, A., {H{\"a}gele}, G.~F., {et~al.}
  2010, \mnras, 408, 2234

\bibitem[{{Garc{\'{\i}}a-Benito} {et~al.}(2014){Garc{\'{\i}}a-Benito},
  {Zibetti}, {S{\'a}nchez}, {Husemann}, {de Amorim}, {Castillo-Morales}, {Cid
  Fernandes}, {.~Ellis}, {Falc{\'o}n-Barroso}, {Galbany}, {Gil de Paz},
  {Gonz{\'a}lez Delgado}, {Lacerda}, {L{\'o}pez-Fernandez}, {de
  Lorenzo-C{\'a}ceres}, {Lyubenova}, {Marino}, \& {Mast}}]{Garcia-Benito14}
{Garc{\'{\i}}a-Benito}, R., {Zibetti}, S., {S{\'a}nchez}, S.~F., {et~al.} 2014,
  ArXiv e-prints

\bibitem[{{Garcia-Lorenzo} {et~al.}(2014){Garcia-Lorenzo}, {Marquez},
  {Barrera-Ballesteros}, {Masegosa}, {Husemann}, {Falc{\'o}n-Barroso},
  {Lyubenova}, {Sanchez}, {Walcher}, {Mast}, {Garcia-Benito}, \&
  {Mendez-Abreu}}]{Garcia-Lorenzo14}
{Garcia-Lorenzo}, B., {Marquez}, I., {Barrera-Ballesteros}, J.~K., {et~al.}
  2014, ArXiv e-prints

\bibitem[{{Geller} {et~al.}(2006){Geller}, {Kenyon}, {Barton}, {Jarrett}, \&
  {Kewley}}]{Geller06}
{Geller}, M.~J., {Kenyon}, S.~J., {Barton}, E.~J., {Jarrett}, T.~H., \&
  {Kewley}, L.~J. 2006, \aj, 132, 2243

\bibitem[{{Giuricin} {et~al.}(1994){Giuricin}, {Monaco}, {Mardirossian}, \&
  {Mezzetti}}]{Giuricin94}
{Giuricin}, G., {Monaco}, P., {Mardirossian}, F., \& {Mezzetti}, M. 1994, \apj,
  425, 450

\bibitem[{{Gonz{\'a}lez Delgado} {et~al.}(2005){Gonz{\'a}lez Delgado},
  {Cervi{\~n}o}, {Martins}, {Leitherer}, \& {Hauschildt}}]{Gonzalez-Delgado05}
{Gonz{\'a}lez Delgado}, R.~M., {Cervi{\~n}o}, M., {Martins}, L.~P.,
  {Leitherer}, C., \& {Hauschildt}, P.~H. 2005, \mnras, 357, 945

\bibitem[{{Gonz{\'a}lez Delgado} \& {P{\'e}rez}(1997)}]{Gonzalez-Delgado97}
{Gonz{\'a}lez Delgado}, R.~M. \& {P{\'e}rez}, E. 1997, \apjs, 108, 199

\bibitem[{{Gonzalez-Delgado} {et~al.}(1995){Gonzalez-Delgado}, {Perez}, {Diaz},
  {Garcia-Vargas}, {Terlevich}, \& {Vilchez}}]{Gonzalez-Delgado95}
{Gonzalez-Delgado}, R.~M., {Perez}, E., {Diaz}, A.~I., {et~al.} 1995, \apj,
  439, 604

\bibitem[{{Gonzalez-Delgado} {et~al.}(1994){Gonzalez-Delgado}, {Perez},
  {Tenorio-Tagle}, {Vilchez}, {Terlevich}, {Terlevich}, {Telles},
  {Rodriguez-Espinosa}, {Mas-Hesse}, {Garcia-Vargas}, {Diaz}, {Cepa}, \&
  {Castaneda}}]{Gonzalez-Delgado94}
{Gonzalez-Delgado}, R.~M., {Perez}, E., {Tenorio-Tagle}, G., {et~al.} 1994,
  \apj, 437, 239

\bibitem[{{Graham} \& {Fruchter}(2013)}]{Graham13}
{Graham}, J.~F. \& {Fruchter}, A.~S. 2013, \apj, 774, 119

\bibitem[{{Gu} {et~al.}(2006){Gu}, {Zhao}, {Shi}, {Peng}, \& {Luo}}]{Gu06}
{Gu}, Q., {Zhao}, Y., {Shi}, L., {Peng}, Z., \& {Luo}, X. 2006, \aj, 131, 806

\bibitem[{{Guseva} {et~al.}(2000){Guseva}, {Izotov}, \& {Thuan}}]{Guseva00}
{Guseva}, N.~G., {Izotov}, Y.~I., \& {Thuan}, T.~X. 2000, \apj, 531, 776

\bibitem[{{Habergham} {et~al.}(2012){Habergham}, {James}, \&
  {Anderson}}]{Habergham12}
{Habergham}, S.~M., {James}, P.~A., \& {Anderson}, J.~P. 2012, \mnras, 424,
  2841

\bibitem[{{Hadfield} \& {Crowther}(2006)}]{Hadfield06}
{Hadfield}, L.~J. \& {Crowther}, P.~A. 2006, \mnras, 368, 1822

\bibitem[{{Hadfield} \& {Crowther}(2007)}]{Hadfield07}
{Hadfield}, L.~J. \& {Crowther}, P.~A. 2007, \mnras, 381, 418

\bibitem[{{Hadfield} {et~al.}(2005){Hadfield}, {Crowther}, {Schild}, \&
  {Schmutz}}]{Hadfield05}
{Hadfield}, L.~J., {Crowther}, P.~A., {Schild}, H., \& {Schmutz}, W. 2005,
  \aap, 439, 265

\bibitem[{{Hainich} {et~al.}(2014){Hainich}, {R{\"u}hling}, {Todt}, {Oskinova},
  {Liermann}, {Gr{\"a}fener}, {Foellmi}, {Schnurr}, \& {Hamann}}]{Hainich14}
{Hainich}, R., {R{\"u}hling}, U., {Todt}, H., {et~al.} 2014, \aap, 565, A27

\bibitem[{{Hammer} {et~al.}(2006){Hammer}, {Flores}, {Schaerer},
  {Dessauges-Zavadsky}, {Le Floc'h}, \& {Puech}}]{Hammer06}
{Hammer}, F., {Flores}, H., {Schaerer}, D., {et~al.} 2006, \aap, 454, 103

\bibitem[{{Han} {et~al.}(2010){Han}, {Hammer}, {Liang}, {Flores}, {Rodrigues},
  {Hou}, \& {Wei}}]{Han10}
{Han}, X.~H., {Hammer}, F., {Liang}, Y.~C., {et~al.} 2010, \aap, 514, A24

\bibitem[{{Heckman} {et~al.}(1986){Heckman}, {Beckwith}, {Blitz}, {Skrutskie},
  \& {Wilson}}]{Heckman86}
{Heckman}, T.~M., {Beckwith}, S., {Blitz}, L., {Skrutskie}, M., \& {Wilson},
  A.~S. 1986, \apj, 305, 157

\bibitem[{{Heckman} {et~al.}(1997){Heckman}, {Gonz{\'a}lez-Delgado},
  {Leitherer}, {Meurer}, {Krolik}, {Wilson}, {Koratkar}, \&
  {Kinney}}]{Heckman97}
{Heckman}, T.~M., {Gonz{\'a}lez-Delgado}, R., {Leitherer}, C., {et~al.} 1997,
  \apj, 482, 114

\bibitem[{{Hjorth} {et~al.}(2003){Hjorth}, {Sollerman}, {M{\o}ller}, {Fynbo},
  {Woosley}, {Kouveliotou}, {Tanvir}, {Greiner}, {Andersen}, {Castro-Tirado},
  {Castro Cer{\'o}n}, \& {Fruchter}}]{Hjorth03}
{Hjorth}, J., {Sollerman}, J., {M{\o}ller}, P., {et~al.} 2003, \nat, 423, 847

\bibitem[{{Ho} {et~al.}(1995){Ho}, {Filippenko}, \& {Sargent}}]{Ho95}
{Ho}, L.~C., {Filippenko}, A.~V., \& {Sargent}, W.~L. 1995, \apjs, 98, 477

\bibitem[{{Hunt} \& {Hirashita}(2009)}]{Hunt09}
{Hunt}, L.~K. \& {Hirashita}, H. 2009, \aap, 507, 1327

\bibitem[{{Husemann} {et~al.}(2013){Husemann}, {Jahnke}, {S{\'a}nchez},
  {Barrado}, {Bekerait*error*{\.e}}, {Bomans}, {Castillo-Morales},
  {Catal{\'a}n-Torrecilla}, {Cid Fernandes}, {Falc{\'o}n-Barroso},
  {Garc{\'{\i}}a-Benito}, \& {Gonz{\'a}lez Delgado}}]{Husemann13}
{Husemann}, B., {Jahnke}, K., {S{\'a}nchez}, S.~F., {et~al.} 2013, \aap, 549,
  A87

\bibitem[{{Izotov} \& {Thuan}(1998)}]{Izotov98}
{Izotov}, Y.~I. \& {Thuan}, T.~X. 1998, \apj, 500, 188

\bibitem[{{Izotov} {et~al.}(1997){Izotov}, {Thuan}, \& {Lipovetsky}}]{Izotov97}
{Izotov}, Y.~I., {Thuan}, T.~X., \& {Lipovetsky}, V.~A. 1997, \apjs, 108, 1

\bibitem[{{James} {et~al.}(2009){James}, {Tsamis}, {Barlow}, {Westmoquette},
  {Walsh}, {Cuisinier}, \& {Exter}}]{James09}
{James}, B.~L., {Tsamis}, Y.~G., {Barlow}, M.~J., {et~al.} 2009, \mnras, 398, 2

\bibitem[{{Karthick} {et~al.}(2014){Karthick}, {L{\'o}pez-S{\'a}nchez}, {Sahu},
  {Sanwal}, \& {Bisht}}]{Karthick14}
{Karthick}, M.~C., {L{\'o}pez-S{\'a}nchez}, {\'A}.~R., {Sahu}, D.~K., {Sanwal},
  B.~B., \& {Bisht}, S. 2014, \mnras, 439, 157

\bibitem[{{Kauffmann} {et~al.}(2003){Kauffmann}, {Heckman}, {Tremonti},
  {Brinchmann}, {Charlot}, {White}, {Ridgway}, {Brinkmann}, {Fukugita}, {Hall},
  {Ivezi{\'c}}, {Richards}, \& {Schneider}}]{Kauffmann03}
{Kauffmann}, G., {Heckman}, T.~M., {Tremonti}, C., {et~al.} 2003, \mnras, 346,
  1055

\bibitem[{{Kehrig} {et~al.}(2013){Kehrig}, {P{\'e}rez-Montero},
  {V{\'{\i}}lchez}, {Brinchmann}, {Kunth}, {Garc{\'{\i}}a-Benito}, {Crowther},
  {Hern{\'a}ndez-Fern{\'a}ndez}, {Durret}, {Contini},
  {Fern{\'a}ndez-Mart{\'{\i}}n}, \& {James}}]{Kehrig13}
{Kehrig}, C., {P{\'e}rez-Montero}, E., {V{\'{\i}}lchez}, J.~M., {et~al.} 2013,
  \mnras, 432, 2731

\bibitem[{{Kehrig} {et~al.}(2008){Kehrig}, {V{\'{\i}}lchez}, {S{\'a}nchez},
  {Telles}, {P{\'e}rez-Montero}, \& {Mart{\'{\i}}n-Gord{\'o}n}}]{Kehrig08}
{Kehrig}, C., {V{\'{\i}}lchez}, J.~M., {S{\'a}nchez}, S.~F., {et~al.} 2008,
  \aap, 477, 813

\bibitem[{{Kelly} {et~al.}(2008){Kelly}, {Kirshner}, \& {Pahre}}]{Kelly08}
{Kelly}, P.~L., {Kirshner}, R.~P., \& {Pahre}, M. 2008, \apj, 687, 1201

\bibitem[{{Kelz} {et~al.}(2006){Kelz}, {Verheijen}, {Roth}, {Bauer}, {Becker},
  {Paschke}, {Popow}, {S{\'a}nchez}, \& {Laux}}]{Kelz06}
{Kelz}, A., {Verheijen}, M.~A.~W., {Roth}, M.~M., {et~al.} 2006, \pasp, 118,
  129

\bibitem[{{Kennicutt}(1984)}]{Kennicutt84}
{Kennicutt}, Jr., R.~C. 1984, \apj, 287, 116

\bibitem[{{Kennicutt} {et~al.}(1987){Kennicutt}, {Roettiger}, {Keel}, {van der
  Hulst}, \& {Hummel}}]{Kennicutt87}
{Kennicutt}, Jr., R.~C., {Roettiger}, K.~A., {Keel}, W.~C., {van der Hulst},
  J.~M., \& {Hummel}, E. 1987, \aj, 93, 1011

\bibitem[{{Kewley} {et~al.}(2001){Kewley}, {Dopita}, {Sutherland}, {Heisler},
  \& {Trevena}}]{Kewley01a}
{Kewley}, L.~J., {Dopita}, M.~A., {Sutherland}, R.~S., {Heisler}, C.~A., \&
  {Trevena}, J. 2001, \apj, 556, 121

\bibitem[{{Kiminki} \& {Kobulnicky}(2012)}]{Kiminki12}
{Kiminki}, D.~C. \& {Kobulnicky}, H.~A. 2012, \apj, 751, 4

\bibitem[{{Kobulnicky} \& {Fryer}(2007)}]{Kobulnicky07}
{Kobulnicky}, H.~A. \& {Fryer}, C.~L. 2007, \apj, 670, 747

\bibitem[{{Kobulnicky} \& {Zaritsky}(1999)}]{Kobulnicky99a}
{Kobulnicky}, H.~A. \& {Zaritsky}, D. 1999, \apj, 511, 118

\bibitem[{{Kudritzki}(2002)}]{Kudritzki02}
{Kudritzki}, R.~P. 2002, \apj, 577, 389

\bibitem[{{Kunth} \& {Joubert}(1985)}]{kunth85}
{Kunth}, D. \& {Joubert}, M. 1985, \aap, 142, 411

\bibitem[{{Kunth} \& {Sargent}(1981)}]{Kunth81}
{Kunth}, D. \& {Sargent}, W.~L.~W. 1981, \aap, 101, L5

\bibitem[{{Kunth} \& {Sargent}(1983)}]{Kunth83}
{Kunth}, D. \& {Sargent}, W.~L.~W. 1983, \apj, 273, 81

\bibitem[{{Lamareille} {et~al.}(2004){Lamareille}, {Mouhcine}, {Contini},
  {Lewis}, \& {Maddox}}]{Lamareille04}
{Lamareille}, F., {Mouhcine}, M., {Contini}, T., {Lewis}, I., \& {Maddox}, S.
  2004, \mnras, 350, 396

\bibitem[{{Leloudas} {et~al.}(2010){Leloudas}, {Sollerman}, {Levan}, {Fynbo},
  {Malesani}, \& {Maund}}]{Leloudas10}
{Leloudas}, G., {Sollerman}, J., {Levan}, A.~J., {et~al.} 2010, \aap, 518, A29

\bibitem[{{Levesque} {et~al.}(2010{\natexlab{a}}){Levesque}, {Berger},
  {Kewley}, \& {Bagley}}]{Levesque10a}
{Levesque}, E.~M., {Berger}, E., {Kewley}, L.~J., \& {Bagley}, M.~M.
  2010{\natexlab{a}}, \aj, 139, 694

\bibitem[{{Levesque} {et~al.}(2011){Levesque}, {Berger}, {Soderberg}, \&
  {Chornock}}]{Levesque11}
{Levesque}, E.~M., {Berger}, E., {Soderberg}, A.~M., \& {Chornock}, R. 2011,
  \apj, 739, 23

\bibitem[{{Levesque} {et~al.}(2010{\natexlab{b}}){Levesque}, {Kewley},
  {Berger}, \& {Zahid}}]{Levesque10b}
{Levesque}, E.~M., {Kewley}, L.~J., {Berger}, E., \& {Zahid}, H.~J.
  2010{\natexlab{b}}, \aj, 140, 1557

\bibitem[{{Lopez} {et~al.}(2011){Lopez}, {Krumholz}, {Bolatto}, {Prochaska}, \&
  {Ramirez-Ruiz}}]{Lopez11}
{Lopez}, L.~A., {Krumholz}, M.~R., {Bolatto}, A.~D., {Prochaska}, J.~X., \&
  {Ramirez-Ruiz}, E. 2011, \apj, 731, 91

\bibitem[{{L{\'o}pez-S{\'a}nchez}(2010)}]{Lopez-Sanchez10c}
{L{\'o}pez-S{\'a}nchez}, {\'A}.~R. 2010, \aap, 521, A63

\bibitem[{{L{\'o}pez-S{\'a}nchez} {et~al.}(2012){L{\'o}pez-S{\'a}nchez},
  {Dopita}, {Kewley}, {Zahid}, {Nicholls}, \&
  {Scharw{\"a}chter}}]{Lopez-Sanchez12}
{L{\'o}pez-S{\'a}nchez}, {\'A}.~R., {Dopita}, M.~A., {Kewley}, L.~J., {et~al.}
  2012, \mnras, 426, 2630

\bibitem[{{L{\'o}pez-S{\'a}nchez} \& {Esteban}(2010)}]{Lopez-Sanchez10a}
{L{\'o}pez-S{\'a}nchez}, {\'A}.~R. \& {Esteban}, C. 2010, \aap, 516, A104

\bibitem[{{L{\'o}pez-S{\'a}nchez} {et~al.}(2011){L{\'o}pez-S{\'a}nchez},
  {Mesa-Delgado}, {L{\'o}pez-Mart{\'{\i}}n}, \& {Esteban}}]{Lopez-Sanchez11}
{L{\'o}pez-S{\'a}nchez}, {\'A}.~R., {Mesa-Delgado}, A.,
  {L{\'o}pez-Mart{\'{\i}}n}, L., \& {Esteban}, C. 2011, \mnras, 411, 2076

\bibitem[{{Marino} {et~al.}(2013){Marino}, {Rosales-Ortega}, {S{\'a}nchez},
  {Gil de Paz}, {V{\'{\i}}lchez}, {Miralles-Caballero}, {Kehrig},
  {P{\'e}rez-Montero}, {Stanishev}, {Iglesias-P{\'a}ramo}, \&
  {D{\'{\i}}az}}]{Marino13}
{Marino}, R.~A., {Rosales-Ortega}, F.~F., {S{\'a}nchez}, S.~F., {et~al.} 2013,
  \aap, 559, A114

\bibitem[{{Mart{\'{\i}}n-Manj{\'o}n} {et~al.}(2010){Mart{\'{\i}}n-Manj{\'o}n},
  {Garc{\'{\i}}a-Vargas}, {Moll{\'a}}, \& {D{\'{\i}}az}}]{Martin-Manjon10}
{Mart{\'{\i}}n-Manj{\'o}n}, M.~L., {Garc{\'{\i}}a-Vargas}, M.~L., {Moll{\'a}},
  M., \& {D{\'{\i}}az}, A.~I. 2010, \mnras, 403, 2012

\bibitem[{{Massey}(2003)}]{Massey03}
{Massey}, P. 2003, \araa, 41, 15

\bibitem[{{Massey} {et~al.}(2004){Massey}, {Bresolin}, {Kudritzki}, {Puls}, \&
  {Pauldrach}}]{Massey04}
{Massey}, P., {Bresolin}, F., {Kudritzki}, R.~P., {Puls}, J., \& {Pauldrach},
  A.~W.~A. 2004, \apj, 608, 1001

\bibitem[{{Massey} \& {Hunter}(1998)}]{Massey98}
{Massey}, P. \& {Hunter}, D.~A. 1998, \apj, 493, 180

\bibitem[{{Meynet}(1995)}]{Meynet95}
{Meynet}, G. 1995, \aap, 298, 767

\bibitem[{{Meynet} \& {Maeder}(2005)}]{Meynet05}
{Meynet}, G. \& {Maeder}, A. 2005, \aap, 429, 581

\bibitem[{{Miller} \& {Rudie}(2008)}]{Miller08}
{Miller}, B.~W. \& {Rudie}, G. 2008, in IAU Symposium, Vol. 245, IAU Symposium,
  ed. M.~{Bureau}, E.~{Athanassoula}, \& B.~{Barbuy}, 311--312

\bibitem[{{Miralles-Caballero} {et~al.}(2011){Miralles-Caballero}, {Colina},
  {Arribas}, \& {Duc}}]{Miralles-Caballero11}
{Miralles-Caballero}, D., {Colina}, L., {Arribas}, S., \& {Duc}, P.-A. 2011,
  \aj, 142, 79

\bibitem[{{Miralles-Caballero}
  {et~al.}(2014{\natexlab{a}}){Miralles-Caballero}, {D{\'{\i}}az},
  {Rosales-Ortega}, {P{\'e}rez-Montero}, \&
  {S{\'a}nchez}}]{Miralles-Caballero14}
{Miralles-Caballero}, D., {D{\'{\i}}az}, A.~I., {Rosales-Ortega}, F.~F.,
  {P{\'e}rez-Montero}, E., \& {S{\'a}nchez}, S.~F. 2014{\natexlab{a}}, \mnras,
  440, 2265

\bibitem[{{Miralles-Caballero}
  {et~al.}(2014{\natexlab{b}}){Miralles-Caballero}, {Rosales-Ortega},
  {D{\'{\i}}az}, {Ot{\'{\i}}-Floranes}, {P{\'e}rez-Montero}, \&
  {S{\'a}nchez}}]{Miralles-Caballero14b}
{Miralles-Caballero}, D., {Rosales-Ortega}, F.~F., {D{\'{\i}}az}, A.~I.,
  {et~al.} 2014{\natexlab{b}}, \mnras, 445, 3803

\bibitem[{{Modjaz} {et~al.}(2011){Modjaz}, {Kewley}, {Bloom}, {Filippenko},
  {Perley}, \& {Silverman}}]{Modjaz11}
{Modjaz}, M., {Kewley}, L., {Bloom}, J.~S., {et~al.} 2011, \apjl, 731, L4

\bibitem[{{Modjaz} {et~al.}(2006){Modjaz}, {Stanek}, {Garnavich}, {Berlind},
  {Blondin}, {Brown}, {Calkins}, {Challis}, \& {Diamond-Stanic}}]{Modjaz06}
{Modjaz}, M., {Stanek}, K.~Z., {Garnavich}, P.~M., {et~al.} 2006, \apjl, 645,
  L21

\bibitem[{{Moll{\'a}} {et~al.}(2009){Moll{\'a}}, {Garc{\'{\i}}a-Vargas}, \&
  {Bressan}}]{Molla09}
{Moll{\'a}}, M., {Garc{\'{\i}}a-Vargas}, M.~L., \& {Bressan}, A. 2009, \mnras,
  398, 451

\bibitem[{{Monreal-Ibero} {et~al.}(2010){Monreal-Ibero}, {V{\'{\i}}lchez},
  {Walsh}, \& {Mu{\~n}oz-Tu{\~n}{\'o}n}}]{Monreal10}
{Monreal-Ibero}, A., {V{\'{\i}}lchez}, J.~M., {Walsh}, J.~R., \&
  {Mu{\~n}oz-Tu{\~n}{\'o}n}, C. 2010, \aap, 517, A27+

\bibitem[{{Monreal-Ibero} {et~al.}(2012){Monreal-Ibero}, {Walsh}, \&
  {V{\'{\i}}lchez}}]{Monreal12}
{Monreal-Ibero}, A., {Walsh}, J.~R., \& {V{\'{\i}}lchez}, J.~M. 2012, \aap,
  544, A60

\bibitem[{{Monreal-Ibero} {et~al.}(2013){Monreal-Ibero}, {Walsh},
  {Westmoquette}, \& {V{\'{\i}}lchez}}]{Monreal13}
{Monreal-Ibero}, A., {Walsh}, J.~R., {Westmoquette}, M.~S., \&
  {V{\'{\i}}lchez}, J.~M. 2013, \aap, 553, A57

\bibitem[{{Neugent} \& {Massey}(2011)}]{Neugent11}
{Neugent}, K.~F. \& {Massey}, P. 2011, \apj, 733, 123

\bibitem[{{Neugent} {et~al.}(2012){Neugent}, {Massey}, \& {Georgy}}]{Neugent12}
{Neugent}, K.~F., {Massey}, P., \& {Georgy}, C. 2012, \apj, 759, 11

\bibitem[{{Osterbrock}(1978)}]{Osterbrock78}
{Osterbrock}, D.~E. 1978, Proceedings of the National Academy of Science, 75,
  540

\bibitem[{{Osterbrock}(1989)}]{Osterbrock89}
{Osterbrock}, D.~E. 1989, {Astrophysics of gaseous nebulae and active galactic
  nuclei}, ed. {Osterbrock, D.~E.}

\bibitem[{{Osterbrock} \& {Cohen}(1982)}]{Osterbrock82}
{Osterbrock}, D.~E. \& {Cohen}, R.~D. 1982, \apj, 261, 64

\bibitem[{{P{\'e}rez-Montero} \& {D{\'{\i}}az}(2007)}]{Perez-Montero07b}
{P{\'e}rez-Montero}, E. \& {D{\'{\i}}az}, {\'A}.~I. 2007, \mnras, 377, 1195

\bibitem[{{P{\'e}rez-Montero} {et~al.}(2010){P{\'e}rez-Montero},
  {Garc{\'{\i}}a-Benito}, {H{\"a}gele}, \& {D{\'{\i}}az}}]{Perez-Montero10}
{P{\'e}rez-Montero}, E., {Garc{\'{\i}}a-Benito}, R., {H{\"a}gele}, G.~F., \&
  {D{\'{\i}}az}, {\'A}.~I. 2010, \mnras, 404, 2037

\bibitem[{{Perez-Montero} {et~al.}(2013){Perez-Montero}, {Kehrig},
  {Brinchmann}, {Vilchez}, {Kunth}, \& {Durret}}]{Perez-Montero13}
{Perez-Montero}, E., {Kehrig}, C., {Brinchmann}, J., {et~al.} 2013, ArXiv
  e-prints

\bibitem[{{Pilyugin} {et~al.}(2012){Pilyugin}, {Grebel}, \&
  {Mattsson}}]{Pilyugin12b}
{Pilyugin}, L.~S., {Grebel}, E.~K., \& {Mattsson}, L. 2012, \mnras, 424, 2316

\bibitem[{{Plauchu-Frayn} {et~al.}(2012){Plauchu-Frayn}, {Del Olmo}, {Coziol},
  \& {Torres-Papaqui}}]{Plauchu-Frayn12}
{Plauchu-Frayn}, I., {Del Olmo}, A., {Coziol}, R., \& {Torres-Papaqui}, J.~P.
  2012, \aap, 546, A48

\bibitem[{{Rodr{\'{\i}}guez Zaur{\'{\i}}n} {et~al.}(2010){Rodr{\'{\i}}guez
  Zaur{\'{\i}}n}, {Tadhunter}, \& {Gonz{\'a}lez Delgado}}]{Rodriguez-Zaurin10}
{Rodr{\'{\i}}guez Zaur{\'{\i}}n}, J., {Tadhunter}, C.~N., \& {Gonz{\'a}lez
  Delgado}, R.~M. 2010, \mnras, 403, 1317

\bibitem[{{Rosales-Ortega} {et~al.}(2010){Rosales-Ortega}, {Kennicutt},
  {S{\'a}nchez}, {D{\'{\i}}az}, {Pasquali}, {Johnson}, \&
  {Hao}}]{Rosales-Ortega10}
{Rosales-Ortega}, F.~F., {Kennicutt}, R.~C., {S{\'a}nchez}, S.~F., {et~al.}
  2010, \mnras, 405, 735

\bibitem[{{Roth} {et~al.}(2005){Roth}, {Kelz}, {Fechner}, {Hahn}, {Bauer},
  {Becker}, {B{\"o}hm}, {Christensen}, {Dionies}, {Paschke}, {Popow}, {Wolter},
  {Schmoll}, {Laux}, \& {Altmann}}]{Roth05}
{Roth}, M.~M., {Kelz}, A., {Fechner}, T., {et~al.} 2005, \pasp, 117, 620

\bibitem[{{Rupke} {et~al.}(2010){Rupke}, {Kewley}, \& {Barnes}}]{Rupke10a}
{Rupke}, D.~S.~N., {Kewley}, L.~J., \& {Barnes}, J.~E. 2010, \apjl, 710, L156

\bibitem[{{Sana} {et~al.}(2012){Sana}, {de Mink}, {de Koter}, {Langer},
  {Evans}, {Gieles}, {Gosset}, {Izzard}, {Le Bouquin}, \& {Schneider}}]{Sana12}
{Sana}, H., {de Mink}, S.~E., {de Koter}, A., {et~al.} 2012, Science, 337, 444

\bibitem[{{S{\'a}nchez} {et~al.}(2012{\natexlab{a}}){S{\'a}nchez}, {Kennicutt},
  {Gil de Paz}, {van de Ven}, {V{\'{\i}}lchez}, {Wisotzki}, {Walcher}, {Mast},
  {Aguerri}, {Albiol-P{\'e}rez}, {Alonso-Herrero}, {Alves}, {Bakos}, \&
  {Bart{\'a}kov{\'a}}}]{Sanchez12a}
{S{\'a}nchez}, S.~F., {Kennicutt}, R.~C., {Gil de Paz}, A., {et~al.}
  2012{\natexlab{a}}, \aap, 538, A8

\bibitem[{{S{\'a}nchez} {et~al.}(2014){S{\'a}nchez}, {Rosales-Ortega},
  {Iglesias-P{\'a}ramo}, {Moll{\'a}}, {Barrera-Ballesteros}, {Marino},
  {P{\'e}rez}, {S{\'a}nchez-Blazquez}, \& {Gonz{\'a}lez Delgado}}]{Sanchez14}
{S{\'a}nchez}, S.~F., {Rosales-Ortega}, F.~F., {Iglesias-P{\'a}ramo}, J.,
  {et~al.} 2014, \aap, 563, A49

\bibitem[{{S{\'a}nchez} {et~al.}(2012{\natexlab{b}}){S{\'a}nchez},
  {Rosales-Ortega}, {Marino}, {Iglesias-P{\'a}ramo}, {V{\'{\i}}lchez},
  {Kennicutt}, {D{\'{\i}}az}, \& {Mast}}]{Sanchez12b}
{S{\'a}nchez}, S.~F., {Rosales-Ortega}, F.~F., {Marino}, R.~A., {et~al.}
  2012{\natexlab{b}}, \aap, 546, A2

\bibitem[{{S{\'a}nchez-Portal} {et~al.}(2000){S{\'a}nchez-Portal},
  {D{\'{\i}}az}, {Terlevich}, {Terlevich}, {{\'A}lvarez {\'A}lvarez}, \&
  {Aretxaga}}]{Sanchez-Portal00}
{S{\'a}nchez-Portal}, M., {D{\'{\i}}az}, {\'A}.~I., {Terlevich}, R., {et~al.}
  2000, \mnras, 312, 2

\bibitem[{{Sander} {et~al.}(2014){Sander}, {Todt}, {Hainich}, \&
  {Hamann}}]{Sander14}
{Sander}, A., {Todt}, H., {Hainich}, R., \& {Hamann}, W.-R. 2014, \aap, 563,
  A89

\bibitem[{{Sargent} \& {Filippenko}(1991)}]{Sargent91}
{Sargent}, W.~L.~W. \& {Filippenko}, A.~V. 1991, \aj, 102, 107

\bibitem[{{Schaerer} {et~al.}(1999){Schaerer}, {Contini}, \&
  {Kunth}}]{Schaerer99}
{Schaerer}, D., {Contini}, T., \& {Kunth}, D. 1999, \aap, 341, 399

\bibitem[{{Schaerer} \& {Vacca}(1998)}]{Schaerer98}
{Schaerer}, D. \& {Vacca}, W.~D. 1998, \apj, 497, 618

\bibitem[{{Schild} {et~al.}(2003){Schild}, {Crowther}, {Abbott}, \&
  {Schmutz}}]{Schild03}
{Schild}, H., {Crowther}, P.~A., {Abbott}, J.~B., \& {Schmutz}, W. 2003, \aap,
  397, 859

\bibitem[{{Shara} {et~al.}(2013){Shara}, {Bibby}, {Zurek}, {Crowther},
  {Moffat}, \& {Drissen}}]{Shara13}
{Shara}, M.~M., {Bibby}, J.~L., {Zurek}, D., {et~al.} 2013, \aj, 146, 162

\bibitem[{{Shi} {et~al.}(2005){Shi}, {Kong}, {Li}, \& {Cheng}}]{Shi05}
{Shi}, F., {Kong}, X., {Li}, C., \& {Cheng}, F.~Z. 2005, \aap, 437, 849

\bibitem[{{Shirazi} \& {Brinchmann}(2012)}]{Shirazi12}
{Shirazi}, M. \& {Brinchmann}, J. 2012, \mnras, 421, 1043

\bibitem[{{Sidoli} {et~al.}(2006){Sidoli}, {Smith}, \& {Crowther}}]{Sidoli06}
{Sidoli}, F., {Smith}, L.~J., \& {Crowther}, P.~A. 2006, \mnras, 370, 799

\bibitem[{{Smith} {et~al.}(1990){Smith}, {Shara}, \& {Moffat}}]{Smith90}
{Smith}, L.~F., {Shara}, M.~M., \& {Moffat}, A.~F.~J. 1990, \apj, 348, 471

\bibitem[{{Smith} \& {Willis}(1982)}]{Smith82}
{Smith}, L.~J. \& {Willis}, A.~J. 1982, \mnras, 201, 451

\bibitem[{{Stanek} {et~al.}(2003){Stanek}, {Matheson}, {Garnavich}, {Martini},
  {Berlind}, {Caldwell}, {Challis}, {Brown}, {Schild}, {Krisciunas}, {Calkins},
  {Lee}, {Hathi}, {Jansen}, {Windhorst}, {Echevarria}, {Eisenstein}, {Pindor},
  {Olszewski}, {Harding}, {Holland}, \& {Bersier}}]{Stanek03}
{Stanek}, K.~Z., {Matheson}, T., {Garnavich}, P.~M., {et~al.} 2003, \apjl, 591,
  L17

\bibitem[{{Th{\"o}ne} {et~al.}(2014){Th{\"o}ne}, {Christensen}, {Prochaska},
  {Bloom}, {Gorosabel}, {Fynbo}, {Jakobsson}, \& {Fruchter}}]{Thone14}
{Th{\"o}ne}, C.~C., {Christensen}, L., {Prochaska}, J.~X., {et~al.} 2014,
  \mnras, 441, 2034

\bibitem[{{Tremonti} {et~al.}(2004){Tremonti}, {Heckman}, {Kauffmann},
  {Brinchmann}, {Charlot}, {White}, {Seibert}, {Peng}, {Schlegel}, {Uomoto},
  {Fukugita}, \& {Brinkmann}}]{Tremonti04}
{Tremonti}, C.~A., {Heckman}, T.~M., {Kauffmann}, G., {et~al.} 2004, \apj, 613,
  898

\bibitem[{{Tresse} {et~al.}(1999){Tresse}, {Maddox}, {Loveday}, \&
  {Singleton}}]{Tresse99}
{Tresse}, L., {Maddox}, S., {Loveday}, J., \& {Singleton}, C. 1999, \mnras,
  310, 262

\bibitem[{{{\'U}beda} \& {Drissen}(2009)}]{Ubeda09}
{{\'U}beda}, L. \& {Drissen}, L. 2009, \mnras, 394, 1847

\bibitem[{{Vacca}(1994)}]{Vacca94}
{Vacca}, W.~D. 1994, \apj, 421, 140

\bibitem[{{Vacca} \& {Conti}(1992)}]{Vacca92}
{Vacca}, W.~D. \& {Conti}, P.~S. 1992, \apj, 401, 543

\bibitem[{{Van Bever} \& {Vanbeveren}(2003)}]{vanBever03}
{Van Bever}, J. \& {Vanbeveren}, D. 2003, \aap, 400, 63

\bibitem[{{van Bever} \& {Vanbeveren}(2007)}]{vanBever07}
{van Bever}, J. \& {Vanbeveren}, D. 2007, in Astronomical Society of the
  Pacific Conference Series, Vol. 367, Massive Stars in Interactive Binaries,
  ed. N.~{St.-Louis} \& A.~F.~J. {Moffat}, 579

\bibitem[{{van der Hucht}(2001)}]{Vanderhucht01}
{van der Hucht}, K.~A. 2001, VizieR Online Data Catalog, 3215, 0

\bibitem[{{Vazdekis} {et~al.}(2010){Vazdekis}, {S{\'a}nchez-Bl{\'a}zquez},
  {Falc{\'o}n-Barroso}, {Cenarro}, {Beasley}, {Cardiel}, {Gorgas}, \&
  {Peletier}}]{Vazdekis10}
{Vazdekis}, A., {S{\'a}nchez-Bl{\'a}zquez}, P., {Falc{\'o}n-Barroso}, J.,
  {et~al.} 2010, \mnras, 404, 1639

\bibitem[{{Walcher} {et~al.}(2014){Walcher}, {Wisotzki}, {Bekerait{\'e}},
  {Husemann}, {Iglesias-P{\'a}ramo}, {Backsmann}, {Barrera Ballesteros},
  {Catal{\'a}n-Torrecilla}, \& {Cortijo}}]{Walcher14}
{Walcher}, C.~J., {Wisotzki}, L., {Bekerait{\'e}}, S., {et~al.} 2014, \aap,
  569, A1

\bibitem[{{Westmoquette} {et~al.}(2013){Westmoquette}, {James},
  {Monreal-Ibero}, \& {Walsh}}]{Westmoquette13}
{Westmoquette}, M.~S., {James}, B., {Monreal-Ibero}, A., \& {Walsh}, J.~R.
  2013, \aap, 550, A88

\bibitem[{{Wilson} {et~al.}(1986){Wilson}, {Baldwin}, {Sun}, \&
  {Wright}}]{Wilson86}
{Wilson}, A.~S., {Baldwin}, J.~A., {Sun}, S.-D., \& {Wright}, A.~E. 1986, \apj,
  310, 121

\bibitem[{{Woosley} \& {Heger}(2006)}]{Woosley06}
{Woosley}, S.~E. \& {Heger}, A. 2006, \apj, 637, 914

\bibitem[{{Youngblood} \& {Hunter}(1999)}]{Youngblood99}
{Youngblood}, A.~J. \& {Hunter}, D.~A. 1999, \apj, 519, 55

\bibitem[{{Zahid} {et~al.}(2011){Zahid}, {Kewley}, \& {Bresolin}}]{Zahid11}
{Zahid}, H.~J., {Kewley}, L.~J., \& {Bresolin}, F. 2011, \apj, 730, 137

\end{thebibliography}
